\documentclass[twocolumn]{aastex701}
\usepackage{tabularx}
\usepackage{amsmath}
\usepackage{amssymb}
\usepackage{longtable}
\usepackage{multirow}
\usepackage{bm}

\newcounter{photo}

\begin{document}

\title{The Open-Source \texttt{Photochem} Code: A General Chemical and Climate Model for Interpreting (Exo)Planet Observations}

\author[0000-0002-0413-3308]{Nicholas F. Wogan}
\affiliation{SETI Institute, Mountain View, CA 94043}
\affiliation{NASA Ames Research Center, Moffett Field, CA 94035}
\email{nicholas.f.wogan@nasa.gov}

\author[0000-0003-1240-6844]{Natasha E. Batalha}
\affiliation{NASA Ames Research Center, Moffett Field, CA 94035}
\email{natasha.e.batalha@nasa.gov}

\author{Kevin Zahnle}
\affiliation{NASA Ames Research Center, Moffett Field, CA 94035}
\email{kevin.j.zahnle@nasa.gov}


\author{Joshua Krissansen-Totton}
\affiliation{Dept. Earth and Space Scienes, University of Washington, Seattle, WA 98103}
\email{joshkt@uw.edu}

\author{David C. Catling}
\affiliation{Dept. Earth and Space Scienes, University of Washington, Seattle, WA 98103}
\email{dcatling@uw.edu}

\author[0000-0002-7188-1648]{Eric T. Wolf}
\affiliation{Laboratory for Atmospheric and Space Physics, University of Colorado Boulder, Boulder, CO, USA}
\affiliation{Blue Marble Space Institute of Science, Seattle, WA, USA}
\email{eric.wolf@colorado.edu}

\author[0000-0002-3196-414X]{Tyler D. Robinson}
\affiliation{Lunar \& Planetary Laboratory, University of Arizona, Tucson, AZ 85721 USA}
\affiliation{NASA Nexus for Exoplanet System Science Virtual Planetary Laboratory, University of Washington, Box 351580, Seattle, WA 98195, USA}
\email{tdrobin@arizona.edu}

\author{Victoria Meadows}
\affiliation{Dept. Astronomy, University of Washington, Seattle, WA 98103}
\email{meadows@uw.edu}

\author{Giada Arney}
\affiliation{NASA Goddard Space Flight Center, Greenbelt, MD 20771, USA}
\email{giada.n.arney@nasa.gov}

\author{Shawn Domagal-Goldman}
\affiliation{NASA Goddard Space Flight Center, Greenbelt, MD 20771, USA}
\email{shawn.goldman@nasa.gov}

\begin{abstract}
  With the launch of the James Webb Space Telescope, we are firmly in the era of exoplanet atmosphere characterization. Understanding exoplanet spectra requires atmospheric chemical and climate models that span the diversity of planetary atmospheres. Here, we present a more general chemical and climate model of planetary atmospheres. Specifically, we introduce the open-source, one-dimensional photochemical and climate code \texttt{Photochem}, and benchmark the model against the observed compositions and climates of Venus, Earth, Mars, Jupiter and Titan with a single set of kinetics, thermodynamics and opacities. We also model the chemistry of the hot Jupiter exoplanet WASP-39b. All simulations are open-source and reproducible. To first order, \texttt{Photochem} broadly reproduces the gas-phase chemistry and pressure-temperature profiles of all six planets. The largest model-data discrepancies are found in Venus's sulfur chemistry, motivating future experimental work on sulfur kinetics and spacecraft missions to Venus. We also find that clouds and hazes are important for the energy balance of Venus, Earth, Mars and Titan, and that accurately predicting aerosols with \texttt{Photochem} is challenging. Finally, we benchmark \texttt{Photochem} against the popular \texttt{VULCAN} and \texttt{HELIOS} photochemistry and climate models, finding excellent agreement for the same inputs; we also find that \texttt{Photochem} simulates atmospheres 2 to $\sim10^{2}\times$ more efficiently. These results show that \texttt{Photochem} provides a comparatively general description of atmospheric chemistry and physics that can be leveraged to study Solar System worlds or interpret telescope observations of exoplanets.
\end{abstract}

\section{Introduction}

In the current era of the James Webb Space Telescope (JWST), our understanding of exoplanets has come from reconciling spectra with atmospheric chemistry and climate models. Researchers routinely rely upon photochemical and climate simulations to interpret JWST detections of disequilibrium species like SO$_2$, CH$_4$ and CO$_2$ in hot Jupiter, sub-Neptune and brown dwarf atmospheres \citep[e.g.,][]{Tsai2023,Benneke2024,Beiler2024}. Models of planetary climate and coupled photochemical-climate models have also become integral to the search for atmospheres of warm rocky exoplanets orbiting M stars. The models predict the expected planetary thermal emission under different atmospheric compositions or thicknesses which are then used to contextualize JWST emission spectra and photometry \citep[e.g.,][]{Ih2023,Lincowski2023}.

The search for exoplanetary life with the next generation Extremely Large Telescopes (ELTs) and NASA's future Habitable Worlds Observatory (HWO) will also depend upon our knowledge of atmospheric chemistry and physics \citep{Catling2018, Schwieterman2018}. The life-detection strategy of these missions is predicated on the circumstellar habitable zone \citep{Kasting1993}, a theoretical concept that emerged in part from climate calculations of early Venus and Mars \citep{Kasting1988,Kasting1991}. Photochemistry can enhance \citep[e.g.,][]{Segura2005}, suppress \citep[e.g.,][]{DomagalGoldman2011} or mimic biosignatures \citep[e.g.,][and references therein]{Meadows2018}, and so ELT and HWO detections of CH$_4$ and O$_2$ will be significantly more compelling biosignatures if the photochemical context on the planet is well understood \citep[e.g.,][]{Currie2025}. For example, if photochemical calculations suggest the lifetime of CH$_4$ is short, this would imply a significant surface flux that is more credibly explained by metabolic methanogenesis, rather than geological processes \citep{Thompson2022}. Similarly, a detection of atmospheric O$_2$ can be made more robust by assessing and ruling out the likelihood that it is generated by abiotic photolyisis of H$_2$O and CO$_2$ \citep[e.g.,][]{Meadows2018}.


The photochemical and climate models used to interpret exoplanet observations should ideally capture the large possible diversity of planetary atmospheres. For some JWST sub-Neptune targets, we do not know whether the planets have primordial Jupiter-like atmospheres, a secondary atmosphere like Venus or Earth's, or no detectable atmosphere \citep[e.g., LHS 1140b;][]{Cadieux2024}. Temperate rocky worlds detected by HWO may have Venus-, Earth- or Mars-like atmospheres, or perhaps an entirely unanticipated composition. Evaluating spectra of unexplored worlds will require modeling tools that can reliably accommodate the wide range of possibilities.


Here we describe and validate a more general model of planetary atmospheres: the open-source 1-D photochemical and climate model \texttt{Photochem}. Using a single set of kinetics, thermodynamics and opacities, we benchmark \texttt{Photochem} against the observed compositions and climates of many Solar System atmospheres: Venus, Earth, Mars, Titan, and Jupiter. We also address the hot Jupiter exoplanet WASP-39b, which resides in a region of parameter space very different from what is found in the Solar System. Our goal is not to perfectly reproduce any single atmosphere. Instead, we aim to demonstrate a description of atmospheric chemistry and physics that broadly captures a wide diversity of planets to enable exoplanet research. A byproduct will be to shed light on poorly understood aspects of Solar System atmospheres, motivating the work needed to improve simulations. All of our calculations are open-source and reproducible to best facilitate future research in this area (\url{https://doi.org/10.5281/zenodo.16509802}). \texttt{Photochem} has previously been used to study the early Earth \citep{Wogan2023}, sub-Neptunes \citep{Wogan2024,Benneke2024,Schmidt2025}, hot Jupiters \citep{Mukherjee2024,Mukherjee2025}, and brown dwarfs \citep{Wogan2025,Beiler2024}. Despite these applications of the code, there is no paper that fully describes and benchmarks \texttt{Photochem}. This article also serves that purpose.

\section{Methods}

The \texttt{Photochem} software package contains three modules: 1-D photochemistry (Section \ref{sec:methods_photochemistry}), 1-D climate (Section \ref{sec:methods_climate}), and an equilibrium chemistry solver (Section \ref{sec:methods_equilibrium}). The model is written in Fortran 2008 and modern C and it has a Python interface enabled by Cython \citep{Behnel2011}. The source code is available on 
GitHub\footnote{\url{https://github.com/Nicholaswogan/photochem}} 
and Zenodo (Section \ref{sec:methods_reproduce}), and pre-built binaries can be installed from the Conda-Forge servers using the Conda package 
manager.\footnote{\url{https://conda-forge.org/}} 
In the following subsections we first describe the heritage of \texttt{Photochem} (Section \ref{sec:methods_history}), then detail the photochemical, climate and equilibrium chemistry modules (Sections \ref{sec:methods_photochemistry}--\ref{sec:methods_equilibrium}), and finally give our overall approach for benchmarking the code against Solar System atmospheres (Section \ref{sec:methods_approach}).

\subsection{History and Motivation} \label{sec:methods_history}

The photochemical and climate models in \texttt{Photochem} are motivated in part by the \texttt{Atmos} code. \texttt{Atmos} was originally developed to study the photochemistry and climate of early Earth, Venus, and Mars \citep{Kasting1979,Zahnle1986,Kasting1988,Kasting1991}, and more recently \texttt{Atmos} has been further developed, and used to model exoplanet atmospheres \citep[e.g.,][]{Kopparapu2013,Arney2016,Lincowski2018}. \citet{Wogan2022} made a Python wrapper to the photochemical model in \texttt{Atmos}, called \texttt{PhotochemPy}, which they used to study Earth's Great Oxidation Event. \texttt{Atmos} and \texttt{PhotochemPy} are written in Fortran 77 and have decades of technical debt, making them challenging to build upon. Therefore, \citet{Wogan2023} developed a new photochemical and climate model from the ground up: \texttt{Photochem}. \citet{Wogan2023} used \texttt{Photochem} version 0.3.14 and the present article describes version 0.6.7.

While \texttt{Photochem} is inspired by \texttt{Atmos}, the models now have fundamental differences many of which are listed below:
\begin{itemize}
    \item \texttt{Photochem} solves a different derivation of the 1-D photochemical equations than \texttt{Atmos} (Section \ref{sec:methods_photochemistry}). The \texttt{Atmos} derivation assumes a hard-coded N$_2$ background filler gas while \texttt{Photochem}'s implementation does not require a background gas to be specified.
    \item The \texttt{Atmos} photochemical model is effectively limited to temperate rocky planets, while \texttt{Photochem} can consider rocky planets and gas giants over a wide range of temperatures ($\sim 50$--2000 K).
    \item The chemical network shipped with \texttt{Photochem} is applicable to a wide range of atmospheric temperatures and compositions (Section \ref{sec:results}). \texttt{Atmos} has several separate chemical networks built for specific planets (e.g., the modern and Archean Earth).
    \item The photochemical model in \texttt{Photochem} can optionally reverse chemical reactions using thermodynamics, whereas \texttt{Atmos} cannot. Reversing chemical reactions is important for modeling warm planets ($\gtrsim 500$ K).
    \item The photochemical model in \texttt{Photochem} can accommodate reactions with an arbitrary number of reactants and products, while \texttt{Atmos} can only consider reactions of a specific format.
    \item The photochemical model in \texttt{Atmos} implements a centered finite volume scheme for particle transport and molecular diffusion which can be unstable because both of these processes involve advection. \texttt{Photochem} instead implements an upwind scheme to avoid these instabilities (Section \ref{sec:methods_photochemistry}).
    \item The photochemical model in \texttt{Photochem} uses the state-of-the-art CVODE integrator \citep{Hindmarsh2005} capable of accurately time-evolving atmospheres. \texttt{Atmos} uses a backward Euler method.
    \item The photochemical model in \texttt{Photochem} can have an arbitrary number of condensibles while \texttt{Atmos} has a limited number of hard-coded condensibles.
    \item The climate model in \texttt{Photochem} mixes k-distributions using the ``resort rebin'' method \citep{Amundsen2017} permitting the code to include many absorbers (e.g., 10) and still be computationally efficient. The climate model in \texttt{Atmos} does not implement a k-distributions mixing method, and therefore is limited to $\lesssim 4$ absorbers. If the \texttt{Atmos} climate model uses more than four absorbers, then the code is prohibitively slow.
    \item The climate model in \texttt{Photochem} uses a novel root-finding method to iterate to radiative-convective equilibrium (RCE; Section \ref{sec:methods_climate}). \texttt{Atmos} uses a less efficient time-integration approach.
    \item The climate model in \texttt{Photochem} can consider an arbitrary number of condensibles and their latent heating impact on the convective lapse rate, while \texttt{Atmos} only considers the latent heating from H$_2$O and CO$_2$.
    \item \texttt{Photochem} can be easily installed with the Conda package manager, without the need for compilers. Using \texttt{Atmos} requires a Fortran compiler.
    \item \texttt{Photochem} has a seamless Python wrapper enabled by Cython while \texttt{Atmos} does not have an effective Python wrapper.
    \item The photochemical or climate models in \texttt{Photochem} can be used interactively in a Jupyter Notebook or in a Python script, while \texttt{Atmos} must be used via its driver Fortran programs.
    \item \texttt{Photochem} can readily perform multiple photochemical or climate calculations in parallel using Python multiprocessing. \texttt{Atmos} is much harder to parallelize.
    \item In \texttt{Atmos}, many array sizes are fixed at compile time (e.g., the number of species). Changing the number of species requires you to recompile the code. In contrast \texttt{Photochem} dynamically allocates arrays and only needs to be compiled once.
\end{itemize}


\texttt{Photochem} is also based, in part, on the open-source \texttt{Cantera} chemical engineering tool \citep{Cantera2024}. \texttt{Photochem} uses an object-oriented coding approach that is inspired by \texttt{Cantera}, and adopts a similar file format for specifying chemical reaction rates and thermodynamics. \texttt{Cantera} uses the CVODE integrator \citep{Hindmarsh2005} for evolving chemical kinetics, which influenced its use in \texttt{Photochem}.

\subsection{Photochemistry} \label{sec:methods_photochemistry}

\subsubsection{Model equations} \label{sec:methods_photochemistry_eqns}

Here, we provide a brief description of the fundamental equations solved in the photochemistry module of \texttt{Photochem} while Appendix \ref{sec:photochem_eqns} gives a complete derivation. The photochemistry code is governed by a one-dimensional continuity equation
\begin{equation} \label{eq:continuity_main} 
  \frac{\partial n_{i}}{\partial t} = - \frac{\partial}{\partial z}\Phi_{i} + \sigma_{i}
\end{equation}
Here $n_i$ is the number density of a species (atom, molecule, or particle), $i$ per cubic centimeter, $t$ is time in seconds, $z$ is altitude in cm, $\Phi_{i}$ is the flux of molecules cm$^{-2}$ s$^{-1}$, and $\sigma_i$ is the source or sink of the species (cm$^{-3}$ s$^{-1}$). Sources and sinks include chemical reactions, rainout in droplets of water, and condensation or evaporation. Section \ref{sec:reactions} details our chemical reaction network. As is usual in 1-D atmospheric chemistry code, the flux of gases ($\Phi_{i}$) is determined by molecular and eddy diffusion (Equation \eqref{eq:phi_gas}). We assume that particles are transported by a fall velocity (Equation \eqref{eq:phi_particle}).

To numerically approximate Equation \eqref{eq:continuity_main}, we discretize altitude with a finite volume method. We use finite volumes because the approximation guarantees molecule conservation, which is not always the case for other discretization techniques (e.g., finite difference or finite element). Conservation is desirable because it allows the model to accurately determine, for example, the fluxes of gases going into and out of the surface of a planet. We use a second-order centered scheme for all spatial derivatives except for the advective terms that arise from molecular diffusion and particle transport, where we use first-order upwind schemes for stability. ``Molecular diffusion'' is a misnomer because the processes includes both advective and diffusive transport. For a heavy molecule in a light background gas, molecular advection can dominate transport, warranting the stability of an upwind discretization.

Our finite volume discretization converts the Equation \eqref{eq:continuity_main} system of partial differential equations (PDEs) to a larger system of ordinary differential equations (ODEs) of the form
\begin{equation} \label{eq:dfdt_general_main}
  \frac{d n_{i}^j}{d t} = \Omega_{i}^{j} n_{i}^{j+1} 
  + \Gamma_{i}^{j} n_{i}^{j}
  + \Lambda_{i}^{j} n_{i}^{j-1}
  + \sigma_i^j(n_{1}^{j},n_{2}^{j},...,n_{N}^{j})
\end{equation}
Here, the superscript $j$ indexes each altitude grid cell, and $N$ is the total number of species. The $\Omega_{i}^{j}$, $\Gamma_{i}^{j}$, and $\Lambda_{i}^{j}$ coefficients arise from our finite volume method (Appendix \ref{sec:finite_volume}). Chemical reactions cause some gases to change concentration much more rapidly than others, which makes the system of ODEs stiff. Therefore, to evolve Equation \eqref{eq:dfdt_general_main}, we use the CVODE stiff integrator \citep{Hindmarsh2005} that implements the backward differential formulas.

Any stiff integrator like CVODE requires the Jacobian of Equation \eqref{eq:dfdt_general_main}. Given that $d n_{i}^j/d t$ largely depends on gas densities in the $j+1$, $j$ and $j-1$ atmospheric layers, the Jacobian is approximately banded with a bandwidth of $2N + 1$. We compute the transport Jacobian terms analytically and estimate the chemistry and other source Jacobian terms ($\sigma_i^j$) using forward mode automatic differentiation \citep[e.g.,][]{Revels2016}. In brief, automatic differentiation efficiently computes derivatives by systematically applying the chain rule at the elementary operation level, avoiding the numerical errors of finite differences and the complexity of symbolic differentiation. There was previously no freely available modern Fortran automatic differentiation package that met our needs, so we wrote our own for \texttt{Photochem}, called \texttt{Differentia}, taking inspiration from the \texttt{DNAD} and \texttt{ForwardDiff} packages \citep{Revels2016,Yu2013}. \texttt{Differentia} is open-source and can be used in other Fortran applications \citep{Differentia2025}. In older versions of \texttt{Photochem} we computed chemistry Jacobian terms using finite differences, but we have found that using \texttt{Differentia} results in more stable integrations.

In this article, we model all atmospheres by computing steady-state solutions of Equation \eqref{eq:dfdt_general_main} (i.e., $d n_{i}^j/d t = 0$). This is a reasonable approach because atmospheres are often in quasi-steady-states over geologic time. In our models of Venus, Earth, Mars, Titan, we assume a steady-state is achieved after integrating Equation \eqref{eq:dfdt_general_main} to $\sim 3$ billion years. Our simulations of Jupiter and WASP-39b (and other gas-giants) follow the \texttt{VULCAN} convergence criteria \citep{Tsai2021} and assume a steady-state when the maximum change of all volume mixing ratios is less than $1\%$ over the last $t_n/2$ seconds, where $t_n$ is the current integration time. While we consider steady-state atmospheres here, \texttt{Photochem} can also accurately evolve atmospheres over time. \citet{Wogan2023} used this feature to study transient atmospheres on the early Earth after massive asteroid impacts.

\subsubsection{Chemical Network} \label{sec:reactions} \label{sec:methods_photochemistry_network}

Our full photochemical network is an updated version of chemistry used in \citet{Zahnle2016}, \citet{Wogan2023} and \citet{Wogan2024}. We consider 98 neutral gas-phase species composed of H, O, C, N, S, Cl, and He, connected by 610 forward chemical reactions and 95 photolysis reactions. All 614 non-photolysis reactions are reversed using the principle of microscopic reversibility \citep{Visscher2011}. The network does not include ions. Photolysis rates are computed using quadrature two-stream radiative transfer \citep{Toon1989}. The Zenodo archive \url{https://doi.org/10.5281/zenodo.15785405} contains the rate coefficients, citations and notes for each non-photolysis reaction. Most of the rates and thermodynamics from the NIST Kinetic Database and WebBook, although there are exceptions. Table \ref{tab:photolysis} details the source of each photolysis cross section and quantum yield. Many of the photolysis cross sections and yields come from the Leiden and PHIDRATES databases \citep{Heays2017,Huebner2015}.

The network includes 16 condensed-phase species. Most aerosols like condensed H$_2$O or NH$_3$ are produced and destroyed through condensation and evaporation (Appendix \ref{sec:condense_evap}), while hydrocarbon haze particles form by gas-phase reactions. In reality, in hazy atmosphere like Titan's and others (e.g., early Earth at certain times), hydrocarbon particles form by the condensation of gas phase molecules that are often larger than what we can model explicitly in our chemical scheme. Therefore, we approximate haze formation by reacting two large hydrocarbons (e.g., $\mathrm{C_2H} + \mathrm{C_4H_2}$) assuming the products ultimately undergo further polymerization and condensation to haze \citep{Lavvas2008}. Once particles form, we make no attempt to simulate their detailed microphysical evolution. Particles have a radius prescribed by the user, and fall according to Stokes' law (Appendix \ref{sec:fall_velocity}) until they leave the bottom of the model domain or evaporate.

\subsection{Climate} \label{sec:methods_climate}




The 1-D climate module in \texttt{Photochem} solves for atmospheric radiative and convective energy balance governed by conservation of energy,
\begin{equation} \label{eq:climate_energy_balance}
  \rho c_p \frac{\partial T}{\partial t} = \frac{\partial F_\mathrm{rad}}{\partial z} + \frac{\partial F_\mathrm{conv}}{\partial z}
\end{equation}
Here, $\rho$ is the gas density in g cm$^{-3}$, $c_{p}$ is the specific heat at constant pressure in erg g$^{-1}$ K$^{-1}$, $T$ is temperature in Kelvin, and $F_\mathrm{rad}$ and $F_\mathrm{conv}$ are the net radiative and convective energy fluxes, respectively, in ergs cm$^{-2}$ s$^{-1}$. We compute $F_\mathrm{rad}$ with standard two-steam radiative transfer \citep{Toon1989} with optical properties for photolysis, Rayleigh scattering, collision-induced absorption, and aerosols. The user must specify the aerosol densities, radii, and optical properties as a function of pressure (i.e., aerosols must be ``painted'' onto the atmosphere). We approximate line absorption with k-distributions. The code mixes k-distributions using the ``resort rebin'' method \citep{Amundsen2017}. All opacities used in this article are detailed in Table \ref{tab:kdistributions}, \ref{tab:cia_rayleigh} and \ref{tab:photolysis}. Our current set of k-distributions are primarily designed for rocky planets, and are not applicable for many gas giants, especially hot Jupiters like WASP-39b. The next several paragraphs describe how we account for convection ($F_\mathrm{conv}$).

Applying a centered finite volume scheme to Equation \eqref{eq:climate_energy_balance} yields
\begin{equation} \label{eq:climate_finite_volume}
  \left(\rho c_p \frac{d T}{d t}\right)^j = \frac{F_\mathrm{rad}^{j+1/2} - F_\mathrm{rad}^{j-1/2}}{\Delta z^j} + \frac{F_\mathrm{conv}^{j+1/2} - F_\mathrm{conv}^{j-1/2}}{\Delta z^j}
\end{equation}
Like in Section \ref{sec:methods_photochemistry}, the super script $j$ denotes an atmospheric grid cell with thickness $\Delta z^j$.


Most 1-D climate codes \citep{Selsis2023,Arney2016,Malik2017,Robinson2018} numerically integrate Equation \eqref{eq:climate_finite_volume} forward in time to a steady-state climate ($\frac{\partial T}{\partial t} = 0$). However, this approach can be prohibitively slow because the upper and lower atmospheric layers often have significantly different rates of temperature change. Small time steps are needed for stability in the low density upper atmosphere, while the deep atmosphere has a large thermal inertia that takes substantially more time to reach an equilibrium. To avoid this problem, \texttt{Photochem} instead solves for the steady-state of Equation \eqref{eq:climate_finite_volume} with a root finding algorithm that permits rapid convergence. Several other climate codes use root finding \citep[e.g.,][]{Mukherjee2023,McKay1989}, however, to our knowledge, these models do not allow for an arbitrary convective pattern in the atmosphere or account for the impact of latent heat on the convective lapse rate. The approach we present below remedies both shortcomings.

\begin{figure}
  \centering
  \includegraphics[width=\linewidth]{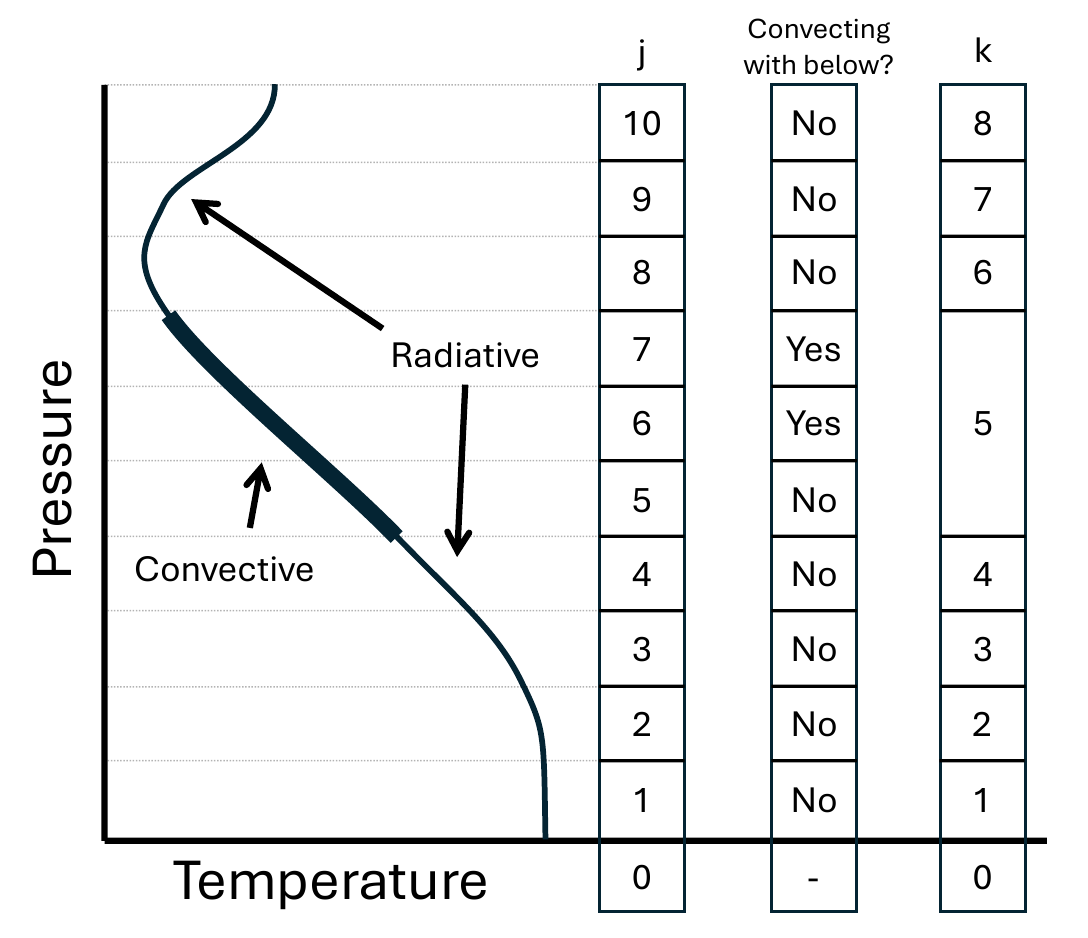}
  \caption{A schematic pressure-temperature profile to aid explanation of our climate model. On the right-hand-side, $j$ indexes each atmospheric layer, and the column labeled ``Convecting with below'' indicates if a given $j$ layer is convecting with the layer below it. Note that layer $j = 5$ is convecting with the layer above it ($j = 6$), but not with the layer below it ($j = 4$), so it is labeled ``No'' in the ``Convecting with below'' column. A new atmospheric grid $k$ can be defined that groups the convecting layers, simplifying the iteration to radiative-convective equilibrium (see text).}
  \label{fig:climate_explanation}
\end{figure}

To solve Equation \eqref{eq:climate_finite_volume} with root-finding, we reformulate it to the following:
\begin{equation} \label{eq:climate_root}
  g^j(T^j) = (F_\mathrm{rad}^{j+1/2} - F_\mathrm{rad}^{j-1/2}) + (F_\mathrm{conv}^{j+1/2} - F_\mathrm{conv}^{j-1/2})
\end{equation}
The roots of $g^j(T^j)$ are a temperature profile where the radiative and convecting energy entering and leaving each layer are balanced. The expression ignores all variables relevant to the atmosphere's thermal inertia ($\rho$, $c_p$, and $\Delta z$).

Now, consider the example temperature profile in Figure \ref{fig:climate_explanation} divided into 10 atmospheric layers and the ground. Suppose we know which layers are convective and radiative. In Figure \ref{fig:climate_explanation}, layer $j = 7$ is convecting with layer $j = 6$, and $j = 6$ is convecting with $j = 5$. Index $k$ defines a new set of layers where convecting zones are grouped together. For the $k$ grouping, all layers are exchanging energy by radiation only, therefore Equation \ref{eq:climate_root} simplifies to,
\begin{equation} \label{eq:climate_finite_volume_k}
  g^k(T^k) = F_\mathrm{rad}^{k+1/2} - F_\mathrm{rad}^{k-1/2}
\end{equation}
We define $T^k$ as the temperature in the lowest $j$ atmospheric layer of the the convecting zone, so in Figure \ref{fig:climate_explanation}, the $k = 5$ and $j = 5$ temperatures are equal. To calculate the radiative fluxes (e.g., $F_\mathrm{rad}^{k+1/2}$), we need the temperature of the $j = 6$ and $j = 7$ layers. Our code computes these temperatures by integrating the multispecies pseudo-adiabat in \citet{Graham2021} upward from the temperature at the bottom of the convective zone. This approach accounts for the latent heating of an arbitrary number of condensibles. Also, Figure \ref{fig:climate_explanation} has a single convective zone but the algorithm works with any number of convective zones.

Our code solves for the roots of Equation \eqref{eq:climate_finite_volume_k} using the ``hybrj'' routine in \texttt{MINPACK} \citep{More1980}, a modified version of Newton's method. The algorithm requires the Jacobian of Equation \eqref{eq:climate_finite_volume_k} which we compute with finite differences.

The text above describes how to solve for radiative-convective equilibrium if it is known which atmospheric layers are convecting, but it does not explain how to determine the convecting pattern in the first place. To determine this initial pattern, consider the following reorganization of Equation \ref{eq:climate_finite_volume} that ignores convection:
\begin{equation} \label{eq:climate_convec}
  h^j(T^j) = \frac{d T^j}{d t} = \left(\frac{1}{\rho c_p}\right)^j \frac{F_\mathrm{rad}^{j+1/2} - F_\mathrm{rad}^{j-1/2}}{\Delta z^j} 
\end{equation}
Suppose we have a temperature profile $T_0^j$ which can be expressed as the vector $\bm{T}_0$. Using a single damped Newton iteration, we can perturb $\bm{h}(\bm{T}_0)$ towards radiative equilibrium. The Newton perturbation ($\Delta \bm{T}$) is given by solving the linear equation,
\begin{equation} \label{eq:climate_convec_newton_1}
  \bm{J_h}(\bm{T}_0) \Delta \bm{T} = \bm{h}(\bm{T}_0)
\end{equation}
Here, $\bm{J_h}(\bm{T}_0)$ is the Jacobian ($d \bm{h}/d \bm{T}$) of $\bm{h}$ at $\bm{T}_0$. $\bm{J_h}(\bm{T}_0)$ is a square matrix with a width and height equal to the number of atmospheric layers plus the surface layer. The new temperature profile ($\bm{T}_1$) becomes
\begin{equation} \label{eq:climate_convec_newton_2}
  \bm{T}_1 = \bm{T}_0 + \epsilon \Delta \bm{T}
\end{equation}
Where, $\epsilon$ is a dampening term for the Newton iteration that we set to $\sim 0.1$. Next, we compare the lapse rate of $\bm{T}_1$, defined as $\Gamma = \frac{\partial \ln T}{\partial \ln P}$, to the lapse rate of the multispecies pseudo-adiabat ($\Gamma_a$) derived in \citet{Graham2021}. We assume convection occurs in layers where the lapse rate exceeds the adiabat ($\Gamma > \Gamma_a$). The above convection scheme is very similar to the ``convective adjustment'' approaches employed in other codes \citep[e.g.,][]{Selsis2023,Arney2016,Malik2017}. Other models with ``convective adjustment'' time integrate Equation \eqref{eq:climate_convec} to a steady-state and periodically adjust the temperature profile so that the lapse rate does not exceed an adiabat. Our approach modifies this standard ``convective adjustment'' by replacing the time integration with a root-finding scheme.

Our algorithm finds radiative-convective equilibrium by repeating the Equation \eqref{eq:climate_finite_volume_k} root solve and the convection update described above. We consider the temperature structure converged when the convection update does not change the convecting pattern. The scheme works remarkably well, even for fairly crude initial guesses of the temperature profile and convecting regions of the atmosphere. The benchmark in Section \ref{sec:discussion_climate} shows how our root-finding scheme compared to the accelerated time-stepping method in the \texttt{HELIOS} climate model.

The climate code takes input gas concentrations as either surface partial pressures or as volume mixing ratios. For a set surface partial pressure, the code distributes the gases throughout the atmospheric column accounting for its saturation vapor pressure, cold trapping, and its latent heating impact on convection. Any specified volume mixing ratio can vary as a function of pressure.


\texttt{Photochem} also contains a so-called ``inverse'' climate model similar to the ones described in \citet{Kopparapu2013} and \citet{Kasting1988}. We do not use this simple model in this paper, but we briefly describe it here because it is a valuable feature. Given the surface temperature and the reservoir of each atmospheric volatile in terms of partial pressures, the code draws a pseudo-adiabat temperature profile \citep{Graham2021} upward until it intersects an assumed isothermal stratosphere. The code then performs radiative transfer on the newly constructed atmosphere to estimate the outgoing longwave and absorbed shortwave radiation. To find a quasi-equilibrium climate, the model solves a non-linear equation for the surface temperature that balances the absorbed and emitted radiation. \citet{Wogan2023} and \citet{Wogan2024} previously used this simpler algorithm to reproduce the \citet{Kopparapu2013} runway greenhouse limit as well as estimate the climate of early Earth and the K2-18b exoplanet.

\subsection{Equilibrium Chemistry} \label{sec:methods_equilibrium}

\texttt{Photochem} contains a module that solves for the chemical composition that minimizes the Gibbs energy for a fixed temperature and pressure (i.e., chemical equilibrium). The module is a modified version of \texttt{easyCHEM} \citep{Lei2024} which itself is a clone of the NASA Chemical Equilibrium Analysis tool \citep{Gordon1994}. The code considers ideal gases and solid-phase species. \citet{Gordon1994} gives a full description of the numerical method, although in brief, the approach uses the method of Lagrange multipliers to reformulate the Gibbs minimization problem into a system of non-linear equations and solves the system with a variation of Newton's method. We primarily use the equilibrium chemistry solver to set lower boundary conditions for photochemical simulations of gas giants (e.g., Jupiter and WASP-39b; Sections \ref{sec:jupiter_chem} and \ref{sec:wasp39b}) in which the model domain extends to deep regions of the atmosphere where equilibrium is a valid assumption.

\subsection{Overall Modeling Approach} \label{sec:methods_approach}

We validate \texttt{Photochem} against six planetary atmospheres: Venus, Earth, Mars, Titan, Jupiter, and WASP-39b. For each world, we first simulate the atmosphere's photochemistry using a temperature profile and other inputs from the literature, and compare the predicted steady-state composition to observations (Table \ref{tab:obs}). Next, we compute the planet's steady-state climate with the composition predicted by the photochemical model, and compare the resulting pressure-temperature profile to measured global averages or model predictions from the literature. We do not compute the climate of WASP-39b as our climate model does not include all the opacities relevant to such a hot atmosphere.

To be clear, for our benchmarks, we do not iterate the photochemical and climate model to a completely self-consistent composition and temperature because fully coupled models are hard to interpret. For example, suppose we simulated Venus with a fully coupled model and the resulting deep-atmosphere temperature profile did not match the global average and several chemical species did not agree with observed values. In this case, it would be hard to know if the mismatch is due to a failure in our climate or photochemical models. Independently considering photochemistry and climate avoids this ambiguity. Note that \texttt{Photochem} is capable of computing a self-consistent composition and climate, which we illustrate and discuss in Section \ref{sec:discussion_couple}.

All nominal simulations use the same chemical network (Section \ref{sec:reactions}), thermodynamics, opacities, and other model data (Appendix \ref{sec:model_data}). However, for some planets we omit species containing chlorine and/or sulfur because we assume these molecules are absent or play a minor role. Our simulations of Venus do not omit any molecules. Models of Mars, Jupiter and Titan only consider H, N, O, C and He species and neglect Cl- and S-containing molecules. Our models of Earth and WASP-39b include S but exclude Cl. Climate simulations use the same set of opacities detailed in Table \ref{tab:kdistributions}, \ref{tab:cia_rayleigh} and \ref{tab:photolysis}. All Solar System atmospheres use the ATLAS 3 reference spectrum of the Sun \citep{Thuillier2004} scaled to each world. Our photochemical simulations of WASP-39b use the stellar spectrum presented in \citet{Tsai2023}. 

\subsection{Reproducibility} \label{sec:methods_reproduce}

Below, we summarize the Zenodo archives that reproduce Section \ref{sec:results}, contain the \texttt{Photochem} v0.6.7 source code and model data, and archive many of the Solar System atmospheric observations used in this article:

\begin{itemize}
  \item Reproduce Section \ref{sec:results} calculations: \url{https://doi.org/10.5281/zenodo.16509802} \citep{SolarSystemExploration2025}
  \item Main \texttt{Photochem} v0.6.7 source: \url{https://doi.org/10.5281/zenodo.16322107} \citep{Photochem2025}
  \item \texttt{Photochem} v0.6.7 climate model source: \url{https://doi.org/10.5281/zenodo.15786151} \citep{Clima2025}
  \item \texttt{Photochem} v0.6.7 equilibrium solver source: \url{https://doi.org/10.5281/zenodo.16508386} \citep{Equilibrate2025}
  \item \texttt{Photochem} v0.6.7 model data: \url{https://doi.org/10.5281/zenodo.15785405} \citep{PhotochemClimaData2025}
  \item \texttt{Differentia} v0.1.4: \url{https://doi.org/10.5281/zenodo.15678369} \citep{Differentia2025}
  \item Solar System observations (Table \ref{tab:obs}): \url{https://doi.org/10.5281/zenodo.16509950} \citep{PlanetaryAtmosphereObservations2025}
\end{itemize}

\section{Results} \label{sec:results}

In the following subsections we benchmark \texttt{Photochem} against the observed compositions and climates of Venus, Earth, Mars, Titan, Jupiter and WASP-39b. All of our work is open-source and can be reproduced with the following Zenodo archive: \url{https://doi.org/10.5281/zenodo.16509802}.

\subsection{Venus}

\subsubsection{Photochemistry} \label{sec:venus_chem}

Our approach to modeling Venus's photochemistry is unique compared to previous works \citep{Bierson2020,Dai2024,Wunderlich2023,Rimmer2021}. Many past models have fixed the concentration of key species (e.g. OCS and SO$_2$) at the surface to match observations. Fixing a gas concentration implicitly assumes that there are surface gas fluxes into or out of the planet that maintain the specified abundance. These implicit fluxes can sometimes be justified by geologic processes, such as volcanism or dry deposition, but in other cases the imposed fluxes are unphysical and are merely compensating for flaws in a chemical network. 

Here, to avoid surface gas fluxes, we simulate Venus's atmosphere as a closed system where the surface and top-of-atmosphere have zero-flux boundary conditions. We initialize the atmosphere with $\sim$ 96.5\% CO$_2$, 3.5\% N$_2$, 100 ppm SO$_2$, 35 ppm H$_2$O, 5 ppm OCS, 10 ppm CO, 500 ppb HCl and 4.5 ppb H$_2$ \citep[loosely following Table 3 in][]{Rimmer2021}, which yields the following atomic column abundances in mol cm$^{-2}$: 0.17 H, 168 N, 4632 O, 2316 C, 0.25 S and 0.0012 Cl. We then evolve the atmosphere to a steady-state with zero-flux boundary conditions for every species, predicting gas concentrations at all altitudes. This approach assumes that surface gas sources and sinks as well as atmospheric escape have minor impacts on Venus's composition compared to gas-phase chemistry. Our model uses the Venus International Reference Atmosphere (VIRA) temperature profile at 45 degrees latitude \citep{Seiff1985} and the \citet{Rimmer2021} eddy diffusion profile (Figure \ref{fig:t_and_kzz}). In addition to gas absorption and scattering, our radiative transfer model also includes H$_2$SO$_4$ cloud opacity using optical properties from \citet{Palmer1975} and accounts for the unknown UV absorber using Equation 11 in \citet{Rimmer2021} except that we assume the UV absorber has zero optical depth for $\lambda > 400$ nm following \citet{Krasnopolsky2012}. The model accounts for H$_2$SO$_4$ condensation, forming droplets with an assumed 2 $\mu$m radius. H$_2$SO$_4$ droplets form above $\sim 50$ km and fall to the warm lower atmosphere where they all evaporate before reaching the surface.

\begin{figure*}
  \centering
  \includegraphics[width=\textwidth]{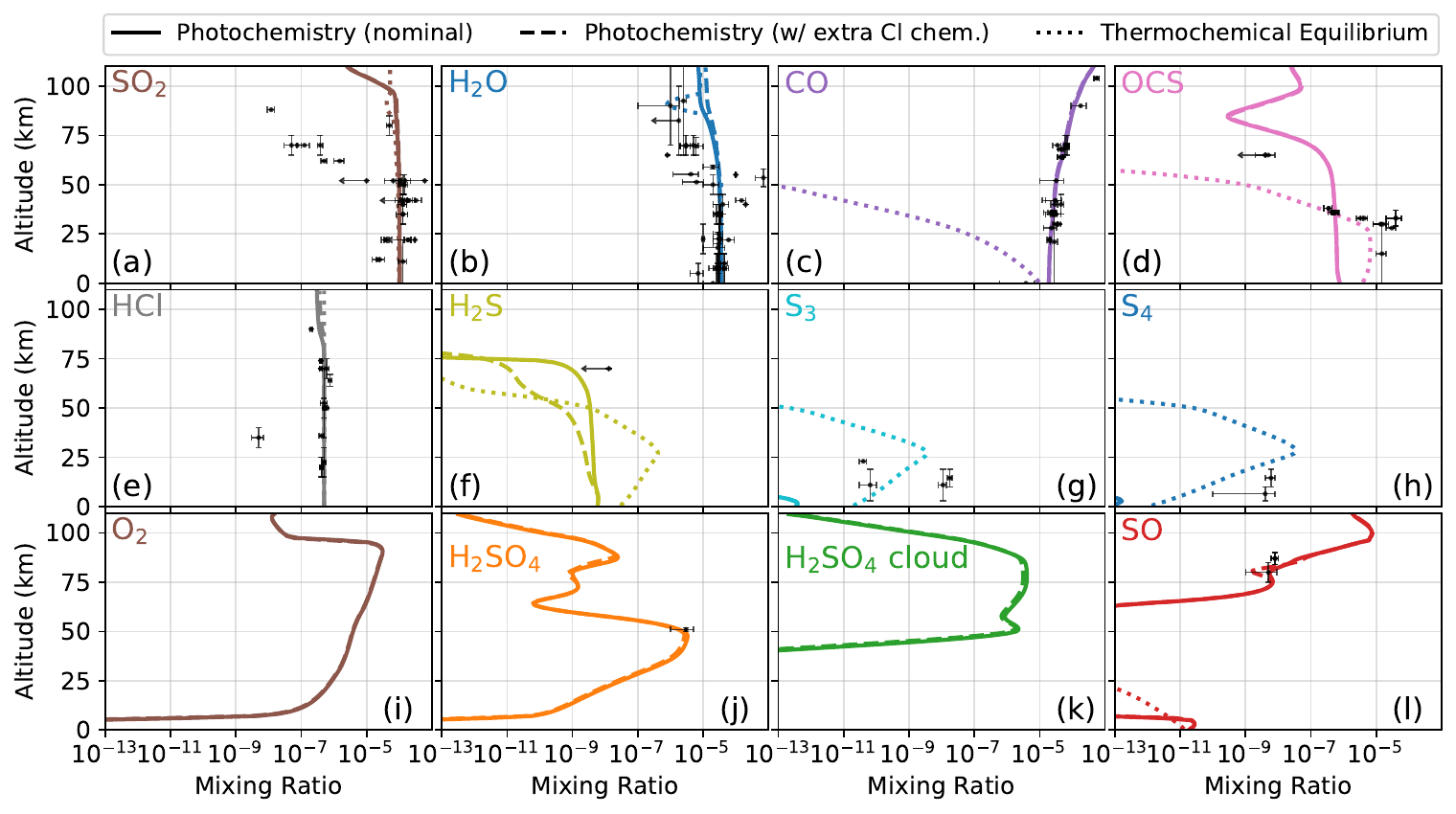}
  \caption{Photochemical simulations of Venus's atmosphere as a closed system (lines) compared to observations (black dots). Table \ref{tab:obs} lists the sources for all observations. The solid lines are the nominal model, while the dashed lines are a sensitivity test that adds additional chlorine chemistry as described in Section \ref{sec:venus_chem}. The dotted lines are the thermochemical equilibrium of each atmospheric layer in the nominal model. Panel (k) is the H$_2$SO$_4$ cloud mixing ratio computed with $f_{\mathrm{H_2OSO_4}(l)} = n_{\mathrm{H_2OSO_4}(l)}/n$, where $n_{\mathrm{H_2OSO_4}(l)}$ is the molecules cm$^{-3}$ in the condensed phase and $n$ is the total gas phase number density. Overall, our nominal model reproduces the observed concentrations of CO, HCl, H$_2$SO$_4$, H$_2$S, and SO at all altitudes, but fails to explain the measured abundances of SO$_2$, H$_2$O, OCS, S$_3$ and S$_4$. Most of these model-data discrepancies are long standing unsolved problems in Venus photochemistry \citep{Rimmer2021,Bierson2020}.}
  \label{fig:venus}
\end{figure*}

Figure \ref{fig:venus} shows the predicted steady-state composition for key species in our nominal model compared to observations. For CO, HCl, H$_2$S, H$_2$SO$_4$ and SO, we adequately reproduce observed values at all altitudes. On the other hand,  there are some model-data discrepancies for SO$_2$, H$_2$O, OCS, S$_3$, and S$_4$.

\underline{\textit{SO$_2$} and \textit{H$_2$O}:} Like other photochemical models in the literature \citep[e.g.,][]{Rimmer2021}, we do not capture the observed decrease in SO$_2$ abundance in and above Venus's clouds \citep[$\gtrsim 50$ km;][]{Zasova1993}. In our model, the main SO$_2$ and H$_2$O depletion mechanism is the formation of sulfuric acid droplets:
\begin{subequations} \label{eq:so2_dest}
\begin{align}
  \mathrm{SO_2} + h\nu &\rightarrow \mathrm{SO} + \mathrm{O}
  \\
  \mathrm{SO_2} + \mathrm{O} + \mathrm{M} &\rightarrow \mathrm{SO_3} + \mathrm{M}
  \\
  \mathrm{SO_3} + 2\:\mathrm{H_2O} &\rightarrow \mathrm{H_2SO_4} + \mathrm{H_2O}
  \\
  \mathrm{H_2SO_4} &\rightarrow \mathrm{H_2SO_4}(l)
  \\
  \cline{1-2}
  2\:\mathrm{SO_2} + \mathrm{H_2O} &\rightarrow \mathrm{SO} + \mathrm{H_2SO_4}(l) \tag{\ref*{eq:so2_dest}, net}
\end{align}
\end{subequations}

Reaction \ref{eq:so2_dest} requires 1 H$_2$O molecule to capture 2 SO$_2$ molecules. In our nominal model, H$_2$O is 35 ppm and SO$_2$ is 101 ppm below the clouds consistent with observations \citep{Bezard2011,Chamberlain2013,Arney2014,Marcq2008}, hence Reaction \ref{eq:so2_dest} mechanism can only decrease SO$_2$ to 31 ppm, limited by availability of H$_2$O. Given this mass conservation, it is understandable that our code cannot deplete SO$_2$ to the $< 1$ ppm observed near $\sim 60$ - 80 km \citep{Zasova1993,Encrenaz2012}. A solution to this ``SO$_2$ problem'' may demand a more sophisticated treatment of Venus' clouds. For example, \citet{Rimmer2021} proposed that SO$_2$ might dissolve in H$_2$SO$_4$ droplets, a loss process that we do not account for. \citet{Rimmer2021} argue that dissolution requires a neutralizing base, maybe in the form of hydroxide salts, to increase droplet pH, enhancing its ability to store SO$_2$.

\underline{\textit{OCS, S$_3$ and S$_4$:}} Our modeled abundances of OCS, S$_3$ and S$_4$ do not match observations in Venus's lower atmosphere ($< 50$ km). Observations \citep{Bezard2007,Arney2014} find that OCS is $\sim 15$ ppm near the surface, which drops to less than 1 ppm at 35 km altitude. By contrast, our model predicts a constant mixing ratio of $\sim 0.6$ ppm below 50 km. Also, \texttt{Photochem} under predicts S$_3$ and S$_4$ in the lower atmosphere compared to observations \citep{Krasnopolsky2013}, species that are closely coupled to OCS (e.g., $\mathrm{CO} + \mathrm{S_3} \leftrightarrow \mathrm{S_2} + \mathrm{OCS}$). This model-data disagreement is unsurprising because the relevant sulfur kinetics are poorly understood (Section \ref{sec:discussion_venus}). The dotted lines in Figure \ref{fig:venus} are the thermochemical equilibrium of each layer in our nominal photochemical model. The equilibrium abundances of OCS, S$_3$ and S$_4$ are in closer agreement with the deep-atmosphere observations, suggesting that we are missing chemical pathways or that our existing kinetics are too slow.

\underline{\textit{O$_2$:}} Historically, photochemical models have predicted too much O$_2$ above the clouds compared to the observational upper-limits \citep[e.g.,][]{Marcq2018}. Our model does not resolve this mystery. Our nominal simulation predicts an O$_2$ column of $30 \times 10^{18}$ molecules cm$^{-2}$ above 62 km, while observational upper-limits range between $< 0.8 \times 10^{18}$ and $< 15 \times 10^{18}$ molecules cm$^{-2}$. 

\citet{Krasnopolsky1981} and \citet{Yung1982} suggested that catalytic cycles involving ClCO and ClCO$_3$ could draw down O$_2$ by oxidizing CO to CO$_2$ (e.g., Equation 4a in \citet{Yung1982}), chemistry that our nominal model does not account for. The dashed lines in Figure \ref{fig:venus} illustrate an additional test that includes this relevant chlorine chemistry. Specifically, we added the species ClCO, ClCO$_3$, COCl$_2$, then added all the reactions in Table 5 of \citet{Rimmer2021} involving species in our network (see ``input/Venus/rimmer2021.yaml'' in \url{https://doi.org/10.5281/zenodo.16509802}). The additional chlorine chemistry does not substantially change upper-atmospheric O$_2$ (Figure \ref{fig:venus}i). The reason is likely because we adopted the mean enthalpy for ClCO reported by \citet{Nicovich1990} ($-21.8 \pm 2.5$ kJ/mol at 298 K), making the gas thermally unstable in Venus's atmosphere, inhibiting the catalytic cycle. \citet{Mills2007} found the chlorine catalyzed O$_2$ destruction is most effective for smaller ClCO enthalpy and a colder upper atmosphere than we used here.

\subsubsection{Climate} \label{sec:venus_climate}

\begin{figure*}
  \centering
  \includegraphics[width=1\textwidth]{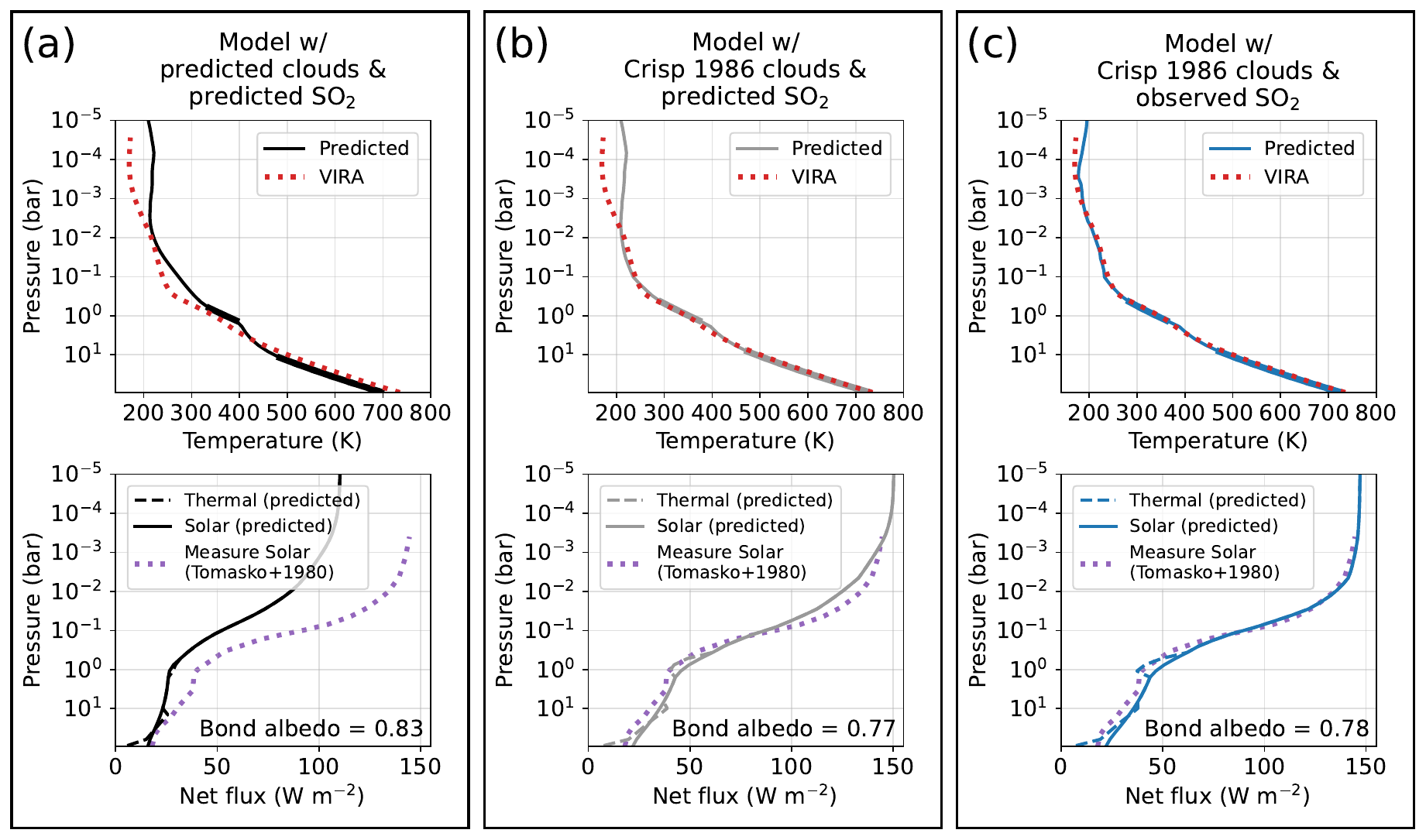}
  \caption{Climate simulations of Venus's atmosphere. (a) A climate model that uses the H$_2$SO$_4$ clouds and gas concentrations predicted by the nominal photochemical simulation in Figure \ref{fig:venus}. (b) The same as (a), except the model uses observed cloud optical properties from \citet{Crisp1986}. (c) The same as (b), except the simulation uses Venus's observed SO$_2$ profile. In (a), (b) and (c), the top panels shows predicted temperature profiles compared to the Venus International Reference Atmosphere \citep[VIRA,][]{Seiff1985}. Thick portions of the predicted P-T profile indicate convecting layers. The bottom panels are the net thermal and solar fluxes compare to the solar fluxes measured by Pioneer Venus \citep{Tomasko1980}. Our climate model can reproduce Venus's energy balance (panel (c)) when we correct for inaccurate clouds and SO$_2$ predicted by the photochemical model.}
  \label{fig:venus_climate}
\end{figure*}

Figure \ref{fig:venus_climate} shows several climate simulations of Venus's atmosphere. Our first model (Figure \ref{fig:venus_climate}a, black lines) adopts the composition and sulfuric acid cloud density from our nominal photochemical simulation shown in Figure \ref{fig:venus}. Like in our photochemical simulations, the climate model uses the \citet{Rimmer2021} parameterization for the unknown UV absorber with zero opacity for $\lambda > 400$ nm. Key radiatively active gases include CO$_2$, H$_2$O, SO$_2$, OCS and HCl. The calculation predicts a temperature profile that disagrees with VIRA by 10s of Kelvin at most altitudes (Figure \ref{fig:venus_climate}a top). Also, the model's net solar fluxes (i.e., shortwave fluxes) are smaller than those measured by Pioneer Venus \citep[Figure \ref{fig:venus_climate}a, bottom;][]{Tomasko1980}. These model-data discrepancies can be attributed to the simplified treatment of Venus's clouds. In Figure \ref{fig:venus_climate}a, we assume the H$_2$SO$_4$ cloud aerosols predicted by the photochemical model (Figure \ref{fig:venus}). The predicted cloud has a total column mass of 5.5 mg cm$^{-2}$, a value slightly smaller than the 18 mg cm$^{-2}$ measured by Pioneer Venus \citep{Knollenberg1980}. However, the predicted cloud has unrealistic optical properties because we assume all aerosols have a 2 $\mu$m radius, but, in reality, Venus's clouds have a complex size distribution ranging from 0.5 to nearly 4 microns that changes with altitude \citep{Crisp1986}.

Model b in Figure \ref{fig:venus_climate} documents a climate simulation where we have instead adopted the cloud optical properties of \citet{Crisp1986} which are based on a range of spacecraft and ground-based observations. The predicted temperate profile agrees with VIRA at altitudes below $10^{-2}$ bar (Figure \ref{fig:venus_climate}b, top). Additionally, the modeled net solar fluxes match measured values from \citet{Tomasko1980}. This result demonstrates that our photochemical model does not predict an H$_2$SO$_4$ cloud layer that reproduces Venus's energy balance.

Both of the Figure \ref{fig:venus_climate}a and \ref{fig:venus_climate}b climate simulations use the SO$_2$ profile from our nominal photochemical model (Figure \ref{fig:venus}a), which overestimates SO$_2$ above $\sim 50$ km. To test the impact of these inaccurate SO$_2$ concentrations on Venus's energy balance, we performed a third climate simulation (Figure \ref{fig:venus_climate}c) with the SO$_2$ profile adjusted so that it matches the data points in Figure \ref{fig:venus}a. Correcting the SO$_2$ cools the predicted $\sim 10^{-4}$ bar temperature from $\sim 220$ K down to $\sim 180$ K, a temperature more in line with the VIRA P-T profile (Figure \ref{fig:venus_climate}c).

Our model has also revealed that the planet's energy balance depends strongly on the adopted CO$_2$-CO$_2$ collision induced absorption (CIA) opacity. In the Figure \ref{fig:venus_climate} simulations that reproduce Venus's climate (and all other climate simulations shown in this paper), we use the same CIA as \citet{Robinson2018}, except for $\lambda < 2.5$ $\mu$m, where use the \citet{Lee2016} recommendations. When we instead use the HITRAN CO$_2$-CO$_2$ CIA cross sections \citep{Karman2019} the model under-predicts the deep atmosphere opacity leading to surface temperatures that are far too cold ($\lesssim 700$ K; not shown). Overall, the ``HITRAN2020'' CIA does not reproduce Venus largely because it does not include any $\lambda < 2.5$ $\mu$m opacity (see \citet{hitran2022} Table 11 lists bands only from 1-3250 cm${^-1}$), a region important for capturing the observed emission at the 2.3 $\mu$m and 1.7 $\mu$m atmospheric windows \citep[e.g.,][]{Lee2016}. However, the \citet{Lee2016} NIR CIA cross sections are not based on laboratory measurements. The NIR CO$_2$-CO$_2$ opacities are instead tuned by researchers to permit agreement between Venus observations and radiative transfer models \citep[e.g.,][]{Marcq2006,Meadows1996}. 

Overall, the Figure \ref{fig:venus_climate} simulations demonstrate that our climate model can reproduce Venus's energy balance, provided that we use the observed clouds, the observed SO$_2$ high altitude mixing ratio, and the NIR CO$_2$-CO$_2$ opacity recommended by \citet{Lee2016}.

\subsection{Earth} \label{sec:earth}

\subsubsection{Photochemistry} \label{sec:earth_chem}

Our photochemical simulation of modern Earth generally follows the model setup of \citet{Hu2012}, with some modifications. The model adopts the \citet{Massie1981} eddy diffusion profile, and, for a temperature profile, we use the COSPAR International Reference Atmosphere (CIRA) 1986 at the equator in January (Figure \ref{fig:t_and_kzz}). Table \ref{tab:earthbc} shows the surface boundary conditions. We fix the surface pressure of N$_2$, O$_2$ and CO$_2$ to 0.79, 0.213 and $4.05 \times 10^{-4}$ bar, respectively. The model condenses out H$_2$O at the input 40\% relative humidity, which sets the tropospheric concentration. Earth's average relative humidity varies between 77\% near the surface and 20\% at the tropopause \citep{Manabe1967}. Our model only accepts a single altitude-independent relative humidity, so we choose 40\% as an average tropospheric value. Water condensation in the atmosphere forms droplets with prescribed 10 $\mu$m radii that fall to the surface. The calculation uses a range of surface gas fluxes and deposition velocities (Table \ref{tab:earthbc}) from the literature (e.g., \citet{Seinfeld2016} and \citet{Hauglustaine1994}), that represent biologic production (e.g., N$_2$O from denitrification), anthropogenic sources (e.g., CO from burning fossil fuels), and dry deposition. Gases are also removed from the atmosphere using the \citet{Giorgi1985} parameterization for rainout in droplets of H$_2$O assuming modern Earth's rainfall rate of $1.1 \times 10^{17}$ molecules cm$^{-2}$ s$^{-1}$ \citep{Giorgi1985}. We turn on diffusion-limited H and H$_2$ escape to space. \texttt{Photochem} simulates diffusion-limited H and H$_2$ escape with an effusion velocity at the upper boundary determined by their rate of molecular diffusion through the background gas.

\begin{table}
  \caption{Modern Earth lower boundary conditions}
  \label{tab:earthbc}
  \begin{center}
  \begin{tabularx}{0.95\linewidth}{p{0.1\linewidth} >{\centering\arraybackslash}p{0.18\linewidth} >{\centering\arraybackslash}p{0.25\linewidth} >{\centering\arraybackslash}p{0.15\linewidth} >{\centering\arraybackslash}p{0.1\linewidth}} 
    \hline \hline
    \multirow{2}{*}{Species} & Pressure & Surface Flux                & $v_\mathrm{dep}$ & \multirow{2}{*}{Ref.} \\
                             & (bar)    & (molec. cm$^{-2}$ s$^{-1}$) & (cm s$^{-1}$)    &                            \\
    \hline
    H$_2$O       & 40\% RH               & -                    & -         & -    \\
    N$_2$        & 0.79                  & -                    & -         & -    \\
    O$_2$        & 0.213                 & -                    & -         & -    \\
    CO$_2$       & $4.05 \times 10^{-4}$ & -                    & -         & -    \\
    CO           & -                     & $3.7 \times 10^{11}$ & 0.03      & a, b \\
    CH$_4$       & -                     & $1.4 \times 10^{11}$ & 0         & c    \\
    NO           & -                     & $3 \times 10^9$      & 0.01      & a, b \\
    N$_2$O       & -                     & $1.5 \times 10^9$    & 0         & d, b \\
    NH$_3$       & -                     & $1.5 \times 10^{10}$ & 1         & a, e \\
    NO$_2$       & -                     & 0                    & 0.05      & b    \\
    NO$_3$       & -                     & 0                    & 0.05      & b    \\
    SO$_2$       & -                     & $9 \times 10^9$      & 1         & a    \\
    H$_2$S       & -                     & $2 \times 10^8$      & 0.015     & a, e \\
    OCS          & -                     & $5.4 \times 10^7$    & 0.003     & a, f \\
    H$_2$SO$_4$  & -                     & $7 \times 10^8$      & 1         & a, e \\
    HCN          & -                     & $1.7 \times 10^8$    & 0.13      & g, f \\
    CH$_3$CN     & -                     & $1.3 \times 10^8$    & 0.13      & g, f \\
    HNO$_3$      & -                     & 0                    & 4         & f    \\
    \hline
    \multicolumn{5}{p{0.95\linewidth}}{
      \textbf{Notes.} a: \citet{Seinfeld2016}; b: \citet{Hauglustaine1994}; c: \citet{Jackson2020}; d: \citet{Schwieterman2022}; e: \citet{Hu2012}; f: \citet{Tsai2021}; g: \citet{Li2003}
    }
  \end{tabularx}
  \end{center}
\end{table}

\begin{figure*}
  \centering
  \includegraphics[width=1\textwidth]{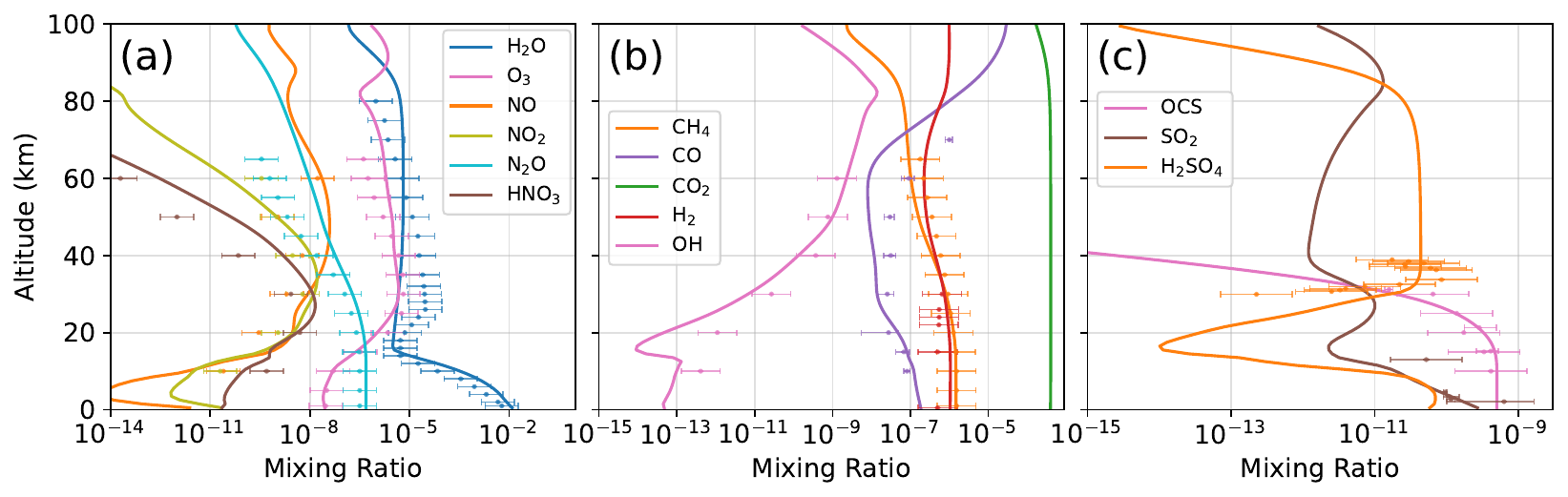}
  \caption{Photochemical simulation of modern Earth's atmosphere (lines) compared to observations (dots). Table \ref{tab:obs} lists the sources for all observations. For observations that do not give an error bar, we assume a one order-of-magnitude error to account for diurnal and spatial variations following \citet{Tsai2021}. \texttt{Photochem} broadly reproduces the observed composition of Earth's atmosphere.}
  \label{fig:earth}
\end{figure*}

Figure \ref{fig:earth} shows that our \texttt{Photochem} simulation of Modern Earth broadly reproduces much of the observed trace gases in the atmosphere. For the observations in Figure \ref{fig:earth} that do not report an error, we follow \citet{Tsai2021} and assume one order-of-magnitude error bars to account for diurnal and spatial variations. Therefore, many error bars in Figure \ref{fig:earth} do not represent one standard deviation uncertainty in a measurement. Our predicted gas abundances result from well-established chemistry in the atmospheric science literature \citep[e.g., see chapter 5 and 6 of][]{Seinfeld2016}. For example, stratospheric ozone is produced and destroyed by the Chapman mechanism \citep[e.g.,][]{Seinfeld2016} as well as several catalytic cycles. One important cycle involves NO and NO$_2$, where the NO is largely sourced from a surface flux boundary condition (Table \ref{tab:earthbc}):
\begin{subequations} \label{eq:earth_nox_cycle}
  \begin{align}
    \mathrm{NO} + \mathrm{O_3} &\rightarrow \mathrm{NO_2} + \mathrm{O_2}
    \\
    \mathrm{NO_2} + \mathrm{O} &\rightarrow \mathrm{NO} + \mathrm{O_2}
    \\
    \cline{1-2}
    \mathrm{O_3} + \mathrm{O} &\rightarrow 2\:\mathrm{O_2} \tag{\ref*{eq:earth_nox_cycle}, net}
  \end{align}
\end{subequations}
OH and HO$_2$ also destroy ozone with the same net reaction:
\begin{subequations} \label{eq:earth_hox_cycle}
  \begin{align}
    \mathrm{OH} + \mathrm{O_3} &\rightarrow \mathrm{HO_2} + \mathrm{O_2}
    \\
    \mathrm{HO_2} + \mathrm{O} &\rightarrow \mathrm{OH} + \mathrm{O_2}
    \\
    \cline{1-2}
    \mathrm{O_3} + \mathrm{O} &\rightarrow 2\:\mathrm{O_2} \tag{\ref*{eq:earth_hox_cycle}, net}
  \end{align}
\end{subequations}
The hydroxyl radicals (OH) in Reaction \ref{eq:earth_hox_cycle} are supplied by the net photochemical reaction of ozone with water vapor: $\mathrm{O_3} + h\nu \rightarrow \mathrm{O(^1D)} + \mathrm{O_2}$ followed by $\mathrm{O(^1D)} + \mathrm{H_2O} \rightarrow \mathrm{OH} + \mathrm{OH}$. Our model ignores catalytic loss of O$_3$ from chlorine chemistry. As a result of ozone production from the Chapman mechanism and the destruction paths listed above, our model estimates a total O$_3$ column of 193 Dobson Units (DU), a value consistent with the $\sim 190$ to 500 DU range observed on modern Earth \citep{Salawitch2022}.

Hydroxyl plays a critical role in several systems beyond ozone photochemistry. One example is the oxidation of CH$_4$, beginning with $\mathrm{CH_4} + \mathrm{OH} \rightarrow \mathrm{CH_3} + \mathrm{H_2O}$. This reaction determines methane's lifetime in the atmosphere, which we find to be $\sim 10$ years, consistent with literature estimates \citep{Catling2017}. Including the CH$_4$ + OH reaction, the carbon in CH$_4$ is ultimately oxidized to CO$_2$ through a series of $\sim 6$ reactions via the following intermediates: CH$_3$, CH$_3$O$_2$, CH$_3$O, H$_2$CO and CO.

Like in Venus's atmosphere, SO$_2$ in Earth's atmosphere is photochemically processed to sulfuric acid, although the mechanism is different because of Earth's more plentiful hydroxyl. On Earth, SO$_3$ forms by $\mathrm{SO_2} + \mathrm{OH} + \mathrm{M} \rightarrow \mathrm{HSO_3} + \mathrm{M}$ followed by $\mathrm{HSO_3} + \mathrm{O_2} \rightarrow \mathrm{HO_2} + \mathrm{SO_3}$, then sulfuric acid is produced from Reaction \ref{eq:so2_dest}c. In the troposphere, H$_2$SO$_4$ dissolves in water droplets and rains out of the atmosphere. In the upper troposphere and lower stratosphere, H$_2$SO$_4$ condenses to sulfuric acid aerosols, which, in our model, fall until they evaporate at $\lesssim 10$ km altitude.

To summarize, \texttt{Photochem} works well for Earth. The code captures the important chemistry in Earth's atmosphere, reproducing its observed composition.

\subsubsection{Climate} \label{sec:earth_climate}

Our climate simulations of Earth use most of the gas abundances predicted by our photochemical model shown in Figure \ref{fig:earth}. One exception is the water profile, which the climate code calculates to be self-consistent with the temperature profile for a 40\% relative humidity. Another exception is that we consider two total ozone column abundances: one with 193 dobson units (DU) from our Figure \ref{fig:earth} photochemical model, and the other with 500 DU. These ozone concentrations span typical values for Earth \citep{Salawitch2022}. Following many previous 1-D models of Earth \citep[e.g.,][]{Kasting1984}, we account for water clouds by using a surface albedo of 0.2, which is larger than Earth's true value of $\sim 0.15$. This approach effectively ``paints'' clouds on the surface. We take this approach because our model cannot yet account for patchy clouds. We reserve a more complete treatment of clouds to future work (Section \ref{sec:discussion_cloud}).

\begin{figure}
  \centering
  \includegraphics[width=\linewidth]{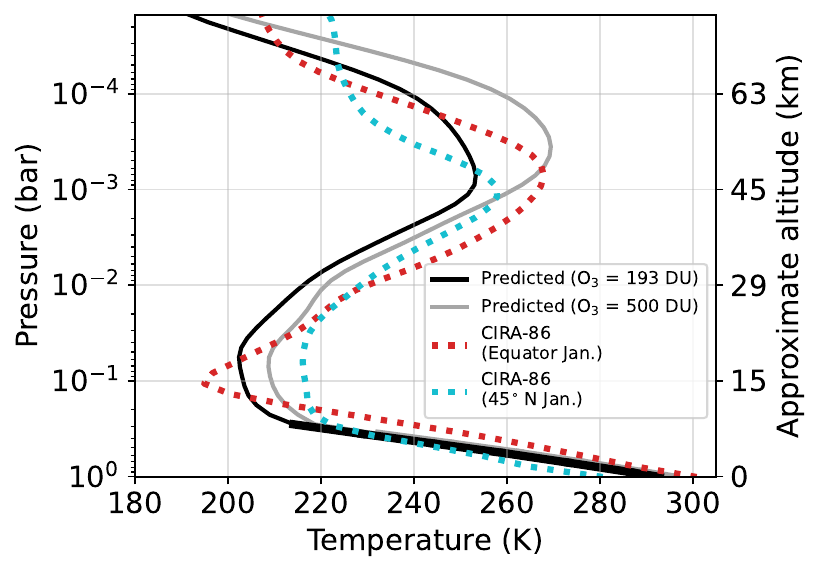}
  \caption{Climate simulations of modern Earth's atmosphere with 193 and 500 DU of ozone (black and grey lines, respectively) compared to temperature profiles at the equator and $45^{\circ}$ N (red and blue lines, respectively) in January from the COSPAR International Reference Atmosphere 1986 (CIRA-86). Thick portions of the predicted P-T profiles indicate convection. Our model broadly reproduces Earth's pressure-temperature profile.}
  \label{fig:earth_climate}
\end{figure}

Figure \ref{fig:earth_climate} shows computed P-T profiles for two ozone columns (193 DU and 500 DU for the black and gray lines, respectively), compared to January equator and $45^{\circ}$ N temperature profiles from the COSPAR International Reference Atmosphere 1986 (CIRA-86). In both simulations, we compute a surface temperature near $\sim 290$ K, a tropopause at 15 km with a $\sim 210$ K temperature, and a stratopause near 50 km with a $\sim 250$ and $270$ K temperature, all values consistent with Earth's global average P-T profile (e.g., Chapter 1 of \citet{Catling2017}).

\subsection{Mars}

\subsubsection{Photochemistry} \label{sec:mars_chem}

Our photochemical simulation of Mars is largely motivated by the model setup of \citet{Zahnle2008}. We use their temperature and eddy diffusion profiles (Figure \ref{fig:t_and_kzz}). For boundary conditions, the we fix the surface CO$_2$ and N$_2$ pressures to 6.0 and 0.03 mbar, respectively, and sets the H$_2$O relative humidity of condensation to 10\%. Note however, that Mars' relative humidity is highly variable \citep{Fischer2019}. We include 0.02 cm/s deposition velocities for H$_2$O$_2$ and O$_3$ that represent the atmospheric oxidation of Mars's surface. H$_2$ and H escape the model domain at the diffusion limited rate. Our model ignores oxygen escape and assumes that H$_2$O$_2$ and O$_3$ deposition represent the main loss of oxidants from the atmosphere. \citet{Zahnle2008} showed that Mars' simulated atmospheric composition is insensitive to whether oxidants leave the atmosphere from surface deposition or from escape.
 
\begin{figure*}
  \centering
  \includegraphics[width=0.95\textwidth]{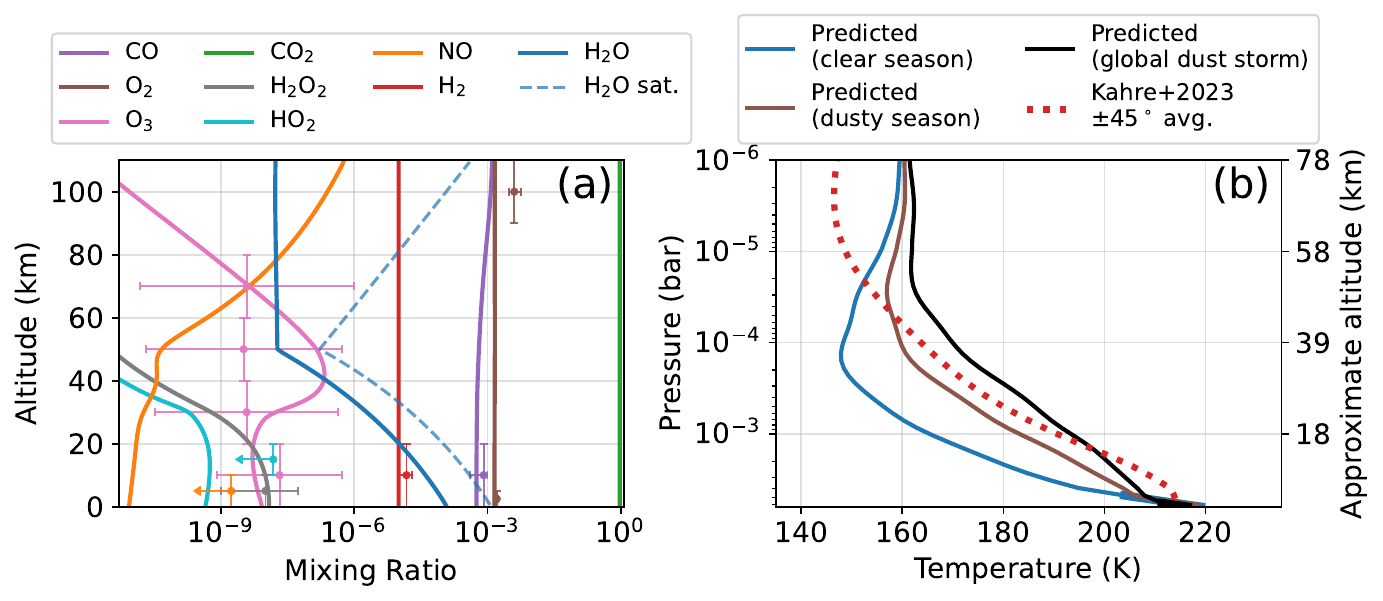}
  \caption{(a) A photochemical model of Mars (lines) compared to observations (dots). Table \ref{tab:obs} lists the sources for all observations. The dashed blue line is the saturation mixing ratio of H$_2$O. (b) Three climate simulations of Mars representing the clear season (blue line), dusty season (brown line), and a global dust storm (black line). The red dotted line is a $\pm45^\circ$ latitude average from the Ames Legacy Mars GCM \citet{Kahre2023}. Thick portions of the predicted P-T profiles indicate convection. The \texttt{Photochem} code generally reproduces Mars composition (panel (a)) and climate (panel (b)).}
  \label{fig:mars}
\end{figure*}

The computed steady-state composition, shown in Figure \ref{fig:mars}a, is generally consistent with a wide range of atmospheric observations and with several past models of Mars's chemistry \citep{Zahnle2008,Krasnopolsky2006b}. The trace abundances of CO, O$_2$, H$_2$, O$_3$, and H$_2$O$_2$ are the result of CO$_2$ and H$_2$O photochemistry. CO$_2$ photolysis produces CO and atomic oxygen, and ultimately O$_2$ and O$_3$ after the O atoms combine. Water vapor photolysis yields OH and atomic hydrogen. Hydrogen atoms liberated by water photolysis react quickly with O$_2$ to form HO$_2$ ($\mathrm{H} + \mathrm{O_2} + \mathrm{M} => \mathrm{HO_2} + \mathrm{M}$). When HO$_2$ reacts with O, it creates OH that speeds CO$_2$ recombination:
\begin{subequations} \label{eq:mars_co2}
  \begin{align}
    \mathrm{H} + \mathrm{O_2} + \mathrm{M} &\rightarrow \mathrm{HO_2} + \mathrm{M}
    \\
    \mathrm{HO_2} + \mathrm{O} &\rightarrow \mathrm{OH} + \mathrm{O_2}
    \\
    \mathrm{CO} + \mathrm{OH} &\rightarrow \mathrm{CO_2} + \mathrm{H}
    \\
    \cline{1-2}
    \mathrm{CO} + \mathrm{O} &\rightarrow \mathrm{CO_2} \tag{\ref*{eq:mars_co2}, net}
  \end{align}
\end{subequations}
Direct CO$_2$ recombination by $\mathrm{CO} + \mathrm{O} + \mathrm{M} \rightarrow \mathrm{CO_2} + \mathrm{M}$ is slow because it is spin-forbidden. 
HO$_2$ can also react with a hydrogen atom to make H$_2$, or with itself to form H$_2$O$_2$. The H$_2$ escapes to space, while H$_2$O$_2$ is lost to surface deposition.

\subsubsection{Climate} \label{sec:mars_climate}

Mars's climate depends on two main opacity sources: CO$_2$ gas and dust aerosols. The planet has a clear season with little atmospheric dust during 0--180 Solar longitude (northern spring and summmer), and a relatively dusty atmosphere for the remainder of the year because perihelion occurs then at 251 solar longitude \citep[e.g.,][]{Montabone2015}. Also, every several years, Mars has global dust storms. We construct three climate models corresponding to the three levels of dust-loading. The simulations use the global and annual mean dust radii and density profiles from the Ames Legacy Mars GCM \citet{Kahre2023} and, also following \citet{Kahre2023}, we use dust optical properties from \citet{Wolff2009} based on Mars Reconnaissance Orbiter observations. We scale the dust density such that 9.3 $\mu$m dust optical depths of 0.075, 0.375, and 1 represent the clear season, the dusty season, and a global dust storm. These optical depths are based on a compilation of spacecraft observations analyzed in \citet{Montabone2015} (see their Figure 21). The models use gas abundances from our photochemical model (Figure \ref{fig:mars}a) and uses a 0.2 surface albedo.

Figure \ref{fig:mars}b shows that the three climate simulations encompass the $\pm45^\circ$ latitude average from the \citet{Kahre2023} GCM. More dust increases sunlight absorption and warms much of the atmosphere above 5 km and has a minor cooling effect on the surface. In the ``clear'' and ``dusty'' models, near infrared CO$_2$ sunlight absorption causes a temperature inversion at $P \lesssim 10^{-4}$ bar. Our models compute a warmer temperature than the \citet{Kahre2023} average for $P < 10^{-5}$ bar, although we do not consider this an issue. Mars Trace Gas Orbiter observations and other spacecraft have shown that Mars's $\sim 10^{-6}$ bar temperature varies between $\sim120$ and 180 K (e.g., see Figure 8 in \citet{LopezValverde2023}). We conclude that the $\sim160$ K at $\sim 10^{-6}$ bar computed by \texttt{Photochem} is reasonable. 

\subsection{Titan}

\subsubsection{Photochemistry} \label{sec:titan_chem}

To model Titan's photochemistry, we use the P-T profile from Figure 3 in \citet{Moreno2012} and the \citet{Loison2015} eddy diffusion coefficient (Figure \ref{fig:t_and_kzz}). For boundary conditions, the model fixes the surface N$_2$ to 1.5 bar and CO to $76.5 \times 10^{-6}$ bar (i.e., $\sim 5.1 \times 10^{-5}$ mixing ratio). We assume a 100\% CH$_4$ relative humidity of condensation. We add $5 \times 10^{6}$ molecules cm$^{-2}$ s$^{-1}$ downward fluxes of O and OH at the top of the atmosphere, representing in-fall from micrometeorite ablation \citep{Loison2015}. Mass leaves the atmosphere from diffusion limited H$_2$ escape, and when aerosols (haze or condensed CH$_4$, C$_2$H$_6$, HCN, etc.) fall to the surface. Haze particles have an assumed constant 0.5 micron radius based loosely on \citet{Tomasko2008a} with \citet{Khare1984} Mie optical properties. We model the production and transport of a number of other condensate aerosols (i.e., C$_2$H$_6$, C$_2$H$_2$, HCN, etc.) with an assumed 10 $\mu$m radii, although we do not account for how they absorb or scatter sunlight. The model accounts for the influence of Galactic Cosmic Rays (GCR) by prescribing altitude-dependent N$_2$ destruction and N and N($^2$D) production rates given by \citet{Lavvas2008} (their Figure 3). We assume 82.5\% of N$_2$ GCR destruction forms grounds state N and 17.5\% makes N($^2$D). Titan is the only planet in this article where we considering GCR effects.

\begin{figure*}
  \centering
  \includegraphics[width=\textwidth]{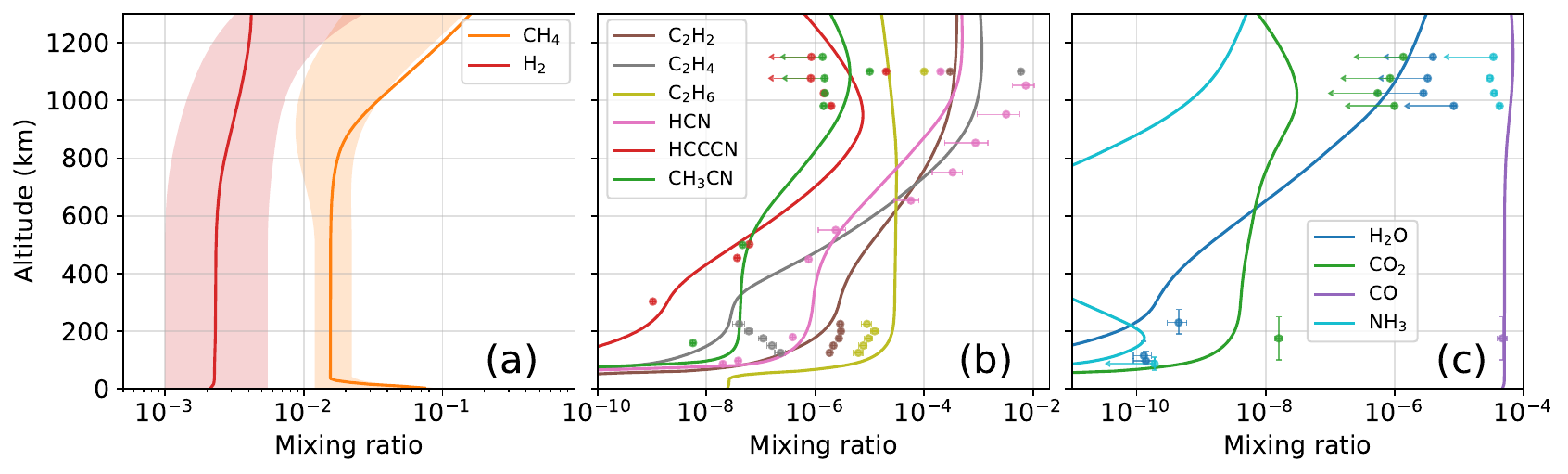}
  \caption{A photochemical simulation of Titan's atmosphere (lines) compared to observations (dots and shading). Some observations do not have error bars because no error was reported. The shading in panel (a) outlines the minimum and maximum H$_2$ and CH$_4$ concentrations from the \citet{Waite2013} empirical model of Titan. Table \ref{tab:obs} lists the sources for all observations. For all species, except NH$_3$, the model is within a factor of several of observed concentrations.}
  \label{fig:titan}
\end{figure*}

Our computed steady-state composition generally agrees with observations to within a factor of several (Figure \ref{fig:titan}). One exception is that we underpredict the observed NH$_3$ abundances near 1000 km by about four orders of magnitude. According to \citet{Loison2015}, ion chemistry is important to NH$_3$ formation in the upper atmosphere, a process that we ignore. 


Like in previous work \citep[e.g.,][]{Lavvas2008,Loison2015,Pearce2020}, we find that CH$_3$ is an important radical in Titan's atmosphere. At high altitudes CH$_3$ forms from CH$_4$ photolysis, while at lower altitudes, it is created catalytically from acetylene:
\begin{subequations} \label{eq:titan_ch3}
  \begin{align}
    \mathrm{C_2H_2} + \mathrm{hv} &\rightarrow \mathrm{C_2H} + \mathrm{H}
    \\
    \mathrm{C_2H} + \mathrm{CH_4} &\rightarrow \mathrm{C_2H_2} + \mathrm{CH_3}
    \\
    \cline{1-2}
    \mathrm{CH_4} &\rightarrow \mathrm{CH_3} + \mathrm{H} \tag{\ref*{eq:titan_ch3}, net}
  \end{align}
\end{subequations}
Methyl radicals combine to make C$_2$H$_6$ ($\mathrm{CH_3} + \mathrm{CH_3} + \mathrm{M} \rightarrow \mathrm{C_2H_6} + \mathrm{M}$) and also help create C$_2$H$_4$ through several intermediate steps. C$_2$H$_4$ photolysis is the primary source for C$_2$H$_2$. Consistent with previous research \citep[e.g.,][]{Pearce2020}, we find that HCN is primarily created by a sequence of reactions involving CH$_3$:
\begin{subequations} \label{eq:titan_hcn}
  \begin{align}
    \mathrm{N_2} + \mathrm{hv} &\rightarrow \mathrm{N} + \mathrm{N(^2D)} \label{eq:titan_hcn_n2}
    \\
    \mathrm{N} + \mathrm{CH_3} &\rightarrow \mathrm{H} + \mathrm{H_2CN}
    \\
    \mathrm{N(^2D)} + \mathrm{CH_4} &\rightarrow \mathrm{H_2CN} + \mathrm{H_2}
    \\
    \mathrm{H_2CN} + \mathrm{H} &\rightarrow \mathrm{HCN} + \mathrm{H_2}
    \\
    \cline{1-2}
    \mathrm{N_2} + \mathrm{CH_4} + \mathrm{CH_3} + \mathrm{H} &\rightarrow 2\: \mathrm{HCN} + 3\: \mathrm{H_2} \tag{\ref*{eq:titan_hcn}, net}
  \end{align}
\end{subequations}
About 50\% of the N$_2$ dissociation in Reaction \eqref{eq:titan_hcn_n2} is from UV photolysis and the other half is from GCRs. Cyanoacetylene (HCCCN) is created when HCN reacts with C$_2$H: $\mathrm{C_2H} + \mathrm{HCN} \rightarrow \mathrm{HCCCN} + \mathrm{H}$.


The main hydrocarbons and nitriles are destroyed by haze formation throughout the atmosphere or by direct condensation at $\lesssim 100$ km altitude. Following \citet{Lavvas2008}, we parameterize haze formation with three reactions, the most important being $\mathrm{H_2CN} + \mathrm{HCN} \rightarrow \mathrm{haze}$. In total, our code predicts a haze production rate of $7\times 10^{-15}$ g cm$^{-2}$, in agreement with other estimates \citep{McKay2001}. The loss of hydrocarbons and nitriles to condensation leaves behind H$_2$ (e.g., Reactions \eqref{eq:titan_ch3} and \eqref{eq:titan_hcn}) near $\sim 0.2\%$ that escapes at the diffusion limited rate.

\subsubsection{Climate} \label{sec:titan_climate}

\begin{figure*}
  \centering
  \includegraphics[width=\textwidth]{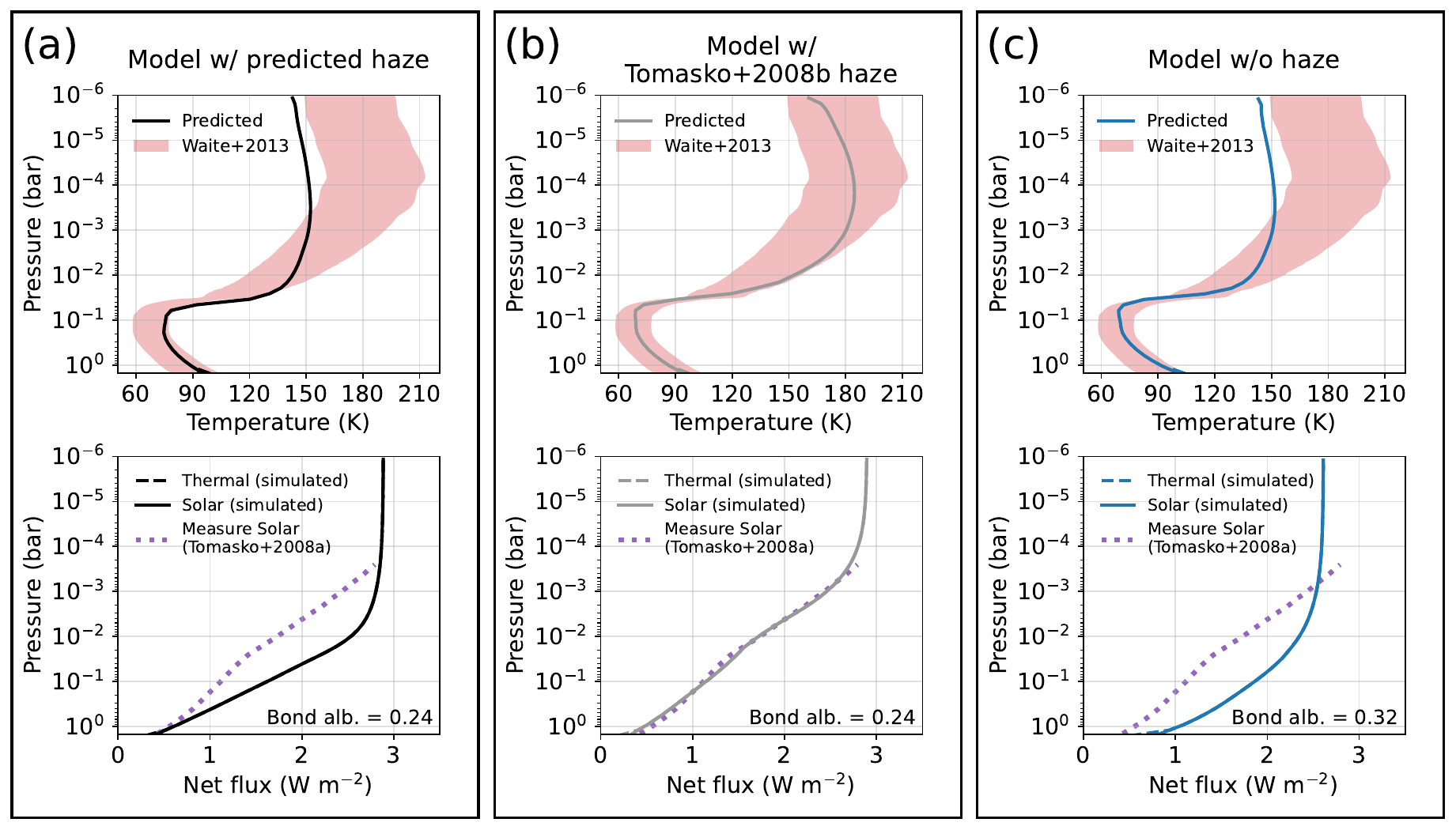}
  \caption{Climate simulations of Titan's atmosphere. (a) A climate model that uses the haze and gas concentrations predicted by the Figure \ref{fig:titan} photochemical model. (b) The same as (a), except that the model replaces the computed haze with the observed  \citet{Tomasko2008a} haze optical properties at solar wavelengths based on the Cassini/Huygens mission. (c) The same as (a), except that the model removes all haze opacity. In (a), (b) and (c), the top panels show predicted temperature profiles compared to the minimum and maximum of the \citet{Waite2013} empirical model (red shading). The bottom panels are the net thermal and solar fluxes compare to solar fluxes from Cassini/Huygens \citep[Figure 3 in][]{Tomasko2008b}. Our climate model can reproduce Titan's energy balance (panel (b)) when we correct for the inaccurate haze opacities predicted by the photochemical model.}
  \label{fig:titan_climate}
\end{figure*}

Figure \ref{fig:titan_climate} shows three climate models of Titan compared to the minimum and maximum temperature profiles from the \citet{Waite2013} empirical model. Our nominal simulation (Figure \ref{fig:titan_climate}a), uses the gas concentrations and haze densities predicted by our photochemical model (Figure \ref{fig:titan}). Important opacities include haze aerosols, CH$_4$, C$_2$H$_6$, C$_2$H$_4$, as well as N$_2$-N$_2$, N$_2$-CH$_4$ and N$_2$-H$_2$ collision induced absorption. The nominal case predicts an acceptable bond albedo of 0.25 \citep{Creecy2021} and a 99 K surface temperature, a value only several degrees warmer than Titan's $\sim 95$ K global average \citep{McKay1989}. However, the computed $10^{-3}$ to $10^{-6}$ bar temperature is 140--150 K, but Titan's observed value is closer to $\sim 180$ K at these altitudes \citep{Waite2013}. 

Titan's stratospheric temperature is largely determined by the balance of CH$_4$, C$_2$H$_6$ and C$_2$H$_4$ infrared emission, and heating when CH$_4$ and haze aerosols absorb sunlight \citep{McKay1989}. The model uses CH$_4$, C$_2$H$_6$ and C$_2$H$_4$ concentrations that generally agree with observations (Figure \ref{fig:titan}), so we expect the code is adequately accounting for their emission. On the other hand, the haze opacity in the Figure \ref{fig:titan_climate}a model is questionable because we crudely assume that all haze particles have a 0.5 $\mu$m radius with the \citet{Khare1984} optical properties using Mie theory. Indeed, the simulated net solar fluxes disagree with fluxes derived from Cassini/Huygens observations \citep[Figure \ref{fig:titan_climate}a, bottom;][]{Tomasko2008b}.

Figure \ref{fig:titan_climate}b shows an additional climate simulation that uses the \citet{Tomasko2008a} haze optical properties inferred from Huygens probe observations. Specifically the model uses the \citet{Tomasko2008a} haze opacities between 350--5000 nm, and the haze predicted by our photochemical model in the infrared ($>5000$ nm). We do not replace our far-infrared haze optical properties with measured values because Huygens did not observe them. The predicted temperature profile and net solar fluxes in Figure \ref{fig:titan_climate}b agree with observations, even in the stratosphere. We estimate a 97 K surface temperature that is consistent with Titan's measured global average \citep[$\sim 95$ K;][]{McKay1989}. This result demonstrates that our photochemical model does not predict a haze that reproduces Titan's energy balance.

To further test the radiative impact of Titan's haze, we conduct an additional simulation that removes all haze opacity (Figure \ref{fig:titan_climate}c). The haze-free model yields a too warm 104 K surface temperature showing, like previous work \citep{McKay1991}, that Titan's haze has an anti-greenhouse effect.

\subsection{Jupiter}

\subsubsection{Photochemistry} \label{sec:jupiter_chem}

We conclude our tour of the Solar System with Jupiter. Our photochemistry setup follows the similar calculation in \citet{Tsai2021}. We use their P-T profile and eddy diffusion coefficient (Figure \ref{fig:t_and_kzz}). The lower boundary of the model is at 10,000 bars where we fix all species to a chemical equilibrium concentration for the following composition: He/H = 0.0785 \citep[$0.92 \times$ solar;][]{Atreya2020}, C/H = $1.2 \times 10^{-3}$ \citep[$4.4 \times$ solar][]{Atreya2020}, O/H = $2.6 \times 10^{-4}$ \citep[$0.5 \times$ solar;][]{Cavalie2023}, and N/H = $2 \times 10^{-4}$ \citep[$2.9 \times$ solar;][]{Moeckel2023}. At the upper boundary, near $\sim 3\times 10^{-9}$ bar, we impose a downward flux of $4 \times 10^4$, $4 \times 10^4$, and $10^4$ molecules cm$^{-2}$ s$^{-1}$ for H$_2$O, CO and CO$_2$, respectively, capturing micrometeorite ablation.

\begin{figure*}
  \centering
  \includegraphics[width=\textwidth]{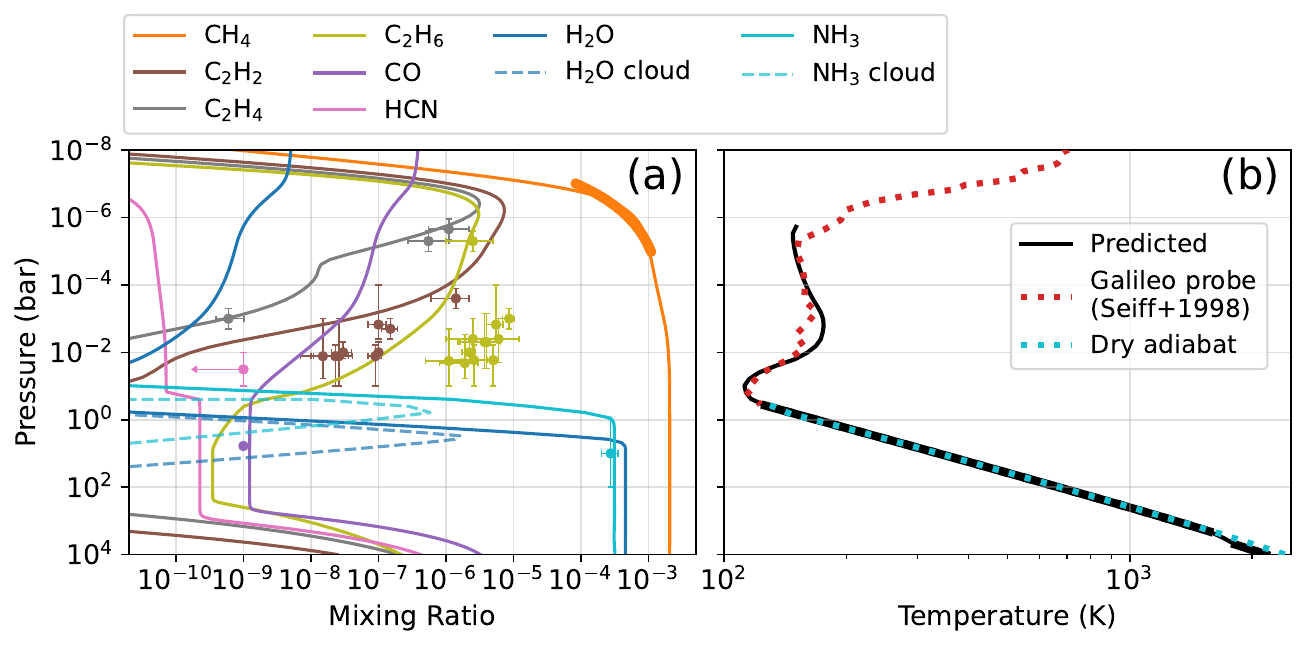}
  \caption{(a) A photochemical model of Jupiter (lines) compared to observations (dots). Table \ref{tab:obs} lists the sources for all observations. (b) A climate simulation of Jupiter's atmosphere (black line). Thick portions of the predicted P-T profiles indicate convection. The red dotted line is from \citet{Seiff1998} profile based on Galileo probe measurements, and the cyan dotted line is a dry He/H$_2$ adiabat. \texttt{Photochem} captures the main chemistry in Jupiter's atmosphere and reproduces the planet's pressure-temperature profile.}
  \label{fig:jupiter}
\end{figure*}

Figure \ref{fig:jupiter} shows that our photochemical simulation matches observations at most altitudes. One notable exception is that our model underpredicts C$_2$H$_6$ near $10^{-2}$ bars by roughly one order of magnitude. Previous photochemical models have been able to fit these C$_2$H$_6$ concentrations with the same P-T-$K_{zz}$ profile that we use here \citep{Moses2005,Tsai2021,Rimmer2016}. However, we believe that \citet{Tsai2021} and \citet{Rimmer2016} were only able to reproduce these data by using CH$_4$ cross sections from PHIDRATES \citep{Huebner2015,Huebner1992}, which incorporates CH$_4$ photolysis up to 237.2 nm. The $\lambda > 160$ nm CH$_4$ cross sections in PHIDRATES ($\le 6 \times 10^{-24}$ cm$^2$ molecule$^{-1}$) appear to be extrapolations of the \citet{Mount1977} experimental values. The more recent \citet{Lee2001} measurements find that methane's cross section sharply decreases with increasing wavelength to $4.5 \times 10^{-24}$ cm$^2$ molecule$^{-1}$ at 152 nm, suggesting the cross section becomes even more negligible at longer wavelengths. \citet{Lee2001} also argued that \citet{Mount1977} over-estimated the CH$_4$ cross section for $\sim150 < \lambda < 160$ nm because the \citet{Mount1977} experiments contained impurities (i.e., their samples contained trace amounts of C$_2$H$_6$, CO$_2$, etc.). Therefore, PHIDRATES likely overpredicts the CH$_4$ cross section for $\lambda > 150$ nm. When we use the PHIDRATES CH$_4$ photolysis cross sections, our model predicts $\sim 1$ ppm C$_2$H$_6$ at $10^{-2}$, in agreement with $\sim 10^{-2}$ bar observations. While our nominal Jupiter model (Figure \ref{fig:jupiter}) underpredicts C$_2$H$_6$ near $10^{-2}$ bars, it uses the most up-to-date methane photolysis cross sections and quantum yields \citep{Heays2017,Lee2001,Gans2011}. We leave a full reconciliation of our code with Jupiter's C$_2$H$_6$ profile to future work.

The main hydrocarbons, including C$_2$H$_6$, C$_2$H$_4$, and C$_2$H$_2$, are produced by similar pathways to those in Titan's atmosphere (Section \ref{sec:titan_chem}). However, unlike Titan, insignificant HCN is created photochemically in Jupiter's atmosphere because there is little available atomic nitrogen \citep{Moses2010}. N$_2$ photolysis is ineffective at delivering N because H$_2$ absorbs and scatters most of the 80--100 nm photons needed to split N$_2$. NH$_3$ sourced from the deep atmosphere cannot supply N because ammonia condenses at $10^{-1}$ bar as it is mixes upward.

Ground-based telescopes have measured $\sim 1$ ppb CO near 6 bars \citep{Bezard2002}. Our model explains this CO concentration with quenching in the deep atmosphere. For pressures $\gtrsim 500$ bar, the atmosphere is hot ($> 1000$ K), and fast reactions enforce a chemical equilibrium composition. As gases mix upward to lower pressures and temperatures, reactions slow, ultimately causing an equilibrium to disequilibrium transition (i.e., quenching). We find that the reactions linking CO to CH$_4$ quench near 500 bars, with a rate limiting step of $\mathrm{CH_3OH} + \mathrm{M} \rightarrow \mathrm{CH_3} + \mathrm{OH} + \mathrm{M}$ (Equation 15 in \citet{Visscher2011}). Researchers have tried to use the observed 1 ppb CO in Jupiter's troposphere, in conjunction with kinetics quenching models like \texttt{Photochem}, to estimate the deep-atmosphere O/H ratio because CO reduction depends on H$_2$O \citep{Visscher2011,Visscher2010,Cavalie2023}. However, CO quenching also depends on the uncertain deep-atmosphere mixing coefficient (i.e., $K_{zz}$), so there remains considerable uncertainty in Jupiter's O/H ratio \citep[e.g., O/H = 0.1--0.8 $\times$ solar;][]{Cavalie2023}.


\subsubsection{Climate} \label{sec:jupiter_climate}

The climate model in Photochem does not have opacities designed for gas-giant atmospheres like Jupiter's. Most of our k-distributions assume air or self-broadening instead of He/H$_2$ broadening. Also, we currently do not include species like TiO, VO, or FeH that can be important opacity sources at high pressures and temperatures (see Table \ref{tab:kdistributions}, \ref{tab:cia_rayleigh} and \ref{tab:photolysis} for our included opacities). With these caveats, we regardless apply our climate code to Jupiter's atmosphere to see if our limited set of opacities can capture Jupiter's temperature profile and bond albedo. We use the gas concentrations from our photochemical simulations of the planet (Figure \ref{fig:jupiter}a) and include a 7.485 W m$^{-2}$ interior heat flow \citep[i.e., a 107 K intrinsic temperature;][]{Li2018}. Our climate simulation does not include any clouds.

Our model (black line in Figure \ref{fig:jupiter}b) broadly reproduces the \citet{Seiff1998} upper atmosphere temperature profile ($P \lesssim 1$ bar) based on Galileo spacecraft observations. Like in Titan's stratosphere, Jupiter's upper atmosphere temperature is largely determined by the balance of heating when CH$_4$ absorbs sunlight, and cooling from CH$_4$, C$_2$H$_6$ and C$_2$H$_4$ infrared emission. The simulation has a bond albedo of 0.39, in general agreement with observed values between $0.34$ and $0.5$ \citep{Li2018}. In the deep atmosphere ($P \gtrsim 10$ bar), where Jupiter's temperature has not been measured, our model predicts the atmosphere is convective, and has a P-T profile that follows a dry He/H$_2$ adiabat.

Despite our incomplete set of opacities, the Figure \ref{fig:jupiter}b simulation broadly capture's Jupiter's expected P-T profile because of the planet's large interior heat flow. In our model, Jupiter's heat flow drives convection throughout the deep atmosphere even though we underestimate its opacity. For a gas-rich planet with a lower interior heat flow \citep[e.g., K2-18b;][]{Wogan2024}, our limited set of absorbers should under predict the deep atmosphere temperature because the model would under estimate its opacity. To reliably use the climate model in \texttt{Photochem} for gas-rich planets, future work should include the relevant deep-atmosphere absorbers (LiCl, FeH, Na, MgH, etc.).


\subsection{WASP-39b} \label{sec:wasp39b}

The final planet that we consider is the hot Jupiter exoplanet WASP-39b. WASP-39b has roughly the mass of Saturn (0.28 $M_\text{Jup}$), Jupiter's radius (1.28 $R_\text{Jup}$), orbits a Sun-like star, and receives $\sim 260$ times Earth's bolometric radiation \citep[zero bond albedo $T_{eq} = 1116$ K;][]{Mancini2018}. The planet was one of the first observed by the James Webb Space Telescope with transmission spectroscopy. The data revealed a $\sim 10 \times$ solar metallicity atmosphere containing H$_2$O, CO$_2$, CO, and SO$_2$ \citep[e.g.,][]{Rustamkulov2023}. The SO$_2$ is generated photochemically and was anticipated by chemical models prior to the launch of JWST \citep{Zahnle2016}. More recently, \citet{Tsai2023} applied four separate photochemical codes to WASP-39b (\texttt{VULCAN}, \texttt{ARGO}, \texttt{KINETICS}, and \texttt{ATMO}) showing that the observed SO$_2$ abundance is consistent with model predictions. Given that WASP-39b's atmosphere has been studied with a wide range of photochemical codes, it is an ideal hot Jupiter benchmark for \texttt{Photochem}.

\begin{figure*}
  \centering
  \includegraphics[width=0.95\textwidth]{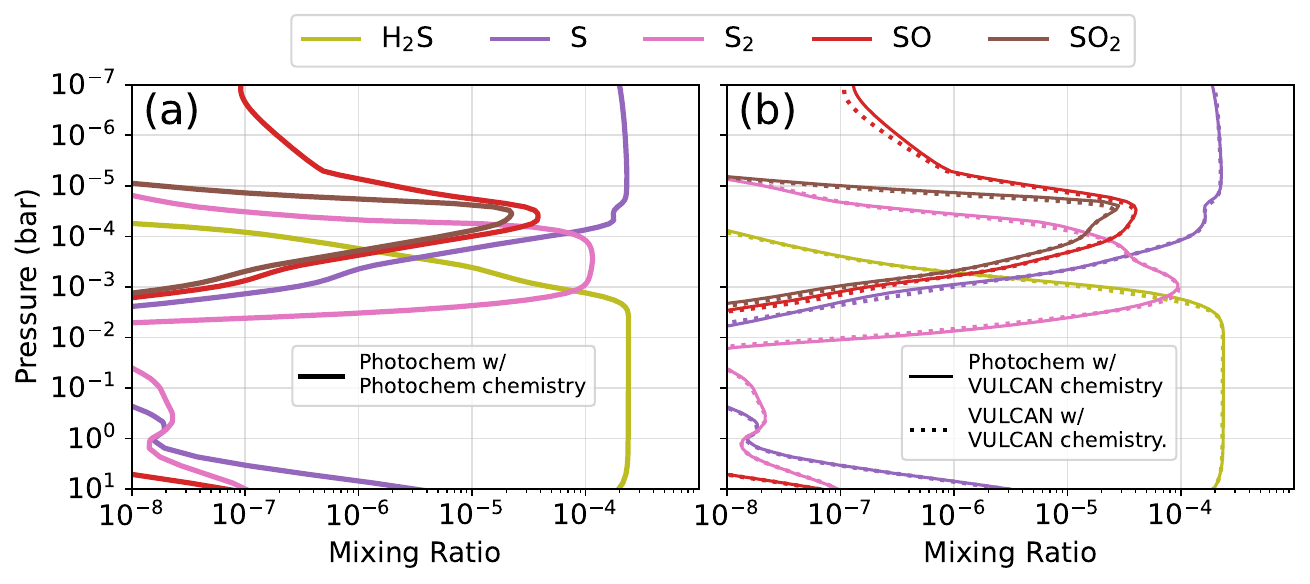}
  \caption{Photochemical simulations of WASP-39b's atmosphere at the evening terminator. (a) A model using \texttt{Photochem} with our nominal chemical network (Section \ref{sec:reactions}). (b) Calculations with \texttt{Photochem} (thin solid lines) and \texttt{VULCAN} (dotted lines) where both codes use a \texttt{VULCAN} chemical network, photolysis cross sections, and thermodynamics. With our nominal chemistry \texttt{Photochem} reproduces the observed SO$_2$ in WASP-39b's atmosphere \citep{Tsai2023}, and \texttt{Photochem} and \texttt{VULCAN} predict a very similar atmospheric composition for the same inputs.}
  \label{fig:wasp39b}
\end{figure*}

We conduct two simulations with \texttt{Photochem} and one with the \texttt{VULCAN} code. For all models we use the \citet{Tsai2023} setup for the evening terminator of WASP-39b. Specifically, we use their P-T profile (Figure \ref{fig:t_and_kzz}), eddy diffusion coefficient (Figure \ref{fig:t_and_kzz}), incident UV spectrum, a 83 degree solar zenith angle, and a 10 times solar metallicity assuming the \citet{Lodders2019} composition for the Sun. Our first simulation, shown in Figure \ref{fig:wasp39b}a, uses \texttt{Photochem} with the same chemical network used for the Solar System planets. We predict a peak $\sim 20$ ppm SO$_2$ abundance at $3\times 10^{-5}$ bar, a result consistent with the four photochemical calculations in \citet{Tsai2023} (see their Figure 1b). Also, other relevant sulfur species, like H$_2$S, S, S$_2$ and SO match the \citet{Tsai2023} ensemble of calculations. Like past work \citep{Zahnle2016,Tsai2023}, we find that SO$_2$ (and H$_2$) is made when H$_2$S is oxidized by H$_2$O, mediated by $\sim 6$ reactions that include water vapor photolysis (e.g., see Equation 1 in \citet{Tsai2023}).

As a further benchmark, Figure \ref{fig:wasp39b}b shows two additional models of WASP-39b: one with \texttt{Photochem} and other with \texttt{VULCAN}. Both calculations use the \texttt{VULCAN} thermodynamics, photolysis cross sections and the same chemical network used by \texttt{VULCAN} in \citet{Tsai2023} (i.e., their ``SNCHO\_photo\_network''). The \texttt{Photochem} software can convert \texttt{VULCAN} chemical networks to the \texttt{Photochem} format, making comparisons like Figure \ref{fig:wasp39b}b straightforward. Given very similar inputs, we find excellent agreement between \texttt{Photochem} and \texttt{VULCAN}. Some minor differences are expected (e.g., SO near $10^{-7}$ bar), because the calculations do not use the same Rayleigh scattering cross sections or molecular diffusion coefficients. Overall, the model consistency validates the numerical schemes in both \texttt{Photochem} and \texttt{VULCAN}.

\section{Discussion}

\subsection{The Data and Modeling Needed to Better Understand Venus} \label{sec:discussion_venus}

In our tour of the Solar System, our photochemical simulations of Earth, Mars, Titan and Jupiter predict compositions within a factor of several of observations in most cases. In contrast, Venus has the some glaring model-data discrepancies, mainly for sulfur species like OCS, S$_3$, S$_4$ and SO$_2$. Here we discuss the new reaction rates, opacities, and modeling approaches that would improve our understanding of the planet.




The reason why \texttt{Photochem} does not reproduce Venus observed OCS, S$_3$ and S$_4$ abundances is because these sulfur species and other relevant molecules have poorly characterized kinetics. One example is S$_3$ and S$_4$ photolysis, which is important in our Figure \ref{fig:venus} simulations. The photolysis cross sections of both species have only sparse and controversial measurements \citep{Billmers1991,Steudel2003}. Another example is the reaction $\mathrm{CO} + \mathrm{S_3} \leftrightarrow \mathrm{S_2} + \mathrm{OCS}$, which is  the most important production and destruction mechanism for OCS and S$_3$ in the deep atmosphere of our model. We use the rate constant $k = 10^{-11} \exp(-10,000/T)$ which is based on \citet{Krasnopolsky2007} and \citet{Krasnopolsky2013}, but they instead adopt a higher activation energy of 20,000 K. There are no laboratory measurements or ab initio calculations for this reaction. \citet{Krasnopolsky2007} chose the rate constant by considering the analogous reactions $\mathrm{CO} + \mathrm{NO_2}$ and $\mathrm{CO} + \mathrm{SO_2}$. A number of other rate constants in our network that are important to sulfur in Venus's deep atmosphere are also, at best, educated guesses. Improved simulations of Venus' sulfur chemistry requires new rate measurements or first-principle estimates for these uncertain reactions.



\texttt{Photochem} also does not capture the SO$_2$ depletion above $\sim 50$ km. As noted in Section \ref{sec:venus_chem}, one hypothesis has SO$_2$ dissolving in H$_2$SO$_4$ cloud droplets, a process we do not account for. \citet{Rimmer2021} used 1-D photochemical models to show that this is perhaps a plausible explanation for Venus' SO$_2$ depletion. Following \citet{Rimmer2021}, \citet{Dai2024} also used droplet chemistry in a separate photochemical model to match the observed SO$_2$ profile. Resolving this mystery would be challenging with only modeling and lab experiments. Spacecraft probes like DAVINCI \citep{Garvin2022} are likely needed to understand the role of Venus's clouds, if any, in sequestering SO$_2$.


No previous 1-D photochemical simulation of Venus's atmosphere from the surface to $\sim 110$ km \citep{Dai2024,Bierson2020,Wunderlich2023,Rimmer2021} used thermodynamics to reverse all chemical reactions. The \citet{Rimmer2021} model comes closest, which reverses all but a handful of reactions using the principle of microscopic reversibility (i.e., using Gibbs free energies). \citet{Rimmer2021} perhaps omitted several reversals because they involve species with uncertain thermodynamics. \citet{Dai2024}, \citet{Bierson2020} and \citet{Wunderlich2023} reversed many (but not all) reactions manually by specifying independent forward and reverse rate constants. The danger of reversing only selected reactions, or choosing forward and reverse rates independently from the literature, is that it can lead to a chemical network that is thermodynamically inconsistent \citep[e.g., Chapter 9 of][]{Kee2003}. To avoid this issue, our nominal model in Figure \ref{fig:venus} reverses all reactions with thermodynamics, and we argue that future models of Venus should do the same. 

Additionally, as discussed in Section \ref{sec:venus_climate}, photochemical models should take care when fixing gases at Venus's surface. Fixed surface abundances imply surface sources or sinks that may not be justified by a geologic process like volcanism. In the same vein, fixing a gas concentration throughout the atmospheric column, as has been done previously \citep{Dai2024,Bierson2020,Wunderlich2023}, is unphysical because it breaks mass conservation. Conservation of mass is important because it is the reason why sulfuric acid cloud formation by itself cannot explain the SO$_2$ depletion above Venus's clouds \citep[Reaction \ref{eq:so2_dest},][]{Rimmer2021}. In our model, we do not fix any gas throughout the atmospheric column. We opt instead to impose zero-flux surface boundary conditions for all molecules. The zero-flux assumption at Venus' surface is equivalent to assuming that atmosphere-surface interactions are not currently important for Venus' today. To be clear, our model is not an argument against surface gas deposition \citep[e.g.,][]{Constantinou2025} or the observational evidence for volcanism on modern Venus \citep[e.g.,][]{Filiberto2020}. Instead, our modeling approach contends that these surface processes do not substantially impact Venus' steady-state atmospheric composition today.

Through our climate simulations (Section \ref{sec:venus_climate}), we found that Venus's energy balance is sensitive to the CO$_2$-CO$_2$ CIA opacity, even for NIR wavelengths ($\lambda < 2.5$ $\mu$m) because the surface emits strongly in the NIR. Previous work has also pointed out the importance of the CO$_2$ continuum to Venus's net thermal flux \citep[e.g.,][]{Lee2016}. Further laboratory measurements of CIA opacity ($\lambda < 2.5$ $\mu$m for $400 < T < 750$ K) would likely improve our understanding of the planet's climate and Venus-like exoplanets.


\subsection{Clouds Are Hard to Predict and Are Important in Many Solar System Climates} \label{sec:discussion_cloud}

Although \texttt{Photochem} predicts the existence of clouds and hazes, it does not predict them accurately enough to reproduce the climates of Venus, Earth, Mars and Titan. To adequately simulate Venus's and Titan's P-T profile we had to replace the sulfuric acid and hydrocarbon hazes calculated by the model with cloud/haze optical properties derived from spacecraft and ground based observations (Section \ref{sec:venus_climate} and \ref{sec:titan_climate}). For Mars, we had to prescribe observed dust aerosol densities and radii to model the planet's climate, because \texttt{Photochem} does not have a scheme for predicting atmospheric dust (Section \ref{sec:mars_climate}). In our climate model of Earth, we parameterized water clouds with an increased surface albedo (Section \ref{sec:earth_climate}).

This shortcoming motivates an improved treatment of aerosols in future versions of \texttt{Photochem}. The current method in the photochemical module is crude: particles form with a prescribed particle radius, then fall until they evaporate or leave the bottom of the model domain. One benefit of this parameterization is that it conserves mass. A better scheme would include more aerosol microphysics. One option is to adopt a modification of the \citet{Ackerman2001} approach, which has previously been used by \citet{Windsor2023} to model water clouds with the \texttt{Atmos} climate code. We leave this update to future work. For now, the cloud and haze densities predicted by \texttt{Photochem} should be treated with caution when using them to predict a planet's climate.

\subsection{Simulating Atmospheres With a Self-Consistent Chemistry and Climate} \label{sec:discussion_couple}

To benchmark \texttt{Photochem} against Solar System atmospheres, Section \ref{sec:results} does not fully couple the photochemical and climate models to compute self-consistent composition and temperature profiles. As described in Section \ref{sec:methods_approach}, we choose to do fairly independent photochemical and climate simulations because they are easier to interpret when there are discrepancies between the model and observations. 

\begin{figure}
  \centering
  \includegraphics[width=\linewidth]{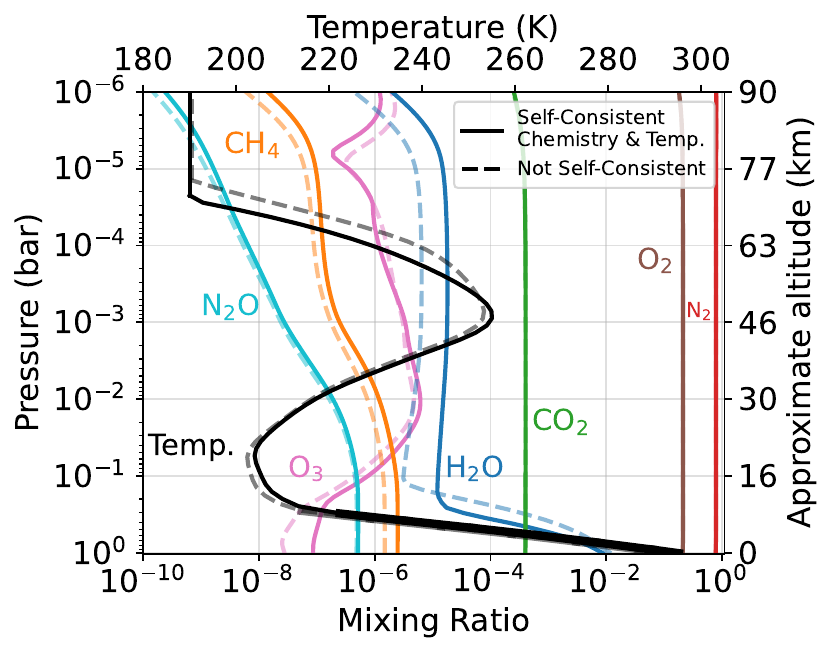}
  \caption{A self-consistent photochemical-climate simulation of the modern Earth (solid lines) compared to photochemical and climate simulations described in Section \ref{sec:earth} that are not self-consistent (dashed lines). Temperature (black lines) are referenced to the top axis, and mixing ratios (colored lines) are referenced to the bottom axis.}
  \label{fig:coupled_earth}
\end{figure}

However, self-consistent simulations are possible in \texttt{Photochem} by iterating between the photochemical and climate models using the approach described in \citet{Arney2016}. In brief, we first compute an estimated climate from the atmosphere's bulk constituents. Next, we feed the resulting P-T profile into the photochemical model and compute a steady-state composition. Then, using the output composition from the photochemical model, we compute a new P-T profile. We iterate back and forth in this manner until the temperature at all altitudes ceases to change to some tolerance (e.g., $< 1$ K). To illustrate this capability, Figure \ref{fig:coupled_earth} shows a self-consistent simulation of Earth (solid lines) with the same model setup and boundary conditions described in Section \ref{sec:earth}. Figure \ref{fig:coupled_earth} also plots the Figure \ref{fig:earth} and \ref{fig:earth_climate} models that are not self-consistent (dashed lines). The self-consistent model has an altered tropospheric H$_2$O concentration given that water vapor on Earth is largely controlled by its saturation vapor pressure (which depends on temperature). The different H$_2$O profile changes photolytic OH production, and OH impacts the concentration of many atmospheric constituents like O$_3$ and CH$_4$ (Section \ref{sec:earth_chem}).

Whether the complexity of a self-consistent simulation is justified depends on the problem. For example, self-consistent simulations were warranted in \citet{Kozakis2022} because the objective was to understand the feedback between ozone chemistry and stratospheric temperature and the resulting impact on an Earth-like planet's emission spectrum. On the other hand, de-coupled models are a more reasonable approach to simulating the photochemistry of the exoplanet K2-18b as a habitable ocean world \citep{Wogan2024}, because a habitable climate on such a planet is uncertain and perhaps not feasible \citep{Wogan2024,Innes2023,Leconte2024}.





\subsection{Comparison With the VULCAN 1-D Photochemical Model}

Here, we compare \texttt{Photochem} to \texttt{VULCAN} which is perhaps the most widely used open-source 1-D photochemical model for interpreting JWST observations. Note that Section \ref{sec:methods_history} already compares \texttt{Photochem} to the open-source \texttt{Atmos} photochemical model for temperate rocky planets. There are several closed-source photochemical models, like \texttt{ARGO} and \texttt{KINETICS} \citep{Rimmer2016,Allen1981}, but we do not address them here because they are hard to compare to without the source code.

\texttt{VULCAN} was first published by \citet{Tsai2017} as a kinetics/diffusion model without photolysis reactions and was later updated by \citet{Tsai2021} to include photons. The model was initially designed for warm gas giants and has since been adapted to rocky planets. \texttt{Photochem} had the opposite trajectory: \citet{Wogan2023} first used \texttt{Photochem} to model the early Earth atmosphere, and later \citet{Wogan2024} modified the code for gas giants. 

\texttt{Photochem} and \texttt{VULCAN} (the version used in \citet{Tsai2023}) differ in their approaches to solving the photochemical equations. For example, at each timestep, \texttt{VULCAN} re-normalizes gas number densities to maintain a hydrostatic atmosphere. This renormalization means \texttt{VULCAN} does not conserve mass. In contrast, \texttt{Photochem} conserves mass and maintains a hydrostatic atmosphere by incorporating the hydrostatic equation in the expressions for eddy and molecular diffusion (Appendix \ref{sec:eddy_diffusion} and \ref{sec:molecular_diffusion}). Another difference is that \texttt{VULCAN} uses a centered finite difference scheme to discretize vertical transport from molecular diffusion. As detailed in Section \ref{sec:methods_photochemistry_eqns}, \texttt{Photochem} instead uses an upwind scheme for molecular diffusion to maintain numerical stability. Regardless of these differences, for the same inputs, we find that \texttt{VULCAN} and \texttt{Photochem} produce nearly identical results for our hot Jupiter WASP-39b test case (Figure \ref{fig:wasp39b}b).

There are also differences between \texttt{VULCAN} and \texttt{Photochem}'s chemical networks and photolysis cross sections. To build both networks, chemical reaction rates were largely sourced from the NIST Kinetic Database, although some chemistry is poorly understood and not listed by NIST (e.g., Venus sulfur chemistry), so the two networks have discrepancies (Figure \ref{fig:wasp39b}). We believe it is valuable to apply both sets of chemical reactions to the same problem (i.e., interpreting a JWST exoplanet spectrum), because the degree to which the results differ should give a sense of the uncertainty in the models attributable to chemical kinetics.

\texttt{Photochem} is generally faster than the version of \texttt{VULCAN} used in \citet{Tsai2023}. \texttt{Photochem} is largely written in compiled Fortran (with a Python interface), while \texttt{VULCAN} is currently written in pure Python. In the Figure \ref{fig:wasp39b}b model of WASP-39b, where both models use nearly the exact same inputs, \texttt{Photochem} took 399 seconds for 3367 integration steps (8.4 steps/second) to reach a steady-state while \texttt{VULCAN} took 972 seconds over 4475 steps (4.6 steps/second). In this scenario, \texttt{Photochem} is about a factor of $\sim 2$ faster when considering integration steps per second.

There are some other practical difference that affect usage. \texttt{Photochem} is a standalone Python package that can be installed with the Conda package manager, then imported using standard Python syntax. Also, \texttt{Photochem} can be used in Python scripts or interactively in Jupyter Notebooks. In contrast, \texttt{VULCAN} cannot be installed and imported as a standard Python package or easily used in a Jupyter Notebook. To use the code, you must download the \texttt{VULCAN} repository from GitHub, then run the model's ``driver'' script.

\subsection{Other Open-Source 1-D Climate Models for Rocky Planets and A Benchmark With HELIOS} \label{sec:discussion_climate}

The climate model in \texttt{Photochem} is one of relatively few open-source 1-D codes designed for rocky planets. Other examples include \texttt{Atmos}, \texttt{Exo\_k}, and \texttt{HELIOS} \citep{Leconte2021,Selsis2023,Malik2017,Malik2019a,Malik2019b}. We consider the climate model in \texttt{Photochem} to be an updated version of the one in \texttt{Atmos} (Section \ref{sec:methods_history}). \texttt{Exo\_k} is a Python based package that was initially designed to manage gas opacities \citep{Leconte2021}, but now includes a 1-D climate module \citep{Selsis2023}. \texttt{Exo\_k} has many similar features to \texttt{Photochem}, such as its ability to consider an arbitrary number of condensibles and their impact on the lapse rate. We anticipate that the two codes will be benchmarked against each other as a part of a CUISINES model inter-comparison \citep{Sohl2024}. \texttt{HELIOS} version 3.1 can simulate both gas giants and rocky planet atmospheres, although it has a limited ability to treat moist convection. Recently, \texttt{HELIOS} has been a popular model used by the community to evaluate JWST emission observations of warm rocky planets orbiting M stars \citep[e.g.,][]{Ih2023,Xue2024,Fortune2025}. \texttt{HELIOS} is unique because it makes use of graphics processing units (GPUs) whereas \texttt{Photochem} and many other climate models use CPUs. GPUs are generally considered to be faster than CPUs for highly parallel problems like radiative transfer where calculations are wavelength independent. 

\begin{figure}
  \centering
  \includegraphics[width=\linewidth]{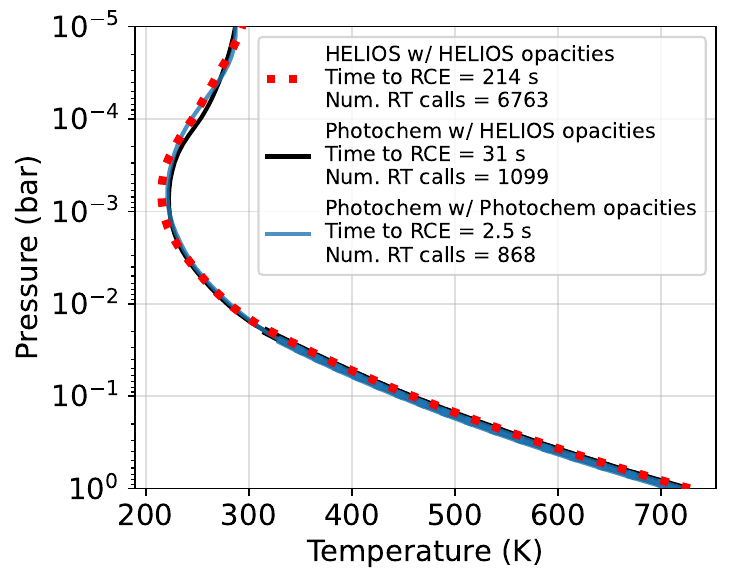}
  \caption{A benchmark between the climate model in \texttt{Photochem} and the \texttt{HELIOS} GPU code for a CO$_2$-H$_2$O atmosphere (50\% each) on an Earth-sized planet orbiting the Sun at 0.5 AU. Three cases are shown: \texttt{HELIOS} with \texttt{HELIOS} k-distributions (red dotted line), \texttt{Photochem} with \texttt{HELIOS} k-distributions (black line), and \texttt{Photochem} with \texttt{Photochem} k-distributions (blue line). See text for more benchmark details. The acronyms ``RCE'' and ``RT'' in the legend stand for ``radiative-convective equilibrium'' and ``radiative transfer'', respectively. Both codes produce very similar P-T profiles, and \texttt{Photochem}'s root-finding approach (Section \ref{sec:methods_climate}) reaches RCE in $\sim 6 \times$ fewer radiative transfer calls.}
  \label{fig:helios_benchmark}
\end{figure}

Here, we conduct a benchmark between \texttt{HELIOS} version 3.1 and \texttt{Photochem} to gauge possible performance differences between a GPU and CPU code. The test computes radiative-convective equilibrium for a planet with Earth's mass, Earth's radius, a 1 bar atmosphere composed of 50\% CO$_2$ and H$_2$O, set 0.5 AU from the Sun. We consider three cases: \texttt{HELIOS} with \texttt{HELIOS} k-distributions, \texttt{Photochem} with \texttt{HELIOS} k-distributions, and \texttt{Photochem} with \texttt{Photochem} k-distributions (Table \ref{tab:kdistributions}). For all scenarios, the code only uses CO$_2$ and H$_2$O k-distributions and omits all other opacity sources (i.e., no Rayleigh scattering or CIA). The tests that use \texttt{HELIOS}'s k-distributions, adopt the k-tables shipped with version 3.1 of the model (385 wavelength bins each with twenty k-terms). Both codes use the same ``resort rebin'' method to mix k-distributions \citep{Amundsen2017}. We ignore condensation and use a dry adiabat. We use \texttt{HELIOS}'s adaptive time stepping scheme and assume an initially isothermal 500 K P-T profile.

Figure \ref{fig:helios_benchmark} shows that the predicted P-T profiles are very similar. With \texttt{HELIOS}'s opacities, \texttt{Photochem} took 31 seconds and 1099 radiative transfer calculations to reach radiative-convective equilibrium using 8 cores of a MacBook M1 Pro. \texttt{HELIOS} took 214 seconds and 6763 radiative transfer calls using an NVIDIA Tesla V100 16 GB GPU. \texttt{Photochem}'s root finding approach (Section \ref{sec:methods_climate}) speeds convergence with $\sim 6 \times$ fewer radiative transfer calls and an overall runtime that is $\sim 7 \times$ shorter (for this hardware and these opacities). In terms of radiative transfer calls per second, \texttt{HELIOS} and \texttt{Photochem} exhibit similar performance for the same opacities. We suspect that \texttt{HELIOS}'s GPU radiative transfer is not faster because the \texttt{HELIOS} k-tables are at fairly low resolution ($R \approx 50$), so the problem may not be parallel enough to benefit significantly from GPUs. 

All climate calculations in Section \ref{sec:results} use even lower resolution k-tables ($2 < R < 60$, with an average $R \approx 15$) and a smaller number of k-terms (eight) than adopted by \texttt{HELIOS}'s opacities. When we repeat the climate benchmark with \texttt{Photochem}'s lower resolution opacities (Table \ref{tab:kdistributions}), \texttt{Photochem} iterates to nearly the same P-T profile in just 2.5 seconds on the same hardware (blue line in Figure \ref{fig:helios_benchmark}), $86\times$ faster than the \texttt{HELIOS} calculation. We were unable to test our low resolution opacities with \texttt{HELIOS} because the latter is limited to k-distributions with twenty terms. This result illustrates that the $R \approx 50$ and 20 term \texttt{HELIOS} k-distributions are excessive for this test case. Adopting \texttt{Photochem} with its nominal opacities for problems similar to this benchmark, such as modeling a warm rocky exoplanet observed by JWST, should be $\sim 10$ to $100\times$ faster than using \texttt{HELIOS} (with its nominal opacities) without sacrificing meaningful accuracy.

\section{Conclusions}

To interpret spectroscopic observations of exoplanet atmospheres, the community needs models of atmospheric chemistry and climate that are applicable to the widest possible diversity of planetary atmospheres. To this end, we present the open-source 1-D photochemical and climate model \texttt{Photochem}. To demonstrate the model's generality, we benchmark it against the observed compositions and climates of Venus, Earth, Mars, Titan and Jupiter using a single set of kinetics, thermodynamics and opacities. We also model the chemistry of the hot-Jupiter WASP-39b. All of these tests are open-source and fully reproducible to best support future model development (\url{https://doi.org/10.5281/zenodo.16509802}). From our tour of the Solar System and one exoplanet our conclusions are as follows:

\begin{itemize}
  \item To first order, \texttt{Photochem} predicts gas concentrations and P-T profiles that are broadly consistent with observations of five Solar System atmospheres (Venus, Earth, Mars, Titan and Jupiter). The code also reproduces the photochemical SO$_2$ observed by JWST in WASP-39b's atmosphere.
  \item Of the Solar System worlds, the most egregious model-data discrepancies involve sulfur species in Venus's atmosphere, motivating an improved understanding of sulfur's chemical kinetics. We also urge that future Venus photochemical models use networks that are thermodynamically self-consistent and explicitly conserve mass (Section \ref{sec:discussion_venus}). Furthermore, near-infrared measurements of CO$_2$-CO$_2$ CIA opacity would improve our understanding of Venus' climate.
  \item \texttt{Photochem} predicts clouds and hazes that do not yet accurately reproduce the climates of planets, motivating an improved treatment of aerosols in future versions of the model. We reproduced these climates by prescribing observed cloud or haze optical properties.
  \item We benchmark the photochemical model in \texttt{Photochem} against the \texttt{VULCAN} code. We find that the two models are in excellent agreement for the same inputs, and we find that \texttt{Photochem} is about twice as fast (Section \ref{sec:wasp39b}).
  \item We also found that our climate model in \texttt{Photochem} agrees with the \texttt{HELIOS} GPU-accelerated climate model for the same problem. In our test case, \texttt{Photochem}'s root-finding approach for computing radiative-convective equilibrium takes $\sim 6 \times$ less radiative-transfer calls than \texttt{HELIOS}'s time-integration scheme. Radiative transfer calculations by the two codes have comparable run times on the hardware we tested, suggesting that the correlated-k radiative transfer in our benchmark case is not sufficiently parallel to benefit from GPUs.
\end{itemize}

Overall, \texttt{Photochem} provides a general description of atmospheric chemistry and physics that captures a wide range of planetary atmospheres. We conclude that \texttt{Photochem} is a good option for future study of exoplanets.

\begin{acknowledgments}

We thank our two reviewers who improved the quality of this article. Conversations with David Crisp, Sonny Harman, Mark Claire, Sandra Bastelberger, Ravi Kopparapu, Andrew Lincowski, Sukrit Ranjan, and Shang-min Tsai benefited the development of \texttt{Photochem}. NFW was supported by the NASA Postdoctoral Program. N.E.B acknowledges support from NASA'S Interdisciplinary Consortia for Astrobiology Research (NNH19ZDA001N-ICAR) and JWST Theory Grant JWST-AR-02232- 001-A. VSM, ETW, TDR, and DCC acknowledge support from the Virtual Planetary Laboratory, a member of the NASA Nexus for Exoplanet System Science (NExSS), funded via the NASA Astrobiology Program grant Nos. 80NSSC23K1398 and 80NSSC18K0829, respectively. DCC and JKT also acknowledge support from the Alfred P. Sloan Foundation under Grant No. 2025-25204.

\end{acknowledgments}

\begin{contribution}

NFW wrote the original paper draft, performed the calculations, wrote and manages the \texttt{Photochem} code. KZ built the reaction and thermodynamic database. NEB and KZ provided numerical and coding advice to NFW as he wrote and developed \texttt{Photochem}. All authors contributed to writing the manuscript.

\end{contribution}

\appendix

\section{The Photochemical Equations} \label{sec:photochem_eqns}

In this Appendix, we derive the fundamental equations solved in the photochemistry module of \texttt{Photochem}. Some of our derivation overlaps with Appendix C.1 in \citet{Wogan2023}, however, we believe that the repetition is justified because some of our assumptions have changed, and the derivation here is more complete than the one in \citet{Wogan2023}.

\subsection{The Continuity Equations}

We begin our derivation of the governing photochemical equations with a statement of the conservation of molecules. Such statements of conservation are often called continuity equations.
\begin{equation} \label{eq:continuity}
  \frac{\partial n_{i}}{\partial t} = - \frac{\partial \Phi_{i}}{\partial z} + \sigma_i
\end{equation}
Here, $n_i$ is the number density of molecule $i$ in units of molecules cm$^{-3}$; $t$ is time in seconds; $z$ is altitude in cm; $\Phi_{i}$ is the flux of species $i$ in units of molecules cm$^{-2}$ s$^{-1}$; and $\sigma_i$ represents the sources and sinks of species $i$ in units of molecules cm$^{-3}$ s$^{-1}$. Equation \eqref{eq:continuity} states that a molecule's concentration changes over time at a point in space in response to molecules entering or leaving the space (i.e., $\Phi_{i}$) and the production or destruction of molecules (i.e., $\sigma_i$). This version of the continuity equation is one-dimensional and assumes that species concentrations change only in the vertical ($z$) direction.

The flux of a gas species $i$ ($\Phi_{i\text{,gas}}$) is determined by eddy and molecular diffusion. The flux of particles ($\Phi_{i\text{,particle}}$) is approximated by how rapidly particles fall through the atmosphere.
\begin{align}
  \Phi_{i\text{,gas}} &= \Phi_{i\text{,eddy}} + \Phi_{i\text{,molecular}} \label{eq:phi_gas} \\
  \Phi_{i\text{,particle}} &= \Phi_{i\text{,fall}} \label{eq:phi_particle}
\end{align}
Here, $\Phi_{i\text{,eddy}}$, $\Phi_{i\text{,molecular}}$ and $\Phi_{i\text{,fall}}$ are approximated by
\begin{align}
  \Phi_{i\text{,eddy}} &= - K_{zz} n_i \left(\frac{1}{n_i} \frac{\partial n_i}{\partial z} + \frac{1}{H_a} + \frac{1}{T} \frac{\partial T}{\partial z}\right) \label{eq:phi_eddy}\\ 
  \Phi_{i\text{,molecular}} &= - D_{i} n_i \left( \frac{1}{n_i} \frac{\partial n_i}{\partial z} + \frac{1}{H_i} + \frac{1}{T} \frac{\partial T}{\partial z} \right) \label{eq:phi_molecular} \\
  \Phi_{i\text{,fall}} &= - w_i n_i \label{eq:phi_fall}
\end{align}
Here, $K_{zz}$ is the eddy diffusion coefficient in cm$^2$ s$^{-1}$; $H_a$ is the scale height in cm; $T$ is temperature in Kelvin; $D_i$ is the molecular diffusion coefficient of species $i$ in cm$^2$ s$^{-1}$; and $H_i$ is scale height of species $i$ in cm. Appendix \ref{sec:eddy_diffusion} derives Equation \eqref{eq:phi_eddy} while Appendix \ref{sec:molecular_diffusion} derives Equation \eqref{eq:phi_molecular} starting with the general molecular diffusion equation. Equations \eqref{eq:phi_eddy} and \eqref{eq:phi_molecular} assume the atmosphere is an ideal gas at hydrostatic equilibrium. Furthermore, Equation \eqref{eq:phi_molecular} assumes that the diffusing gas is a minor atmospheric constituent and ignores thermal diffusion. In Equation \eqref{eq:phi_fall}, $w_i$ is the fall velocity of a particle, which can be estimated with Stokes' law (Appendix \ref{sec:fall_velocity}). 

In a real atmosphere, particles are also transported by turbulent mixing (i.e., parameterized with eddy diffusion), but we choose to ignore this effect because we do not include detailed microphysics. Without microphysics, we have found that particle turbulent mixing can lead to undesirable solutions. For example, a large eddy diffusion coefficient can loft small particles in a way that eliminates a cold trap. While such a process can sometimes occur in a real atmosphere, a credible model of it would require microphysics that we do not account for.

We assume that the sources and sinks of a molecule or particle $i$ are chemical production ($P_{i}$), chemical loss ($L_i$), rainout in droplets of liquid ($R_{i\text{,rainout}}$), and condensation or evaporation ($E_{i}$):
\begin{equation} \label{eq:source_terms}
  \sigma_i = P_{i} - L_{i} - R_{i\text{,rainout}} + E_{i}
\end{equation}
\texttt{Photochem} computes chemical production and loss using the elementary, three-body, and photolysis reactions detailed in Section \ref{sec:methods_photochemistry_network}. 
We model rainout following the \citet{Georgii1980} parameterization with solubility data from \citet{Sander2015}. Appendix \ref{sec:condense_evap} describes our treatment of condensation and evaporation.

To facilitate a finite volume discretization (Appendix \ref{sec:finite_volume}), below, we reorganize Equation \eqref{eq:phi_gas} into advective and diffusive terms:
\begin{gather}
  \Phi_{i\text{,gas}} = - \alpha_i \frac{\partial n_i}{\partial z} - \beta n_i - \gamma_{i} n_i \label{eq:phi_gas1} \\
  \alpha_i = K_{zz} + D_i \\
  \beta = K_{zz} \left(\frac{1}{H_a} + \frac{1}{T} \frac{\partial T}{\partial z}\right) \\
  \gamma_{i} = D_i \left(\frac{1}{H_i} + \frac{1}{T} \frac{\partial T}{\partial z}\right)
\end{gather}
The 1-D continuity equation (Equation \eqref{eq:continuity}) and corresponding fluxes (Equations \eqref{eq:phi_gas1} and \eqref{eq:phi_particle}) and source terms (Equation \eqref{eq:source_terms}) are a system of partial differential equations (PDEs) describing how the number density ($n_{i}$) of each chemical species $i$ changes over altitude and time. In the following Appendix subsection, we discretize this system of PDEs using a finite volume method.

\subsection{Finite Volume Discretization} \label{sec:finite_volume}

We use the finite volume method to discretize Equation \eqref{eq:continuity} allowing an approximate numerical solution. The finite volume method divides the spatial domain ($z$) into grid cells, or finite volumes, in order to approximate the cell-averaged value of a species number density, $n_i$. Approximations to the cell-averaged value of $n_i$ are modified over time by considering the flux of molecules through the edges of the cells, and the sources and sinks of molecules within the cell. The main advantage of this approach is that the approximation conserves molecules, which is not always the case for other PDE discretization methods (e.g., finite difference or finite element methods). Molecule conservation is valuable for understanding whether a model run has reached steady-state, or interrogating the redox fluxes into and out of an atmosphere. Below, we describe the application of the finite volume method to Equation \eqref{eq:continuity}. For an in-depth understanding of the finite volume method, see \citet{Leveque2002}.

Consider an atmosphere that has been vertically divided into $m$ grid-cell (i.e. finite volumes) of thickness $\Delta z^j$. The superscript $j$ refers to a grid cell ($j$ does not refer to a power), where $j = l$ is the lowest altitude grid-cell $j = m$ highest grid-cell. $z^j$ is the altitude at the center of grid-cell $j$, such that the upper and lower edges of the cell are given by $z^{j+1/2} = z^j + \frac{\Delta z^j}{2}$ and $z^{j-1/2} = z^j - \frac{\Delta z^j}{2}$, respectfully.

Consider a single grid cell between altitudes $z = z^{j-1/2}$ and $z = z^{j+1/2}$. Integrating Equation \eqref{eq:continuity} from the bottom to the top of the grid cell gives
\begin{equation} \label{eq:continuity_simple_conservative}
  \frac{\partial}{\partial t} \int_{z^{j-1/2}}^{z^{j+1/2}} n_i dz = - (\Phi_{i}(z^{j+1/2}) - \Phi_{i}(z^{j-1/2})) + \int_{z^{j-1/2}}^{z^{j+1/2}} \sigma_i dz
\end{equation}
The average value of $n_i$ and $\sigma_i$ between $z^{j-1/2}$ and $z^{j+1/2}$ are given by the integrals
\begin{align}
  \overline{n_i}^j = \frac{1}{\Delta z^j} \int_{z^{j-1/2}}^{z^{j+1/2}} n_i dz \\
  \overline{\sigma_i}^j = \frac{1}{\Delta z^j} \int_{z^{j-1/2}}^{z^{j+1/2}} \sigma_i dz
\end{align}
Substituting the above expressions into Equation \eqref{eq:continuity_simple_conservative} yields
\begin{equation} \label{eq:continuity_simple_conservative1}
  \frac{\partial}{\partial t} \overline{n_i}^j(t) = - \frac{\Phi_{i}(z^{j+1/2},t) - \Phi_{i}(z^{j-1/2},t)}{\Delta z^j} + \overline{\sigma_i}^j(t)
\end{equation}
Equation \eqref{eq:continuity_simple_conservative1} states that the average number density of species $i$ in the $j$ atmospheric layer ($\overline{n_i}^j$) changes over time because of the fluxes at the edges of the layer (e.g. $\Phi_{i}(z^{j+1/2},t)$), and the average production or loss of the species in the layer ($\overline{\sigma_i}^j$). The above formula is exact for $\overline{n_i}^j(t)$. However, now we introduce some approximations. First, we assume that $\overline{n_i}^j(t)$, $\overline{\sigma_i}^j(t)$, $\Phi_{i}(z^{j+1/2},t)$ and $\Phi_{i}(z^{j-1/2},t)$ are constant over small segments of time. Also, we approximate $\Phi_{i}(z^{j+1/2},t)$, $\Phi_{i}(z^{j-1/2},t)$ and $\overline{\sigma_i}^j(t)$ in terms of cell-averaged quantities, such as $\overline{n_i}^j$. We denote the approximate quantities in the following way
\begin{equation}
  \begin{aligned}
    n_i^{j} &\approx \overline{n_i}^{j} \\
    \sigma_i^j &\approx \overline{\sigma_i}^j(t)
  \end{aligned}
  \quad\quad\quad
  \begin{aligned}
    \Phi_{i}^{j+1/2} &\approx \Phi_{i}(z^{j+1/2},t) \\
    \Phi_{i}^{j-1/2} &\approx \Phi_{i}(z^{j-1/2},t)
  \end{aligned}
\end{equation}
Substituting the approximate quantities into Equation \eqref{eq:continuity_simple_conservative1} gives
\begin{equation} \label{eq:continuity_simple_approx}
  \frac{\partial n_i^j}{\partial t} = - \frac{\Phi_{i}^{j+1/2} - \Phi_{i}^{j-1/2}}{\Delta z^j} + \sigma_i^j
\end{equation}
We now must derive expressions for the gas and particle fluxes at grid-cell edges that are accurate and stable for our problem. 

\subsubsection{Gases} \label{sec:FV_gases}

The equation for gas flux (Equation \eqref{eq:phi_gas1}) contains the $\gamma_i$ advection term which can be much stronger than the diffusion term ($\alpha_i$), especially high in the atmosphere when a heavy species (e.g., S$_8$) is transported through a light background gas (e.g., H$_2$). The $\beta$ advection term, on the other hand, is generally small compared to the diffusion term. Therefore, for stability, we use a first-order upwind scheme for the $\gamma_i$ term, and a centered scheme for the $\alpha_i$ and $\beta$ terms. To accomplish this, we first separate Equation \eqref{eq:phi_gas1} in the following way
\begin{gather}
  \Phi_{i\text{,gas}} = \Phi_{i\text{,center}} + \Phi_{i\text{,upwind}} \\
  \Phi_{i\text{,center}} = - \alpha_i \frac{\partial n_i}{\partial z} - \beta n_i \\
  \Phi_{i\text{,upwind}} = - \gamma_{i} n_i
\end{gather}
Therefore, we can write Equation \eqref{eq:continuity_simple_approx} as
\begin{equation} \label{eq:continuity_simple_approx_1}
  \frac{\partial n_{i\in\text{gas}}^j}{\partial t} = - \frac{\Phi_{i\text{,center}}^{j+1/2} - \Phi_{i\text{,center}}^{j-1/2}}{\Delta z^j} - \frac{\Phi_{i\text{,upwind}}^{j+1/2} - \Phi_{i\text{,upwind}}^{j-1/2}}{\Delta z^j} + \sigma_i^j
\end{equation}
Applying a centered scheme (Equation 4.10 in \cite{Leveque2002}) to $\Phi_{i\text{,center}}$ yields:
\begin{gather}
  \Phi_{i\text{,center}}^{j+1/2} = - \alpha_i^{j+1/2} \left( \frac{\partial n_{i}}{\partial z} \right)^{j+1/2} - \beta_{i}^{j+1/2} n_{i}^{j+1/2} \\
  \left( \frac{\partial n_{i}}{\partial z} \right)^{j+1/2} \approx \frac{n_i^{j+1} - n_i^{j}}{\Delta z^{j+1/2}}
\end{gather}
Where $\Delta z^{j+1/2} = \frac{1}{2}\Delta z^j + \frac{1}{2} \Delta z^{j+1}$. Also, $n_{i}^{j+1/2}$ can be estimated by linearly interpolating between the two adjacent cells.

$$n_{i}^{j+1/2} \approx \frac{\Delta z^{j}n_i^{j+1} + \Delta z^{j+1} n_i^{j}}{\Delta z^{j+1}+\Delta z^{j}}$$
Therefore,

\begin{equation} \label{eq:phi_gas_p}
  \Phi_{i\text{,center}}^{j+1/2} = - \alpha^{j+1/2} \frac{n_i^{j+1} - n_i^{j}}{\Delta z^{j+1/2}} - \beta_{i}^{j+1/2} \frac{\Delta z^{j}n_i^{j+1} + \Delta z^{j+1} n_i^{j}}{\Delta z^{j+1}+\Delta z^{j}}
\end{equation}
Using a similar procedure, $\Phi_{i\text{,gas}}^{j-1/2}$ is given by

\begin{equation} \label{eq:phi_gas_m}
  \Phi_{i\text{,center}}^{j-1/2} = - \alpha^{j-1/2} \frac{n_i^{j} - n_i^{j-1}}{\Delta z^{j-1/2}} - \beta_{i}^{j-1/2} \frac{\Delta z^{j-1}n_i^{j} + \Delta z^{j} n_i^{j-1}}{\Delta z^{j}+\Delta z^{j-1}}
\end{equation}

To apply a first-order upwind scheme to $\Phi_{i\text{,upwind}}$, we first recognize that $\gamma_i$ is often positive, except under extreme temperature gradients, so information frequently travels from up to down in the model. Therefore, an upwind scheme approximates the advective fluxes with a forward difference
\begin{equation} \label{eq:upwind}
  \frac{\Phi_{i\text{,upwind}}^{j+1/2} - \Phi_{i\text{,upwind}}^{j-1/2}}{\Delta z^j} \approx \frac{\Phi_{i\text{,upwind}}^{j+1} - \Phi_{i\text{,upwind}}^{j}}{\Delta z^j} = \frac{-\gamma_i^{j+1} n_i^{j+1} + \gamma_i^{j} n_i^{j}}{\Delta z^j}
\end{equation}

Substituting Equations \eqref{eq:phi_gas_p}, \eqref{eq:phi_gas_m} and \eqref{eq:upwind} into Equation \eqref{eq:continuity_simple_approx_1} and re-arranging gives an expression of the form

\begin{equation} \label{eq:dfdt_general}
\begin{split} 
  \frac{\partial n_{i}^j}{\partial t} &= (A_{i}^{j\text{,upper}}+ B_{i}^{j\text{,upper}}) n_{i}^{j+1} \\
  &+ (A_{i}^{j\text{,center}} + B_{i}^{j\text{,center}}) n_{i}^{j} \\
  &+ (A_{i}^{j\text{,lower}} + B_{i}^{j\text{,lower}}) n_{i}^{j-1} \\
  &+ \sigma_i^j
\end{split}
\end{equation}

The $A$ and $B$ coefficients for gases correspond to the diffusive and advective approximations to the cell-edge fluxes, respectively:

\begin{equation} \label{eq:a_b_gas}
\begin{aligned}
  A_{i\text{,gas}}^{j\text{,upper}} &= \frac{\alpha_i^{j+1/2}}{\Delta z^j \Delta z^{j+1/2}} \\
  A_{i\text{,gas}}^{j\text{,center}} &= - \frac{\alpha_i^{j+1/2}}{\Delta z^j \Delta z^{j+1/2}} - \frac{\alpha_i^{j-1/2}}{\Delta z^j \Delta z^{j-1/2}} \\
  A_{i\text{,gas}}^{j\text{,lower}} &= \frac{\alpha_i^{j-1/2}}{\Delta z^j \Delta z^{j-1/2}} \\
  B_{i\text{,gas}}^{j\text{,upper}} &= \frac{\beta_{i}^{j+1/2}}{\Delta z^{j+1}+\Delta z^{j}} + \frac{\gamma_i^{j+1}}{\Delta z^{j}} \\
  B_{i\text{,gas}}^{j\text{,center}} &= \beta_{i}^{j+1/2} \frac{\Delta z^{j+1}}{\Delta z^{j}(\Delta z^{j+1}+\Delta z^{j})} - \beta_{i}^{j-1/2} \frac{\Delta z^{j-1}}{\Delta z^{j}(\Delta z^{j}+\Delta z^{j-1})} - \frac{\gamma_i^j}{\Delta z^j}  \\
  B_{i\text{,gas}}^{j\text{,lower}} &= - \frac{\beta_{i}^{j-1/2}}{\Delta z^{j}+\Delta z^{j-1}}
\end{aligned}
\end{equation}

The lower boundary condition is the flux through the lower edge of the lowest-altitude cell, $\Phi_{i}^{l-1/2} = \Phi_{i}^\text{lower}$. Similarly, the upper boundary condition is the flux through through the top edge of the highest-altitude cell, $\Phi_{i}^{m+1/2} = \Phi_{i}^\text{upper}$. It follows from Equation \eqref{eq:continuity_simple_approx} that

\begin{align}
  \frac{\partial n_i^l}{\partial t} &= - \frac{\Phi_{i}^\text{l+1/2} - \Phi_{i}^\text{lower}}{\Delta z^l} + \sigma_i^l \label{eq:lower_bc} \\
  \frac{\partial n_i^m}{\partial t} &= - \frac{\Phi_{i}^\text{upper} - \Phi_{i}^{m-1/2}}{\Delta z^m} + \sigma_i^m \label{eq:upper_bc}
\end{align}
Plugging in Equation \eqref{eq:phi_gas_p}, \eqref{eq:phi_gas_m} and \eqref{eq:upwind} into Equation \eqref{eq:lower_bc} and \eqref{eq:upper_bc} gives the following general expression for the rate densities change at the boundaries

\begin{equation} \label{eq:dfdt_boundary}
\begin{aligned}
  \frac{\partial n_{i}^l}{\partial t} &= (A_{i}^{l\text{,upper}}+ B_{i}^{l\text{,upper}}) n_{i}^{l+1} \\
  &+ (A_{i}^{l\text{,center}} + B_{i}^{l\text{,center}}) n_{i}^{l} \\
  &+ \frac{\Phi_i^\text{lower}}{\Delta z^l} \\
  &+ \sigma_i^l
\end{aligned}
\quad\quad\quad
\begin{aligned}
  \frac{\partial n_{i}^m}{\partial t} &= (A_{i}^{m\text{,center}} + B_{i}^{m\text{,center}}) n_{i}^{m} \\
  &+ (A_{i}^{m\text{,lower}} + B_{i}^{m\text{,lower}}) n_{i}^{m-1} \\
  &- \frac{\Phi_i^\text{upper}}{\Delta z^m} \\
  &+ \sigma_i^m
\end{aligned}
\end{equation}

For gases, the $A$ and $B$ coefficients for the lower boundary are
\begin{equation} \label{eq:a_b_gas_lower}
  \begin{aligned}
  A_{i\text{,gas}}^{l\text{,upper}} &= \frac{\alpha_i^{l+1/2}}{\Delta z^l \Delta z^{l+1/2}} \\
  A_{i\text{,gas}}^{l\text{,center}} &= - \frac{\alpha_i^{l+1/2}}{\Delta z^l \Delta z^{l+1/2}} \\
  \end{aligned}
  \quad\quad
  \begin{aligned}
  B_{i\text{,gas}}^{l\text{,upper}} &= \frac{\beta_{i}^{l+1/2}}{\Delta z^{l+1}+\Delta z^{l}} + \frac{\gamma_i^{l+1}}{\Delta z^{l}} \\
  B_{i\text{,gas}}^{l\text{,center}} &= \beta_{i}^{l+1/2} \frac{\Delta z^{l+1}}{\Delta z^{l}(\Delta z^{l+1}+\Delta z^{l})} \\
  \end{aligned}
\end{equation}
and the upper boundary,
\begin{equation} \label{eq:a_b_gas_upper}
  \begin{aligned}
  A_{i\text{,gas}}^{m\text{,center}} &= - \frac{\alpha_i^{m-1/2}}{\Delta z^m \Delta z^{m-1/2}} \\
  A_{i\text{,gas}}^{m\text{,lower}} &= \frac{\alpha_i^{m-1/2}}{\Delta z^m \Delta z^{m-1/2}}
  \end{aligned}
  \quad\quad
  \begin{aligned}
  B_{i\text{,gas}}^{m\text{,center}} &= - \beta_{i}^{m-1/2} \frac{\Delta z^{m-1}}{\Delta z^{m}(\Delta z^{m}+\Delta z^{m-1})} - \frac{\gamma_i^m}{\Delta z^m} \\
  B_{i\text{,gas}}^{m\text{,lower}} &= - \frac{\beta_{i}^{m-1/2}}{\Delta z^{m}+\Delta z^{m-1}}
  \end{aligned}
\end{equation}

\subsubsection{Particles} \label{sec:FV_particles}

We assume particle transport is governed by the rate they fall through the atmosphere (Equation \eqref{eq:phi_fall}). The flux is an advection equation, so we use a first order upwind scheme for stability. Additionally, the fall velocity $w_i$ is always positive (i.e. particles always fall downward), so an upwind scheme should approximate the advective fluxes with a forward difference. In the previous subsection, we applied the same forward upwind scheme to the $\gamma_i$ advective terms for gases. Therefore, by analogy to the $\gamma_i$ terms in Equation \eqref{eq:a_b_gas}, the $A$ and $B$ coefficients in Equation \eqref{eq:dfdt_general} for particles are,
\begin{equation} \label{eq:a_b_particle}
\begin{aligned}
  A_{i\text{,particle}}^{j\text{,upper}} &= 0 \\
  A_{i\text{,particle}}^{j\text{,center}} &= 0 \\
  A_{i\text{,particle}}^{j\text{,lower}} &= 0 \\
\end{aligned}
\quad\quad\quad
\begin{aligned}
  B_{i\text{,particle}}^{j\text{,upper}} &= \frac{w_i^{j+1}}{\Delta z^{j}} \\
  B_{i\text{,particle}}^{j\text{,center}} &= - \frac{w_i^j}{\Delta z^j}  \\
  B_{i\text{,particle}}^{j\text{,lower}} &= 0
\end{aligned}
\end{equation}
In the same vein, all of the $A$ coefficients in Equation \eqref{eq:dfdt_boundary} for particles are zero. The $B$ coefficients for the boundaries are
\begin{equation} \label{eq:a_b_particle_boundary}
  \begin{aligned}
  B_{i\text{,particle}}^{l\text{,upper}} &= \frac{w_i^{l+1}}{\Delta z^{l}} \\
  B_{i\text{,particle}}^{l\text{,center}} &= 0 \\
  \end{aligned}
  \quad\quad
  \begin{aligned}
  B_{i\text{,particle}}^{m\text{,center}} &= - \frac{w_i^m}{\Delta z^m} \\
  B_{i\text{,particle}}^{m\text{,lower}} &= 0
  \end{aligned}
\end{equation}

\subsection{Solving the Finite Volume Discretization} \label{sec:solve}

Equations \eqref{eq:dfdt_general} and \eqref{eq:dfdt_boundary} and the corresponding coefficients for gases and particles (Equations \eqref{eq:a_b_gas}, \eqref{eq:a_b_gas_lower}, \eqref{eq:a_b_gas_upper}, \eqref{eq:a_b_particle} and \eqref{eq:a_b_particle_boundary}) are a system of ordinary differential equations (ODEs) which are a finite volume approximation of our original system of PDEs (Equation \eqref{eq:continuity}). Section \ref{sec:methods_photochemistry_eqns} describes how \texttt{Photochem} solves these equations. In brief, we evolve Equations \eqref{eq:dfdt_general} and \eqref{eq:dfdt_boundary} using the CVODE stiff ODE integrator \citep{Hindmarsh2005}. In \texttt{Photochem}, CVODE can be used to study the time-evolution of an atmosphere, or alternatively integrate the atmosphere to a steady-state composition ($d n_{i}^j/d t = 0$).

\subsection{Condensation and Evaporation} \label{sec:condense_evap}

\texttt{Photochem}'s scheme for condensation and evaporation accomplishes the following simple goals: Gas-phase species should be converted to particles if they are above a given relative humidity (RH), and particles should be converted to gases if the gas is below a specified RH. Another requirement is that the rate of gas-to-particle or particle-to-gas conversion is smooth such that it does not introduce excessive ``stiffness'' in the system of ODEs. To accomplish these objectives, \texttt{Photochem} computes the rate constant of condensation with
\begin{equation}
  k_\mathrm{cond} =
  \begin{cases}  
  A_\mathrm{cond} \left(\frac{K_{zz}}{H_a^2}\right) \left(\frac{2}{\pi}\right) \arctan\left(\frac{r - r_c}{S r_c}\right) & r > r_c \\
  0 & r \leq r_c
  \end{cases}
\end{equation}
Here, $A_\mathrm{cond}$ is a condensation rate coefficient of order $\sim 100$, $r$ is the relative humidity, $r_c$ is the relative humidity of condensation ($r_c \approx 1$), and $S$ is a smoothing factor ($S \approx 0.5$). Similarly, the rate constant of evaporation is
\begin{equation}
  k_\mathrm{evap} =
  \begin{cases}  
  0 & r < r_c \\
  A_\mathrm{evap} \left(\frac{w_i}{H_a}\right) \left(\frac{2}{\pi}\right) \arctan\left(\frac{1/r - 1/r_c}{S/r_c}\right) & r \leq r_c
  \end{cases}
\end{equation}
Here, $A_\mathrm{evap}$ is an evaporation rate coefficient of order $\sim 10$. By including terms like $K_{zz}/H_a^2$ and $w_i/H_a$, the rate constants condense or evaporate a species more rapidly than the gas or particle is vertically transported. The $\arctan$ term is a sigmoid function that smoothly varies the rate constant from 0 for small relative humidity, and, for example, $A_\mathrm{cond} \left(K_{zz}/H_a^2\right)$ for large relativity humidity (in the case of condensation). Using H$_2$O as an example, the rate gas-phase and condensed-phase H$_2$O are produced and destroyed from condensation and evaporation is
\begin{gather}
  E_{i,\mathrm{H_2O,gas}} = - n_\mathrm{H_2O,gas} k_\mathrm{cond} + n_\mathrm{H_2O,particle} k_\mathrm{evap} \\
  E_{i,\mathrm{H_2O,particle}} = - E_{i,\mathrm{H_2O,gas}}
\end{gather}
Here, $n_\mathrm{H_2O,gas}$ and $n_\mathrm{H_2O,particle}$ are the number density of H$_2$O gas and particles in molecules cm$^{-3}$ s$^{-1}$, respectively.

\subsection{Eddy Diffusion} \label{sec:eddy_diffusion}

Like many other photochemical models, \texttt{Photochem} parametrizes turbulent vertical transport with an eddy diffusion flux (e.g., Equation 5.44 in \citet{Catling2017}): 
\begin{equation} \label{eq:eddy_og}
  \Phi_{i\text{,eddy}} = - K_{zz} n \frac{\partial}{\partial z} \left(\frac{n_i}{n}\right)
\end{equation}
Here, $n$ is the total number density. To account for eddy diffusion, we need to rewrite Equation \eqref{eq:eddy_og} so that there are not any partial derivatives with respect to the total number density. To accomplish this, first consider the ideal gas law and hydrostatic equation,
\begin{gather}
  P = n k T \\
  \frac{\partial P}{\partial z} = -g \rho = \frac{-g P \bar \mu}{N_a k T}
\end{gather}
Substituting the ideal gas law in the hydrostatic equation yields
\begin{gather}
  \frac{\partial}{\partial z} (n T) = \frac{-g n \bar \mu}{N_a k} \\
  n \frac{\partial T}{\partial z} + T \frac{\partial n}{\partial z} = \frac{-g n \bar \mu}{N_a k}
\end{gather}
After rearrangement and substituting the definition of scale height,
\begin{equation}
  \frac{1}{n} \frac{\partial n}{\partial z} = - \frac{1}{H_a} - \frac{1}{T} \frac{\partial T}{\partial z} \label{eq:hydrostatic_ideal_gas}
\end{equation}
Now consider the following expansion using the quotient rule
\begin{equation}
  \frac{\partial }{\partial z} \left(\frac{n_i}{n}\right) = \frac{1}{n} \frac{\partial n_i}{\partial z} - \frac{n_i}{n^2} \frac{\partial n}{\partial z} \label{eq:quotient_expansion}
\end{equation}
Substituting Equation \eqref{eq:hydrostatic_ideal_gas} into Equation \eqref{eq:quotient_expansion} and rearrangement gives
\begin{equation}
  n \frac{\partial }{\partial z} \left(\frac{n_i}{n}\right) = \frac{\partial n_i}{\partial z} + \frac{n_i}{H_a} + \frac{n_i}{T} \frac{\partial T}{\partial z} \label{eq:densub}
\end{equation}
Finally, substituting Equation \eqref{eq:densub} into Equation \eqref{eq:eddy_og} yields the eddy diffusion flux that we use in \texttt{Photochem}:
\begin{equation}
  \Phi_{i\text{,eddy}} = - K_{zz} n_i \left(\frac{1}{n_i} \frac{\partial n_i}{\partial z} + \frac{1}{H_a} + \frac{1}{T} \frac{\partial T}{\partial z}\right)
\end{equation}

\subsection{Molecular Diffusion} \label{sec:molecular_diffusion}

The general molecular diffusion equation giving the relative diffusion velocity of gas $i$ with respect to gas $j$ in one dimension is (Equation 15.1 in \cite{Banks2013}, or Equation 14.1, 1 in \cite{Chapman1990})

\begin{equation} \label{eq:molec_diffusion_general}
  v_i - v_j = -D_{ij} \left( \frac{n^2}{n_i n_j} \frac{\partial}{\partial z} \left(\frac{n_i}{n}\right) + \frac{\mu_j - \mu_i}{\overline{\mu}} \frac{1}{P} \frac{\partial P}{\partial z} + \frac{\alpha_{Ti}}{T} \frac{\partial T}{\partial z} - \frac{\mu_i \mu_j}{\overline{\mu} N_a k T} (a_i - a_j)\right)
\end{equation}
Here, $a_1$ and $a_2$ are external accelerations of each molecule from, for example, a magnetic or electric field if the molecule is an ion. We assume neutral molecules such that $a_i = a_j = 0$. We also assume that the atmosphere is an ideal gas and is in hydrostatic equilibrium:

\begin{align} 
  \frac{\partial P}{\partial z} &= -g \rho = \frac{-g P \overline{\mu}}{N_a k T} \\
  \frac{1}{P}\frac{\partial P}{\partial z} &= \frac{-g \overline{\mu}}{N_a k T} = \frac{-1}{H_a} \label{eq:hydrostatic1}
\end{align}
Substitution of $a_i = a_j = 0$ and Equation \eqref{eq:hydrostatic1} into Equation \eqref{eq:molec_diffusion_general} gives

\begin{equation} \label{eq:molec_diffusion_simplify1}
  v_i - v_j = -D_{ij} \left( \frac{n^2}{n_i n_j} \frac{\partial}{\partial z} \left(\frac{n_i}{n}\right) - \frac{\mu_j - \mu_i}{\overline{\mu}} \frac{1}{H_a} + \frac{\alpha_{Ti}}{T} \frac{\partial T}{\partial z} \right)
\end{equation}
We make two further approximation: We assume gas $i$ is a small abundance compared to gas $j$ which we take to be a stationary background gas ($v_j = 0$, $n_j = n$, $\mu_j = \overline{\mu}$). Also, we neglect thermal diffusion ($\alpha_{Ti} = 0$):
\begin{equation} \label{eq:v_molec_diffusion}
  v_i = - D_{i} \left( \frac{n}{n_i} \frac{\partial}{\partial z} \left(\frac{n_i}{n}\right) - \frac{1}{H_a} + \frac{1}{H_i}\right)
\end{equation}
Finally, substituting Equation \eqref{eq:densub} into Equation \eqref{eq:v_molec_diffusion} yields the diffusive flux that we use in \texttt{Photochem}:
\begin{equation} \label{eq:phi_molec_diffusion}
  \Phi_{i\mathrm{,molecular}} = n_i v_i = - D_{i} n_i \left( \frac{1}{n_i} \frac{\partial n_i}{\partial z} + \frac{1}{H_i} + \frac{1}{T} \frac{\partial T}{\partial z} \right)
\end{equation}

For most atmospheres, we estimate the molecular diffusion coefficient with the formula (Equation 15.29 in \cite{Banks2013})
\begin{equation} \label{eq:molec_diffusion_coeff}
  D_i = 
  \frac{b_i}{n} = \frac{1.52 \times 10^{18}}{n} \sqrt{\left( \frac{1}{\mu_i} + \frac{1}{\overline{\mu}} \right) T}
\end{equation}
For hydrogen-dominated atmospheres we instead compute $D_i$ with Equation 6 in \citet{Gladstone1996}. 

\subsection{Particle Fall Velocity} \label{sec:fall_velocity}

The particle fall velocity ($w_i$) is given by stokes law with a slip correction factor ($C_{c,i}$) (Equation 9.42 in \cite{Seinfeld2006}).

\begin{equation} \label{eq:stokes_law}
  w_i = \frac{2}{9} \frac{(\rho_i - \rho)r_i^2}{\eta} C_{c,i}
\end{equation}
We approximate the dynamic viscosity of air ($\eta$) with the following empirical relation from Equation 1-36 and Table 1-2 in \cite{White2006}. This expression is for modern Earth air.

\begin{equation} \label{eq:dynamic_viscosity}
  \eta = 1.716 \times 10^{-4} \left(\frac{T}{273}\right)^{3/2} \left( \frac{384}{T + 111} \right)
\end{equation}
The slip correction factor ($C_{c,i}$), is given by Equation 9.34 in \cite{Seinfeld2006}.

\begin{equation} \label{eq:slip_correction}
  C_{c,i} = 1 + \frac{\lambda}{r_i}\left( -1.257 + 0.4 \exp \left(\frac{1.1 r_i}{\lambda}\right) \right)
\end{equation}
Here, $\lambda$ is the mean free path, which comes from the kinetic theory of gases (Equation 9.6 in \cite{Seinfeld2006})

\begin{equation}
  \lambda = \frac{2 \eta}{n} \sqrt{\frac{\pi N_a}{8 k T \overline{\mu}}}
\end{equation}

\section{Model Data} \label{sec:model_data}

All nominal simulations of the photochemistry and climate of the solar system atmospheres and WASP-39b use the same chemical reaction rates, thermodynamics, photolysis cross sections, and opacities. All model data, and their source in the literature, is available at the following Zenodo archive: \url{https://doi.org/10.5281/zenodo.15785405}. Furthermore, Tables \ref{tab:kdistributions}, 
\ref{tab:cia_rayleigh} and \ref{tab:photolysis} describe the k-distributions, CIA and Rayleigh opacities, and photolysis reactions, respectively. Table \ref{tab:obs} lists the literature citations for the observations shown in Figures \ref{fig:venus}, \ref{fig:earth}, \ref{fig:mars}, \ref{fig:titan}, and \ref{fig:jupiter}.

\begin{longtable*}{p{0.06\linewidth} p{0.38\linewidth} p{0.13\linewidth} p{0.12\linewidth} p{0.17\linewidth}}
  \caption{k-distributions used in climate calculations$^\text{a}$} \label{tab:kdistributions} \\
  \hline 
  \hline
  Species & Source & Line shape & Cutoff & Broadening \\
  \hline 
  H$_2$O$^\text{b}$ & HITEMP2010, HITRAN2016 \citep{Rothman2010,Gordon2017} & Voigt & 25 cm$^{-1}$ & Earth air \\
  CO$_2$ & HITEMP2010 \citep{Rothman2010} & Sub-Lorentzian$^\text{c}$ & 500 cm$^{-1}$ & Self-broadening \\
  CH$_4$ & HITEMP2020 \citep{Hargreaves2020} & Voigt & 25 cm$^{-1}$ & Earth air \\
  CO & HITEMP2019 \citet{Li2015} & Voigt & 25 cm$^{-1}$ & Self-broadening \\
  O$_2$ & HITRAN2016 \citep{Gordon2017} & Voigt & 25 cm$^{-1}$ & Earth air \\
  O$_3$ & HITRAN2016 \citep{Gordon2017} & Voigt & 25 cm$^{-1}$ & Earth air \\
  NH$_3$ & HITRAN2016 \citep{Gordon2017} & Voigt & 25 cm$^{-1}$ & Earth air \\
  SO$_2$ & HITRAN2016 \citep{Gordon2017} & Voigt & 25 cm$^{-1}$ & Earth air \\
  C$_2$H$_2$ & Downloaded w/ HAPI$^\text{d}$ on 4/23/25 & Voigt & 25 cm$^{-1}$ & Earth air \\
  C$_2$H$_6$ & Downloaded w/ HAPI$^\text{d}$ on 4/23/25 & Voigt & 25 cm$^{-1}$ & Earth air \\
  HCl & Downloaded w/ HAPI$^\text{d}$ on 4/23/25 & Voigt & 25 cm$^{-1}$ & Earth air \\
  N$_2$O & HITEMP2019 \citep{Hargreaves2019} & Voigt & 25 cm$^{-1}$ & Earth air \\
  OCS & Downloaded w/ HAPI$^\text{d}$ on 4/23/25 & Voigt & 25 cm$^{-1}$ & Earth air \\
  \hline
  \multicolumn{5}{p{0.93\textwidth}}{
    $^\text{a}$All k-distributions were computed with HELIOS-K \citep{Grimm2021} for temperatures between 100 and 2000 K, and pressures from 10$^{3}$ to 10$^{-5}$ bar.

    $^\text{b}$Plinth or base is removed because this opacity is combined with MT\_CKD H$_2$O continuum.

    $^\text{c}$\citet{Perrin1989} $\chi$ factors.

    $^\text{d}$HITRAN Application Programming Interface \citep{Kochanov2016}.
  }
\end{longtable*}

\begin{longtable*}{p{0.13\linewidth} p{0.1\linewidth} p{0.4\linewidth}}
  \caption{CIA, Rayleigh and Aerosol opacities} \label{tab:cia_rayleigh} \\
  \hline \hline
  Opacity type & Opacity & Citation \\
  \hline 
  CIA & H$_2$-H$_2$ & \citet{Molliere2019} \\
  & H$_2$-He & \citet{Molliere2019} \\
  & N$_2$-N$_2$ & \citet{Molliere2019} \\
  & CH$_4$-CH$_4$ & \citet{Karman2019} \\
  & N$_2$-O$_2$ & \citet{Karman2019} \\
  & O$_2$-O$_2$ & \citet{Karman2019} \\
  & H$_2$-CH$_4$ & \citet{Karman2019} \\
  & CO$_2$-CO$_2$ & \citet{Robinson2018,Lee2016} \\
  & CO$_2$-CH$_4$ & \citet{Karman2019} \\
  & CO$_2$-H$_2$ & \citet{Karman2019} \\
  & N$_2$-CH$_4$ & \citet{Karman2019} \\
  & N$_2$-H$_2$ & \citet{Karman2019} \\
  \hline
  Rayleigh scattering$^\text{a}$ & N$_2$ & \citet{Keady2002,Penndorf1957} \\
  & CO$_2$ & \citet{Keady2002,Shemansky1972} \\
  & O$_2$ & \citet{Keady2002,Penndorf1957} \\
  & H$_2$O & \citet{Keady2002,Ranjan2017,Murphy1977} \\
  & H$_2$ & \citet{Keady2002} \\
  & He & \citet{Keady2002,Penndorf1957} \\
  & CO & \citet{Keady2002,Penndorf1957} \\
  & CH$_4$ & \citet{Batalha2019} \\
  & NH$_3$ & \citet{Batalha2019} \\
  \hline
  Aerosols$^\text{b}$ & H$_2$SO$_4$ & \citet{Palmer1975} \\
  & Dust & \citet{Wolff2009} \\
  & Haze & \citet{Khare1984} \\
  \hline
  \multicolumn{3}{p{0.63\linewidth}}{
    $^\text{a}$Rayleigh scattering opacities are computed using a parameterization from \citet{Vardavas1984}.

    $^\text{b}$Assumes Mie scattering.
  }
\end{longtable*}

\refstepcounter{photo}\label{P1}%
\begin{longtable}{p{0.05\textwidth} p{0.1\textwidth} p{0.2\textwidth} p{0.29\textwidth} p{0.29\textwidth}}%
\caption{Photolysis reactions} \label{tab:photolysis} \\
\hline%
\hline%
\#&Species&Reactions&Cross Section Ref.&Yield Ref.\\%
\hline%
\endhead%
P\arabic{photo}\refstepcounter{photo}\label{P2}&$\mathrm{C}$&{-}&a&{-}\\%
P\arabic{photo}\refstepcounter{photo}\label{P3}&$\mathrm{C_2}$&$\rightarrow \mathrm{C} + \mathrm{C}$&b, a&b\\%
P\arabic{photo}\refstepcounter{photo}\label{P4}&$\mathrm{C_2H}$&$\rightarrow \mathrm{C_2} + \mathrm{H}$&a&a\\%
P\arabic{photo}\refstepcounter{photo}\label{P5}&$\mathrm{C_2H_2}$&$\rightarrow \mathrm{C_2H} + \mathrm{H}$&b, a&a, \cite{Okabe1983}\\%
P\arabic{photo}\refstepcounter{photo}\label{P6}&$\mathrm{C_2H_4}$&$\rightarrow \mathrm{C_2H_2} + \mathrm{H} + \mathrm{H}$&\multirow[t]{3}{0.29\textwidth}{b, a}&\multirow[t]{3}{0.29\textwidth}{d}\\%
P\arabic{photo}\refstepcounter{photo}\label{P7}&&$\rightarrow \mathrm{C_2H_2} + \mathrm{H_2}$&&\\%
P\arabic{photo}\refstepcounter{photo}\label{P8}&&$\rightarrow \mathrm{C_2H_3} + \mathrm{H}$&&\\%
P\arabic{photo}\refstepcounter{photo}\label{P9}&$\mathrm{C_2H_6}$&$\rightarrow \mathrm{C_2H_2} + \mathrm{H_2} + \mathrm{H_2}$&\multirow[t]{5}{0.29\textwidth}{b, a}&\multirow[t]{5}{0.29\textwidth}{d}\\%
P\arabic{photo}\refstepcounter{photo}\label{P10}&&$\rightarrow \mathrm{C_2H_4} + \mathrm{H} + \mathrm{H}$&&\\%
P\arabic{photo}\refstepcounter{photo}\label{P11}&&$\rightarrow \mathrm{C_2H_4} + \mathrm{H_2}$&&\\%
P\arabic{photo}\refstepcounter{photo}\label{P12}&&$\rightarrow \mathrm{CH_3} + \mathrm{CH_3}$&&\\%
P\arabic{photo}\refstepcounter{photo}\label{P13}&&$\rightarrow \mathrm{CH_4} + \mathrm{^1CH_2}$&&\\%
P\arabic{photo}\refstepcounter{photo}\label{P14}&$\mathrm{C_3H_6}$&$\rightarrow \mathrm{C_2H_2} + \mathrm{CH_4}$&\multirow[t]{4}{0.29\textwidth}{\cite{Walker2007}}&\multirow[t]{4}{0.29\textwidth}{d}\\%
P\arabic{photo}\refstepcounter{photo}\label{P15}&&$\rightarrow \mathrm{C_2H_3} + \mathrm{CH_3}$&&\\%
P\arabic{photo}\refstepcounter{photo}\label{P16}&&$\rightarrow \mathrm{C_2H_4} + \mathrm{^1CH_2}$&&\\%
P\arabic{photo}\refstepcounter{photo}\label{P17}&&$\rightarrow \mathrm{C_3H_4} + \mathrm{H_2}$&&\\%
P\arabic{photo}\refstepcounter{photo}\label{P18}&$\mathrm{C_4H_2}$&$\rightarrow \mathrm{C_2H} + \mathrm{C_2H}$&\multirow[t]{4}{0.29\textwidth}{\cite{Ferradaz2009,Smith1998}}&\multirow[t]{4}{0.29\textwidth}{d}\\%
P\arabic{photo}\refstepcounter{photo}\label{P19}&&$\rightarrow \mathrm{C_2H_2} + \mathrm{C_2}$&&\\%
P\arabic{photo}\refstepcounter{photo}\label{P20}&&$\rightarrow \mathrm{C_4H} + \mathrm{H}$&&\\%
P\arabic{photo}\refstepcounter{photo}\label{P21}&&$\rightarrow \mathrm{C_4H_2}$&&\\%
P\arabic{photo}\refstepcounter{photo}\label{P22}&$\mathrm{C_4H_4}$&$\rightarrow \mathrm{C_2H_2} + \mathrm{C_2H_2}$&\multirow[t]{2}{0.29\textwidth}{\cite{Fahr1996}}&\multirow[t]{2}{0.29\textwidth}{d}\\%
P\arabic{photo}\refstepcounter{photo}\label{P23}&&$\rightarrow \mathrm{C_4H_2} + \mathrm{H_2}$&&\\%
P\arabic{photo}\refstepcounter{photo}\label{P24}&$\mathrm{CH}$&$\rightarrow \mathrm{C} + \mathrm{H}$&b, a&b\\%
P\arabic{photo}\refstepcounter{photo}\label{P25}&$\mathrm{CH_2}$&$\rightarrow \mathrm{CH} + \mathrm{H}$&a&a\\%
P\arabic{photo}\refstepcounter{photo}\label{P26}&$\mathrm{CH_2CO}$&$\rightarrow \mathrm{^1CH_2} + \mathrm{CO}$&Assumed&Assumed\\%
P\arabic{photo}\refstepcounter{photo}\label{P27}&$\mathrm{CH_2N_2}$&$\rightarrow \mathrm{^1CH_2} + \mathrm{N_2}$&e&Assumed\\%
P\arabic{photo}\refstepcounter{photo}\label{P28}&$\mathrm{CH_3}$&$\rightarrow \mathrm{^1CH_2} + \mathrm{H}$&\multirow[t]{2}{0.29\textwidth}{a}&\multirow[t]{2}{0.29\textwidth}{d}\\%
P\arabic{photo}\refstepcounter{photo}\label{P29}&&$\rightarrow \mathrm{CH_2} + \mathrm{H}$&&\\%
P\arabic{photo}\refstepcounter{photo}\label{P30}&$\mathrm{CH_3CHO}$&$\rightarrow \mathrm{CH_3} + \mathrm{HCO}$&\multirow[t]{2}{0.29\textwidth}{a}&\multirow[t]{2}{0.29\textwidth}{b}\\%
P\arabic{photo}\refstepcounter{photo}\label{P31}&&$\rightarrow \mathrm{CH_4} + \mathrm{CO}$&&\\%
P\arabic{photo}\refstepcounter{photo}\label{P32}&$\mathrm{CH_3CN}$&$\rightarrow \mathrm{CH_3} + \mathrm{CN}$&b, a&d\\%
P\arabic{photo}\refstepcounter{photo}\label{P33}&$\mathrm{CH_3OH}$&$\rightarrow \mathrm{CH_3} + \mathrm{OH}$&\multirow[t]{3}{0.29\textwidth}{a}&\multirow[t]{3}{0.29\textwidth}{\cite{Hagege1968}}\\%
P\arabic{photo}\refstepcounter{photo}\label{P34}&&$\rightarrow \mathrm{CH_3O} + \mathrm{H}$&&\\%
P\arabic{photo}\refstepcounter{photo}\label{P35}&&$\rightarrow \mathrm{H_2CO} + \mathrm{H_2}$&&\\%
P\arabic{photo}\refstepcounter{photo}\label{P36}&$\mathrm{CH_4}$&$\rightarrow \mathrm{^1CH_2} + \mathrm{H_2}$&\multirow[t]{4}{0.29\textwidth}{a, \cite{Karkoschka1994,Lee2001}}&\multirow[t]{4}{0.29\textwidth}{a, \cite{Gans2011}}\\%
P\arabic{photo}\refstepcounter{photo}\label{P37}&&$\rightarrow \mathrm{CH} + \mathrm{H_2} + \mathrm{H}$&&\\%
P\arabic{photo}\refstepcounter{photo}\label{P38}&&$\rightarrow \mathrm{CH_2} + \mathrm{H} + \mathrm{H}$&&\\%
P\arabic{photo}\refstepcounter{photo}\label{P39}&&$\rightarrow \mathrm{CH_3} + \mathrm{H}$&&\\%
P\arabic{photo}\refstepcounter{photo}\label{P40}&$\mathrm{CN}$&$\rightarrow \mathrm{C} + \mathrm{N}$&b, a&a\\%
P\arabic{photo}\refstepcounter{photo}\label{P41}&$\mathrm{CO}$&$\rightarrow \mathrm{C} + \mathrm{O}$&b, a&a\\%
P\arabic{photo}\refstepcounter{photo}\label{P42}&$\mathrm{CO_2}$&$\rightarrow \mathrm{CO} + \mathrm{O}$&\multirow[t]{2}{0.29\textwidth}{a, \cite{Schmidt2013}}&\multirow[t]{2}{0.29\textwidth}{b}\\%
P\arabic{photo}\refstepcounter{photo}\label{P43}&&$\rightarrow \mathrm{CO} + \mathrm{O(^1D)}$&&\\%
P\arabic{photo}\refstepcounter{photo}\label{P44}&$\mathrm{CS}$&$\rightarrow \mathrm{C} + \mathrm{S}$&a&a\\%
P\arabic{photo}\refstepcounter{photo}\label{P45}&$\mathrm{CS_2}$&$\rightarrow \mathrm{CS} + \mathrm{S}$&b, a&a\\%
P\arabic{photo}\refstepcounter{photo}\label{P46}&$\mathrm{Cl}$&{-}&b&{-}\\%
P\arabic{photo}\refstepcounter{photo}\label{P47}&$\mathrm{Cl_2}$&$\rightarrow \mathrm{Cl} + \mathrm{Cl}$&b&b\\%
P\arabic{photo}\refstepcounter{photo}\label{P48}&$\mathrm{ClO}$&$\rightarrow \mathrm{Cl} + \mathrm{O}$&\multirow[t]{2}{0.29\textwidth}{\cite{Schmidt1998}}&\multirow[t]{2}{0.29\textwidth}{\cite{Burkholder1990}}\\%
P\arabic{photo}\refstepcounter{photo}\label{P49}&&$\rightarrow \mathrm{Cl} + \mathrm{O(^1D)}$&&\\%
P\arabic{photo}\refstepcounter{photo}\label{P50}&$\mathrm{H}$&{-}&b, a&{-}\\%
P\arabic{photo}\refstepcounter{photo}\label{P51}&$\mathrm{H_2}$&$\rightarrow \mathrm{H} + \mathrm{H}$&b, a&a\\%
P\arabic{photo}\refstepcounter{photo}\label{P52}&$\mathrm{H_2CO}$&$\rightarrow \mathrm{CO} + \mathrm{H} + \mathrm{H}$&\multirow[t]{3}{0.29\textwidth}{b, a}&\multirow[t]{3}{0.29\textwidth}{c, b}\\%
P\arabic{photo}\refstepcounter{photo}\label{P53}&&$\rightarrow \mathrm{CO} + \mathrm{H_2}$&&\\%
P\arabic{photo}\refstepcounter{photo}\label{P54}&&$\rightarrow \mathrm{HCO} + \mathrm{H}$&&\\%
P\arabic{photo}\refstepcounter{photo}\label{P55}&$\mathrm{H_2O}$&$\rightarrow \mathrm{H_2} + \mathrm{O(^1D)}$&\multirow[t]{3}{0.29\textwidth}{b, a, \cite{Ranjan2020}}&\multirow[t]{3}{0.29\textwidth}{c, \cite{Slanger1982,Stief1975}}\\%
P\arabic{photo}\refstepcounter{photo}\label{P56}&&$\rightarrow \mathrm{O} + \mathrm{H} + \mathrm{H}$&&\\%
P\arabic{photo}\refstepcounter{photo}\label{P57}&&$\rightarrow \mathrm{OH} + \mathrm{H}$&&\\%
P\arabic{photo}\refstepcounter{photo}\label{P58}&$\mathrm{H_2O_2}$&$\rightarrow \mathrm{OH} + \mathrm{OH}$&a&c, a\\%
P\arabic{photo}\refstepcounter{photo}\label{P59}&$\mathrm{H_2S}$&$\rightarrow \mathrm{HS} + \mathrm{H}$&b&b, a\\%
P\arabic{photo}\refstepcounter{photo}\label{P60}&$\mathrm{H_2SO_4}$&$\rightarrow \mathrm{SO_3} + \mathrm{H_2O}$&\cite{Lane2008}&\cite{Zhang2012}\\%
P\arabic{photo}\refstepcounter{photo}\label{P61}&$\mathrm{HCCCN}$&$\rightarrow \mathrm{C_2H} + \mathrm{CN}$&b, a&b\\%
P\arabic{photo}\refstepcounter{photo}\label{P62}&$\mathrm{HCN}$&$\rightarrow \mathrm{H} + \mathrm{CN}$&b, a&a\\%
P\arabic{photo}\refstepcounter{photo}\label{P63}&$\mathrm{HCO}$&$\rightarrow \mathrm{H} + \mathrm{CO}$&a&a\\%
P\arabic{photo}\refstepcounter{photo}\label{P64}&$\mathrm{HCl}$&$\rightarrow \mathrm{H} + \mathrm{Cl}$&a&b, a\\%
P\arabic{photo}\refstepcounter{photo}\label{P65}&$\mathrm{HNCO}$&$\rightarrow \mathrm{H} + \mathrm{NCO}$&\multirow[t]{2}{0.29\textwidth}{b, a}&\multirow[t]{2}{0.29\textwidth}{b}\\%
P\arabic{photo}\refstepcounter{photo}\label{P66}&&$\rightarrow \mathrm{NH} + \mathrm{CO}$&&\\%
P\arabic{photo}\refstepcounter{photo}\label{P67}&$\mathrm{HNO_3}$&$\rightarrow \mathrm{OH} + \mathrm{NO_2}$&b&b\\%
P\arabic{photo}\refstepcounter{photo}\label{P68}&$\mathrm{HO_2}$&$\rightarrow \mathrm{OH} + \mathrm{O}$&a&b\\%
P\arabic{photo}\refstepcounter{photo}\label{P69}&$\mathrm{HOCl}$&$\rightarrow \mathrm{OH} + \mathrm{Cl}$&b&b\\%
P\arabic{photo}\refstepcounter{photo}\label{P70}&$\mathrm{HS}$&$\rightarrow \mathrm{H} + \mathrm{S}$&a&a\\%
P\arabic{photo}\refstepcounter{photo}\label{P71}&$\mathrm{HSO}$&$\rightarrow \mathrm{HS} + \mathrm{O}$&Assumed&Assumed\\%
P\arabic{photo}\refstepcounter{photo}\label{P72}&$\mathrm{He}$&{-}&b&{-}\\%
P\arabic{photo}\refstepcounter{photo}\label{P73}&$\mathrm{N}$&{-}&b, a&{-}\\%
P\arabic{photo}\refstepcounter{photo}\label{P74}&$\mathrm{N_2}$&$\rightarrow \mathrm{N} + \mathrm{N(^2D)}$&b, a&a\\%
P\arabic{photo}\refstepcounter{photo}\label{P75}&$\mathrm{N_2H_4}$&$\rightarrow \mathrm{N_2H_3} + \mathrm{H}$&e, \cite{Vaghjiani1993,BiehlStuhl1991}&d\\%
P\arabic{photo}\refstepcounter{photo}\label{P76}&$\mathrm{N_2O}$&$\rightarrow \mathrm{N_2} + \mathrm{O(^1D)}$&a&c, a\\%
P\arabic{photo}\refstepcounter{photo}\label{P77}&$\mathrm{NH}$&$\rightarrow \mathrm{N} + \mathrm{H}$&a&a\\%
P\arabic{photo}\refstepcounter{photo}\label{P78}&$\mathrm{NH_2}$&$\rightarrow \mathrm{NH} + \mathrm{H}$&b, a&b\\%
P\arabic{photo}\refstepcounter{photo}\label{P79}&$\mathrm{NH_3}$&$\rightarrow \mathrm{NH} + \mathrm{H} + \mathrm{H}$&\multirow[t]{3}{0.29\textwidth}{e, b, a, \cite{Cheng2006}}&\multirow[t]{3}{0.29\textwidth}{b}\\%
P\arabic{photo}\refstepcounter{photo}\label{P80}&&$\rightarrow \mathrm{NH} + \mathrm{H_2}$&&\\%
P\arabic{photo}\refstepcounter{photo}\label{P81}&&$\rightarrow \mathrm{NH_2} + \mathrm{H}$&&\\%
P\arabic{photo}\refstepcounter{photo}\label{P82}&$\mathrm{NO}$&$\rightarrow \mathrm{N} + \mathrm{O}$&b, a&a\\%
P\arabic{photo}\refstepcounter{photo}\label{P83}&$\mathrm{NO_2}$&$\rightarrow \mathrm{NO} + \mathrm{O}$&c, b&a\\%
P\arabic{photo}\refstepcounter{photo}\label{P84}&$\mathrm{NO_3}$&$\rightarrow \mathrm{NO} + \mathrm{O_2}$&\multirow[t]{2}{0.29\textwidth}{b}&\multirow[t]{2}{0.29\textwidth}{b}\\%
P\arabic{photo}\refstepcounter{photo}\label{P85}&&$\rightarrow \mathrm{NO_2} + \mathrm{O}$&&\\%
P\arabic{photo}\refstepcounter{photo}\label{P86}&$\mathrm{O}$&{-}&b, a&{-}\\%
P\arabic{photo}\refstepcounter{photo}\label{P87}&$\mathrm{O_2}$&$\rightarrow \mathrm{O} + \mathrm{O}$&\multirow[t]{2}{0.29\textwidth}{a}&\multirow[t]{2}{0.29\textwidth}{c}\\%
P\arabic{photo}\refstepcounter{photo}\label{P88}&&$\rightarrow \mathrm{O} + \mathrm{O(^1D)}$&&\\%
P\arabic{photo}\refstepcounter{photo}\label{P89}&$\mathrm{O_3}$&$\rightarrow \mathrm{O} + \mathrm{O_2}$&\multirow[t]{2}{0.29\textwidth}{b, a}&\multirow[t]{2}{0.29\textwidth}{b, \cite{Matsumi2002}}\\%
P\arabic{photo}\refstepcounter{photo}\label{P90}&&$\rightarrow \mathrm{O(^1D)} + \mathrm{O_2}$&&\\%
P\arabic{photo}\refstepcounter{photo}\label{P91}&$\mathrm{OCS}$&$\rightarrow \mathrm{CO} + \mathrm{S}$&a&c, a\\%
P\arabic{photo}\refstepcounter{photo}\label{P92}&$\mathrm{OClO}$&$\rightarrow \mathrm{ClO} + \mathrm{O}$&\multirow[t]{2}{0.29\textwidth}{b}&\multirow[t]{2}{0.29\textwidth}{b}\\%
P\arabic{photo}\refstepcounter{photo}\label{P93}&&$\rightarrow \mathrm{ClO} + \mathrm{O(^1D)}$&&\\%
P\arabic{photo}\refstepcounter{photo}\label{P94}&$\mathrm{OH}$&$\rightarrow \mathrm{O} + \mathrm{H}$&b, a&a\\%
P\arabic{photo}\refstepcounter{photo}\label{P95}&$\mathrm{S}$&{-}&b, a&{-}\\%
P\arabic{photo}\refstepcounter{photo}\label{P96}&$\mathrm{S_2}$&$\rightarrow \mathrm{S} + \mathrm{S}$&a&a\\%
P\arabic{photo}\refstepcounter{photo}\label{P97}&$\mathrm{S_3}$&$\rightarrow \mathrm{S_2} + \mathrm{S}$&\cite{Billmers1991}&Assumed\\%
P\arabic{photo}\refstepcounter{photo}\label{P98}&$\mathrm{S_4}$&$\rightarrow \mathrm{S_2} + \mathrm{S_2}$&\multirow[t]{2}{0.29\textwidth}{e, \cite{Billmers1991}}&\multirow[t]{2}{0.29\textwidth}{Assumed}\\%
P\arabic{photo}\refstepcounter{photo}\label{P99}&&$\rightarrow \mathrm{S_3} + \mathrm{S}$&&\\%
P\arabic{photo}\refstepcounter{photo}\label{P100}&$\mathrm{S_8}$&{-}&\cite{Bass1953}&{-}\\%
P\arabic{photo}\refstepcounter{photo}\label{P101}&$\mathrm{SO}$&$\rightarrow \mathrm{S} + \mathrm{O}$&b, a&a\\%
P\arabic{photo}\refstepcounter{photo}\label{P102}&$\mathrm{SO_2}$&$\rightarrow \mathrm{S} + \mathrm{O_2}$&\multirow[t]{2}{0.29\textwidth}{a}&\multirow[t]{2}{0.29\textwidth}{b}\\%
P\arabic{photo}\refstepcounter{photo}\label{P103}&&$\rightarrow \mathrm{SO} + \mathrm{O}$&&\\%
P\arabic{photo}\refstepcounter{photo}\label{P104}&$\mathrm{SO_3}$&$\rightarrow \mathrm{SO_2} + \mathrm{O}$&c&Assumed\\%
\hline%
\hline%
\multicolumn{5}{p{0.93\textwidth}}{\textbf{Notes.} a: \cite{Heays2017}; b: \cite{Huebner2015}; c: \cite{Burkholder2020}; d: \cite{La08}; e: \cite{KellerRudek2013}. P\ref{P13}: The \cite{La08} paths making $\mathrm{CH_2CCH_2} + \mathrm{H_2}$, $\mathrm{CH_3C_2H} + \mathrm{H_2}$ and $\mathrm{C_3H_5} + \mathrm{H}$ are all in the $\mathrm{C_3H_4} + \mathrm{H_2}$ branch. P\ref{P25}: Assumed to be identical to OCS. P\ref{P26}: We used the T = 298 K file from \cite{KellerRudek2013}. P\ref{P32}: \cite{Hagege1968} reports two quantum yields, and we average them. P\ref{P41}: We use a low resolution version of the 270 K \cite{Schmidt2013} data. P\ref{P59}: We extrapolated the \cite{Lane2008} cross section beyond 181 nm. P\ref{P70}: Assumed to be identical to HO$_2$. P\ref{P73}: \cite{Heays2017} Figure 10 suggests $\mathrm{N} + \mathrm{N(^2D)}$ is the dominant path. P\ref{P88}: We assume the 250 K quantum yields from \cite{Matsumi2002} between 306 and 328 nm. P\ref{P97}: We assume even split between $\mathrm{S_3} + \mathrm{S}$ and $\mathrm{S_2} + \mathrm{S_2}$ chanel. P\ref{P99}: Extracted from Figure 2 of Bass (1953) by Kevin Zahnle. For S8 rings.}\\%
\end{longtable}%
\begin{longtable*}{p{0.06\textwidth} p{0.07\textwidth} p{0.8\textwidth} }%
  \caption{Sources for the observations shown in Figure \ref{fig:venus}, \ref{fig:earth}, \ref{fig:mars}, \ref{fig:titan} and \ref{fig:jupiter}} \label{tab:obs} \\
  \hline%
  \hline%
  Planet&Species&Citations\\%
  \hline%
  \endhead%
  \multirow[c]{12}{0.06\textwidth}[-7em]{Venus}&$\mathrm{CO}$&\cite{Wilson1981,Young1972,Marcq2015,Marcq2008,Cotton2012,Grassi2014,Oyama1980,Krasnopolsky2010b,Pollack1993,Marcq2006,Bezard2007,Collard1993,Krasnopolsky2008,Fegley2014,Krasnopolsky2014,Gelman1979,Oyama1979,Connes1968,Bezard1990,Tsang2008,Marcq2005}\\%
  &$\mathrm{H_2O}$&\cite{Moroz1990,Donahue1997,Fedorova2015,Evans1969,Fedorova2016,Bezard2011,Sandor2005,Mukhin1982,Chamberlain2013,Encrenaz2015,Arney2014,Marcq2008,Bell1991,Meadows1996,Cottini2012b,Pollack1993,Marcq2006,Fedorova2008,Donahue1992,Bezard2007,Surkov1982,Surkov1987,DeBergh1995,Bezard1990,Moroz1979,Tsang2008}\\%
  &$\mathrm{H_2S}$&\cite{Krasnopolsky2008}\\%
  &$\mathrm{H_2SO_4}$&\cite{Oschlisniok2012}\\%
  &$\mathrm{HCl}$&\cite{Connes1967,Arney2014,Sandor2012,Bezard1990,Iwagami2008,Young1972,Krasnopolsky2010a}\\%
  &$\mathrm{O_2}$&\cite{Oyama1980,Mukhin1982,Oyama1979,Marcq2018}\\%
  &$\mathrm{OCS}$&\cite{Pollack1993,Marcq2006,Arney2014,Marcq2008,Bezard2007,Mukhin1982,Krasnopolsky2010b,Marcq2005}\\%
  &$\mathrm{S_3}$&\cite{Maiorov2005,Bezard2007,Krasnopolsky2013}\\%
  &$\mathrm{S_4}$&\cite{Krasnopolsky2013}\\%
  &$\mathrm{SO}$&\cite{Na1990,Encrenaz2015}\\%
  &$\mathrm{SO_2}$&\cite{Pollack1993,Zasova1993,Encrenaz2012,Gelman1979,Oyama1979,Na1990,Greaves2021,Arney2014,Marcq2008,Bertaux1996,Bezard2007,Oyama1980,Hoffman1980}\\%
  \hline
  \multirow[c]{13}{0.06\textwidth}{Earth}&$\mathrm{CH_4}$&\cite{Massie1981}\\%
  &$\mathrm{CO}$&\cite{Funke2009}\\%
  &$\mathrm{H_2}$&\cite{Ehhalt1975}\\%
  &$\mathrm{H_2O}$&U.S. Standard Atmosphere 1976\\%
  &$\mathrm{H_2SO_4}$&\cite{Viggiano1981}\\%
  &$\mathrm{HNO_3}$&\cite{Massie1981}\\%
  &$\mathrm{N_2O}$&\cite{Massie1981}\\%
  &$\mathrm{NO}$&\cite{Massie1981}\\%
  &$\mathrm{NO_2}$&\cite{Massie1981}\\%
  &$\mathrm{O_3}$&\cite{Massie1981}\\%
  &$\mathrm{OCS}$&\cite{Inn1979,Barkley2008}\\%
  &$\mathrm{OH}$&\cite{Massie1981}\\%
  &$\mathrm{SO_2}$&\cite{Georgii1980,Jaeschke1976}\\%
  \hline
  \multirow[c]{6}{0.06\textwidth}{Mars}&$\mathrm{CO}$&\cite{Bouche2021}\\%
  &$\mathrm{H_2O_2}$&\cite{Encrenaz2019}\\%
  &$\mathrm{HO_2}$&\cite{Alday2024}\\%
  &$\mathrm{NO}$&\cite{Krasnopolsky2006}\\%
  &$\mathrm{O_2}$&\cite{Sandel2015,Trainer2019}\\%
  &$\mathrm{O_3}$&\cite{Patel2021}\\%
  &$\mathrm{H_2}$&\cite{Krasnopolsky2001}\\%
  \hline
  \multirow[c]{10}{0.06\textwidth}{Titan}&$\mathrm{C_2H_2}$&\cite{Vuitton2006,Nixon2013}\\%
  &$\mathrm{C_2H_4}$&\cite{Vuitton2006,Nixon2013}\\%
  &$\mathrm{C_2H_6}$&\cite{Vuitton2006,Nixon2013}\\%
  &$\mathrm{CH_3CN}$&\cite{Cui2009,Vuitton2006,Marten2002}\\%
  &$\mathrm{CO}$&\cite{DeKok2007}\\%
  &$\mathrm{CO_2}$&\cite{DeKok2007,Cui2009}\\%
  &$\mathrm{H_2O}$&\cite{Cottini2012a,Cui2009}\\%
  &$\mathrm{HCCCN}$&\cite{Cui2009,Vuitton2006,Marten2002}\\%
  &$\mathrm{HCN}$&\cite{Adriani2011,Vuitton2006,Marten2002}\\%
  &$\mathrm{NH_3}$&\cite{Cui2009,Teanby2013}\\%
  \hline
  \multirow[c]{7}{0.06\textwidth}{Jupiter}&$\mathrm{C_2H_2}$&\cite{Moses2005,Gladstone1996}\\%
  &$\mathrm{C_2H_4}$&\cite{Romani2008,Bezard2001}\\%
  &$\mathrm{C_2H_6}$&\cite{Gladstone1996,Fouchet2000}\\%
  &$\mathrm{CH_4}$&\cite{Drossart1999}\\%
  &$\mathrm{CO}$&\cite{Bezard2002}\\%
  &$\mathrm{HCN}$&\cite{Bezard1995}\\%
  &$\mathrm{NH_3}$&\cite{Moeckel2023}\\%
  \hline%
  \multicolumn{3}{p{0.93\textwidth}}{\textbf{Notes.} We gathered most of the Venus, Earth and Jupiter observations from \citet{Rimmer2021} (their Table 5), \citet{Tsai2021} (their Figure 9) and \citet{Rimmer2016} (their Figure 12), respectively. All data is available at this Zenodo archive: \url{https://doi.org/10.5281/zenodo.16509950}.}\\
\end{longtable*}%

\begin{figure*}
  \centering
  \includegraphics[width=\textwidth]{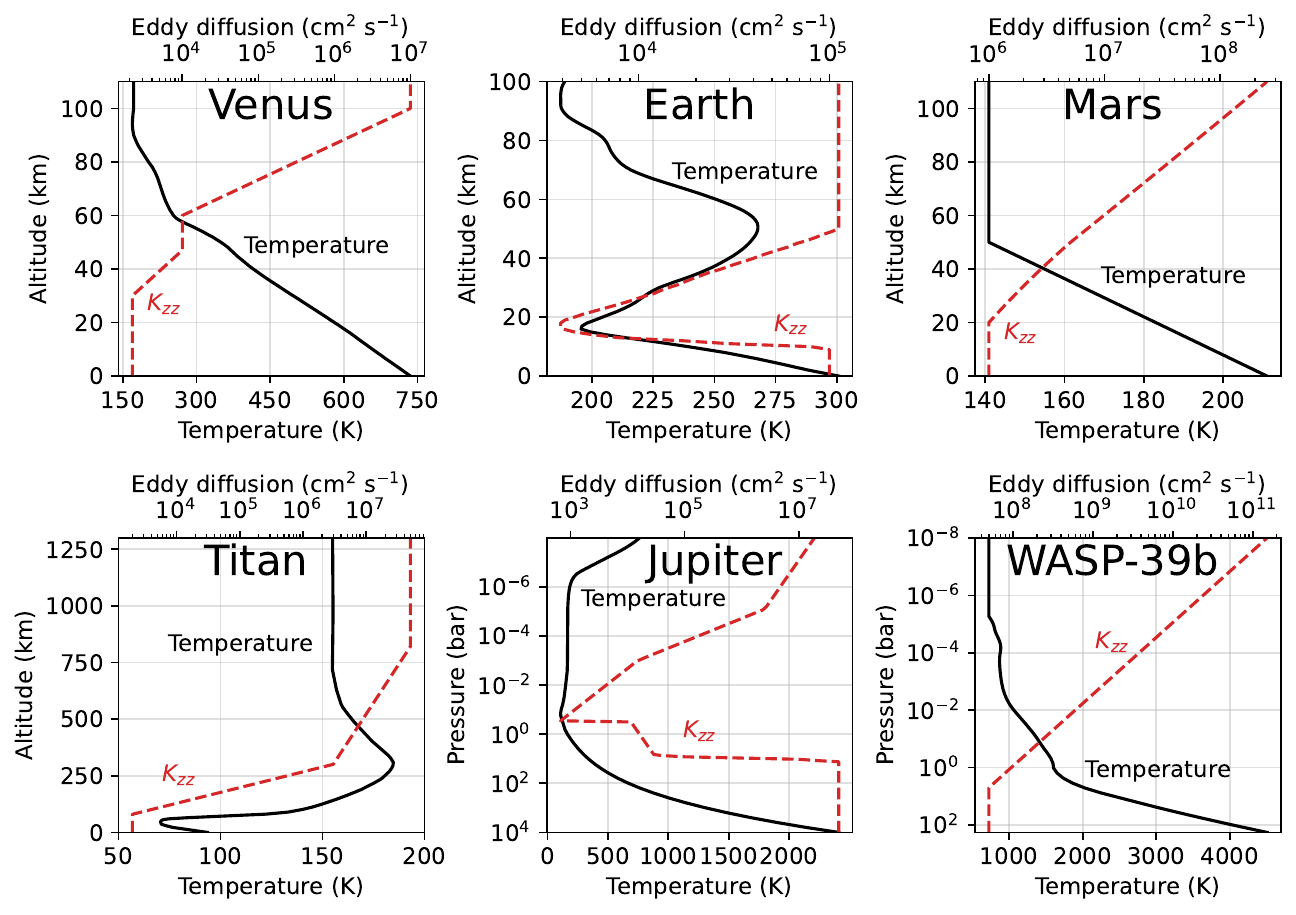}
  \caption{The temperature and eddy diffusion profiles used in our photochemical simulations of Venus, Earth, Mars, Titan, Jupiter and WASP-39b.}
  \label{fig:t_and_kzz}
\end{figure*}

\bibliography{bib, observations_table, reactions}

\begin{thebibliography}{}
\expandafter\ifx\csname natexlab\endcsname\relax\def\natexlab#1{#1}\fi
\providecommand{\url}[1]{\href{#1}{#1}}
\providecommand{\dodoi}[1]{doi:~\href{http://doi.org/#1}{\nolinkurl{#1}}}
\providecommand{\doeprint}[1]{\href{http://ascl.net/#1}{\nolinkurl{http://ascl.net/#1}}}
\providecommand{\doarXiv}[1]{\href{https://arxiv.org/abs/#1}{\nolinkurl{https://arxiv.org/abs/#1}}}

\bibitem[{A.~S. {Ackerman} \& M.~S. {Marley}(2001){Ackerman} \& {Marley}}]{Ackerman2001}
{Ackerman}, A.~S., \& {Marley}, M.~S. 2001, \bibinfo{title}{{Precipitating Condensation Clouds in Substellar Atmospheres},} \apj, 556, 872, \dodoi{10.1086/321540}

\bibitem[{A. {Adriani} {et~al.}(2011){Adriani}, {Dinelli}, {López-Puertas}, {García-Comas}, {Moriconi}, {D’Aversa}, {Funke}, \& {Coradini}}]{Adriani2011}
{Adriani}, A., {Dinelli}, B., {López-Puertas}, M., {et~al.} 2011, \bibinfo{title}{Distribution of HCN in Titan’s upper atmosphere from Cassini/VIMS observations at 3$\mu$m,} Icarus, 214, 584–595, \dodoi{10.1016/j.icarus.2011.04.016}

\bibitem[{J. {Alday} {et~al.}(2024){Alday}, {Trokhimovskiy}, {Belyaev}, {Fedorova}, {Holmes}, {Patel}, {Mason}, {Lefèvre}, {Olsen}, {Montmessin}, {Korablev}, {Baggio}, \& {Patrakeev}}]{Alday2024}
{Alday}, J., {Trokhimovskiy}, A., {Belyaev}, D.~A., {et~al.} 2024, \bibinfo{title}{Upper limits of HO2 in the atmosphere of Mars from the ExoMars Trace Gas Orbiter,} Monthly Notices of the Royal Astronomical Society, 532, 4429–4435, \dodoi{10.1093/mnras/stae1814}

\bibitem[{M. {Allen} {et~al.}(1981){Allen}, {Yung}, \& {Waters}}]{Allen1981}
{Allen}, M., {Yung}, Y.~L., \& {Waters}, J.~W. 1981, \bibinfo{title}{{Vertical transport and photochemistry in the terrestrial mesosphere and lower thermosphere /50-120 km/},} \jgr, 86, 3617, \dodoi{10.1029/JA086iA05p03617}

\bibitem[{D.~S. {Amundsen} {et~al.}(2017){Amundsen}, {Tremblin}, {Manners}, {Baraffe}, \& {Mayne}}]{Amundsen2017}
{Amundsen}, D.~S., {Tremblin}, P., {Manners}, J., {Baraffe}, I., \& {Mayne}, N.~J. 2017, \bibinfo{title}{{Treatment of overlapping gaseous absorption with the correlated-k method in hot Jupiter and brown dwarf atmosphere models},} \aap, 598, A97, \dodoi{10.1051/0004-6361/201629322}

\bibitem[{G. {Arney} {et~al.}(2014){Arney}, {Meadows}, {Crisp}, {Schmidt}, {Bailey}, \& {Robinson}}]{Arney2014}
{Arney}, G., {Meadows}, V., {Crisp}, D., {et~al.} 2014, \bibinfo{title}{Spatially resolved measurements of H$_{2}$O, HCl, CO, OCS, SO$_{2}$, cloud opacity, and acid concentration in the Venus near-infrared spectral windows,} Journal of Geophysical Research (Planets), 119, 1860, \dodoi{10.1002/2014JE004662}

\bibitem[{G. {Arney} {et~al.}(2016){Arney}, {Domagal-Goldman}, {Meadows}, {Wolf}, {Schwieterman}, {Charnay}, {Claire}, {H{\'e}brard}, \& {Trainer}}]{Arney2016}
{Arney}, G., {Domagal-Goldman}, S.~D., {Meadows}, V.~S., {et~al.} 2016, \bibinfo{title}{{The Pale Orange Dot: The Spectrum and Habitability of Hazy Archean Earth},} Astrobiology, 16, 873, \dodoi{10.1089/ast.2015.1422}

\bibitem[{S.~K. {Atreya} {et~al.}(2020){Atreya}, {Hofstadter}, {In}, {Mousis}, {Reh}, \& {Wong}}]{Atreya2020}
{Atreya}, S.~K., {Hofstadter}, M.~H., {In}, J.~H., {et~al.} 2020, \bibinfo{title}{{Deep Atmosphere Composition, Structure, Origin, and Exploration, with Particular Focus on Critical in situ Science at the Icy Giants},} \ssr, 216, 18, \dodoi{10.1007/s11214-020-0640-8}

\bibitem[{P.~M. Banks \& G. Kockarts(2013)Banks \& Kockarts}]{Banks2013}
Banks, P.~M., \& Kockarts, G. 2013, Aeronomy (Elsevier)

\bibitem[{M.~P. {Barkley} {et~al.}(2008){Barkley}, {Palmer}, {Boone}, {Bernath}, \& {Suntharalingam}}]{Barkley2008}
{Barkley}, M.~P., {Palmer}, P.~I., {Boone}, C.~D., {Bernath}, P.~F., \& {Suntharalingam}, P. 2008, \bibinfo{title}{Global distributions of carbonyl sulfide in the upper troposphere and stratosphere,} Geophysical Research Letters, 35, \dodoi{10.1029/2008gl034270}

\bibitem[{A.~M. {Bass}(1953){Bass}}]{Bass1953}
{Bass}, A.~M. 1953, \bibinfo{title}{{The Optical Absorption of Sulfur},} \jcp, 21, 80, \dodoi{10.1063/1.1698629}

\bibitem[{N.~E. {Batalha} {et~al.}(2019){Batalha}, {Marley}, {Lewis}, \& {Fortney}}]{Batalha2019}
{Batalha}, N.~E., {Marley}, M.~S., {Lewis}, N.~K., \& {Fortney}, J.~J. 2019, \bibinfo{title}{{Exoplanet Reflected-light Spectroscopy with PICASO},} \apj, 878, 70, \dodoi{10.3847/1538-4357/ab1b51}

\bibitem[{S. {Behnel} {et~al.}(2011){Behnel}, {Bradshaw}, {Citro}, {Dalcin}, {Seljebotn}, \& {Smith}}]{Behnel2011}
{Behnel}, S., {Bradshaw}, R., {Citro}, C., {et~al.} 2011, \bibinfo{title}{{Cython: The Best of Both Worlds},} Computing in Science and Engineering, 13, 31, \dodoi{10.1109/MCSE.2010.118}

\bibitem[{S.~A. {Beiler} {et~al.}(2024){Beiler}, {Mukherjee}, {Cushing}, {Kirkpatrick}, {Schneider}, {Kothari}, {Marley}, \& {Visscher}}]{Beiler2024}
{Beiler}, S.~A., {Mukherjee}, S., {Cushing}, M.~C., {et~al.} 2024, \bibinfo{title}{{A Tale of Two Molecules: The Underprediction of CO$_{2}$ and Overprediction of PH$_{3}$ in Late T and Y Dwarf Atmospheric Models},} \apj, 973, 60, \dodoi{10.3847/1538-4357/ad6759}

\bibitem[{J. {Bell} {et~al.}(1991){Bell}, {Crisp}, {Lucey}, {Ozoroski}, {Sinton}, {Willis}, \& {Campbell}}]{Bell1991}
{Bell}, J., {Crisp}, D., {Lucey}, P., {et~al.} 1991, \bibinfo{title}{Spectroscopic Observations of Bright and Dark Emission Features on the Night Side of Venus,} Science, 252, 1293, \dodoi{10.1126/science.252.5010.1293}

\bibitem[{B. {Benneke} {et~al.}(2024){Benneke}, {Roy}, {Coulombe}, {Radica}, {Piaulet}, {Ahrer}, {Pierrehumbert}, {Krissansen-Totton}, {Schlichting}, {Hu}, {Yang}, {Christie}, {Thorngren}, {Young}, {Pelletier}, {Knutson}, {Miguel}, {Evans-Soma}, {Dorn}, {Gagnebin}, {Fortney}, {Komacek}, {MacDonald}, {Raul}, {Cloutier}, {Acuna}, {Lafreni{\`e}re}, {Cadieux}, {Doyon}, {Welbanks}, \& {Allart}}]{Benneke2024}
{Benneke}, B., {Roy}, P.-A., {Coulombe}, L.-P., {et~al.} 2024, \bibinfo{title}{{JWST Reveals CH$_4$, CO$_2$, and H$_2$O in a Metal-rich Miscible Atmosphere on a Two-Earth-Radius Exoplanet},} arXiv e-prints, arXiv:2403.03325, \dodoi{10.48550/arXiv.2403.03325}

\bibitem[{J.-L. {Bertaux} {et~al.}(1996){Bertaux}, {Widemann}, {Hauchecorne}, {Moroz}, \& {Ekonomov}}]{Bertaux1996}
{Bertaux}, J.-L., {Widemann}, T., {Hauchecorne}, A., {Moroz}, V., \& {Ekonomov}, A. 1996, \bibinfo{title}{VEGA 1 and VEGA 2 entry probes: An investigation of local UV absorption (220-400 nm) in the atmosphere of Venus (SO$_{2}$, aerosols, cloud structure),} \jgr, 101, 12709, \dodoi{10.1029/96JE00466}

\bibitem[{B. {Bezard} {et~al.}(1990){Bezard}, {de Bergh}, {Crisp}, \& {Maillard}}]{Bezard1990}
{Bezard}, B., {de Bergh}, C., {Crisp}, D., \& {Maillard}, J.~P. 1990, \bibinfo{title}{The deep atmosphere of Venus revealed by high-resolution nightside spectra,} \nat, 345, 508, \dodoi{10.1038/345508a0}

\bibitem[{H. Biehl \& F. Stuhl(1991)Biehl \& Stuhl}]{BiehlStuhl1991}
Biehl, H., \& Stuhl, F. 1991, \bibinfo{title}{Vacuum UV absorption and excitation spectra of N2H4,} Journal of Photochemistry and Photobiology A: Chemistry, 59, 135

\bibitem[{C.~J. {Bierson} \& X. {Zhang}(2020){Bierson} \& {Zhang}}]{Bierson2020}
{Bierson}, C.~J., \& {Zhang}, X. 2020, \bibinfo{title}{{Chemical Cycling in the Venusian Atmosphere: A Full Photochemical Model From the Surface to 110 km},} Journal of Geophysical Research (Planets), 125, e06159, \dodoi{10.1029/2019JE006159}

\bibitem[{R.~I. Billmers \& A.~L. Smith(1991)Billmers \& Smith}]{Billmers1991}
Billmers, R.~I., \& Smith, A.~L. 1991, \bibinfo{title}{Ultraviolet-visible absorption spectra of equilibrium sulfur vapor: Molar absorptivity spectra of S3 and S4,} The Journal of Physical Chemistry, 95, 4242

\bibitem[{J. {Bouche} {et~al.}(2021){Bouche}, {Coheur}, {Giuranna}, {Wolkenberg}, {Nardi}, {Amoroso}, {Vandaele}, {Daerden}, {Neary}, \& {Bauduin}}]{Bouche2021}
{Bouche}, J., {Coheur}, P., {Giuranna}, M., {et~al.} 2021, \bibinfo{title}{Seasonal and Spatial Variability of Carbon Monoxide (CO) in the Martian Atmosphere From PFS/MEX Observations,} Journal of Geophysical Research: Planets, 126, \dodoi{10.1029/2020je006480}

\bibitem[{J. Burkholder {et~al.}(2020)Burkholder, Sander, Abbatt, Barker, Cappa, Crounse, Dibble, Huie, Kolb, Kurylo, {et~al.}}]{Burkholder2020}
Burkholder, J., Sander, S., Abbatt, J., {et~al.} 2020, Chemical kinetics and photochemical data for use in atmospheric studies; evaluation number 19, Tech. rep., Pasadena, CA: Jet Propulsion Laboratory, National Aeronautics and Space~…

\bibitem[{J.~B. {Burkholder} {et~al.}(1990){Burkholder}, {Orlando}, \& {Howard}}]{Burkholder1990}
{Burkholder}, J.~B., {Orlando}, J.~J., \& {Howard}, C.~J. 1990, \bibinfo{title}{Ultraviolet absorption cross sections of chlorine oxide (Cl2O2) between 210 and 410 nm,} The Journal of Physical Chemistry, 94, 687–695, \dodoi{10.1021/j100365a033}

\bibitem[{B. {Bézard} \& C. {de Bergh}(2007){Bézard} \& {de Bergh}}]{Bezard2007}
{Bézard}, B., \& {de Bergh}, C. 2007, \bibinfo{title}{Composition of the atmosphere of Venus below the clouds,} Journal of Geophysical Research: Planets, 112, \dodoi{10.1029/2006je002794}

\bibitem[{B. {Bézard} {et~al.}(2011){Bézard}, {Fedorova}, {Bertaux}, {Rodin}, \& {Korablev}}]{Bezard2011}
{Bézard}, B., {Fedorova}, A., {Bertaux}, J.-L., {Rodin}, A., \& {Korablev}, O. 2011, \bibinfo{title}{The 1.10- and 1.18-m nightside windows of Venus observed by SPICAV-IR aboard Venus Express,} \icarus, 216, 173, \dodoi{10.1016/j.icarus.2011.08.025}

\bibitem[{B. {Bézard} {et~al.}(1995){Bézard}, {Griffith}, {Lacy}, \& {Owen}}]{Bezard1995}
{Bézard}, B., {Griffith}, C., {Lacy}, J., \& {Owen}, T. 1995, \bibinfo{title}{Non-Detection of Hydrogen Cyanide on Jupiter,} Icarus, 118, 384–391, \dodoi{10.1006/icar.1995.1198}

\bibitem[{B. {Bézard} {et~al.}(2002){Bézard}, {Lellouch}, {Strobel}, {Maillard}, \& {Drossart}}]{Bezard2002}
{Bézard}, B., {Lellouch}, E., {Strobel}, D., {Maillard}, J.-P., \& {Drossart}, P. 2002, \bibinfo{title}{Carbon Monoxide on Jupiter: Evidence for Both Internal and External Sources,} Icarus, 159, 95–111, \dodoi{10.1006/icar.2002.6917}

\bibitem[{B. {Bézard} {et~al.}(2001){Bézard}, {Moses}, {Lacy}, {Greathouse}, {Richter}, \& {Griffith}}]{Bezard2001}
{Bézard}, B., {Moses}, J., {Lacy}, J., {et~al.} 2001, \bibinfo{title}{Detection of Ethylene (C$_{2}$H$_{4}$) on Jupiter and Saturn in Non–Auroral Regions,} in AAS/Division for Planetary Sciences Meeting Abstracts, Vol.~33, AAS/Division for Planetary Sciences Meeting Abstracts \#33, 22.07

\bibitem[{C. {Cadieux} {et~al.}(2024){Cadieux}, {Doyon}, {MacDonald}, {Turbet}, {Artigau}, {Lim}, {Radica}, {Fauchez}, {Salhi}, {Dang}, {Albert}, {Coulombe}, {Cowan}, {Lafreni{\`e}re}, {L'Heureux}, {Piaulet-Ghorayeb}, {Benneke}, {Cloutier}, {Charnay}, {Cook}, {Fournier-Tondreau}, {Plotnykov}, \& {Valencia}}]{Cadieux2024}
{Cadieux}, C., {Doyon}, R., {MacDonald}, R.~J., {et~al.} 2024, \bibinfo{title}{{Transmission Spectroscopy of the Habitable Zone Exoplanet LHS 1140 b with JWST/NIRISS},} \apjl, 970, L2, \dodoi{10.3847/2041-8213/ad5afa}

\bibitem[{D.~C. Catling \& J.~F. Kasting(2017)Catling \& Kasting}]{Catling2017}
Catling, D.~C., \& Kasting, J.~F. 2017, Atmospheric evolution on inhabited and lifeless worlds (Cambridge University Press)

\bibitem[{D.~C. Catling {et~al.}(2018)Catling, Krissansen-Totton, Kiang, Crisp, Robinson, DasSarma, Rushby, Del~Genio, William, \& Shawn}]{Catling2018}
Catling, D.~C., Krissansen-Totton, J., Kiang, N.~Y., {et~al.} 2018, \bibinfo{title}{Exoplanet biosignatures: A framework for their assessment,} Astrobiol., 18, 709, \dodoi{10.1089/ast.2017.1737}

\bibitem[{T. {Cavali{\'e}} {et~al.}(2023){Cavali{\'e}}, {Lunine}, \& {Mousis}}]{Cavalie2023}
{Cavali{\'e}}, T., {Lunine}, J., \& {Mousis}, O. 2023, \bibinfo{title}{{A subsolar oxygen abundance or a radiative region deep in Jupiter revealed by thermochemical modelling},} Nature Astronomy, 7, 678, \dodoi{10.1038/s41550-023-01928-8}

\bibitem[{S. {Chamberlain} {et~al.}(2013){Chamberlain}, {Bailey}, {Crisp}, \& {Meadows}}]{Chamberlain2013}
{Chamberlain}, S., {Bailey}, J., {Crisp}, D., \& {Meadows}, V. 2013, \bibinfo{title}{Ground-based near-infrared observations of water vapour in the Venus troposphere,} \icarus, 222, 364, \dodoi{10.1016/j.icarus.2012.11.014}

\bibitem[{S. Chapman \& T.~G. Cowling(1990)Chapman \& Cowling}]{Chapman1990}
Chapman, S., \& Cowling, T.~G. 1990, The mathematical theory of non-uniform gases: an account of the kinetic theory of viscosity, thermal conduction and diffusion in gases (Cambridge university press)

\bibitem[{B.-M. {Cheng} {et~al.}(2006){Cheng}, {Lu}, {Chen}, {Bahou}, {Lee}, {Mebel}, {Lee}, {Liang}, \& {Yung}}]{Cheng2006}
{Cheng}, B.-M., {Lu}, H.-C., {Chen}, H.-K., {et~al.} 2006, \bibinfo{title}{{Absorption Cross Sections of NH$_{3}$, NH$_{2}$D, NHD$_{2}$, and ND$_{3}$ in the Spectral Range 140-220 nm and Implications for Planetary Isotopic Fractionation},} \apj, 647, 1535, \dodoi{10.1086/505615}

\bibitem[{A. {Collard} {et~al.}(1993){Collard}, {Taylor}, {Calcutt}, {Carlson}, {Kamp}, {Baines}, {Encrenaz}, {Drossart}, {Lellouch}, \& {Bezard}}]{Collard1993}
{Collard}, A., {Taylor}, F., {Calcutt}, S., {et~al.} 1993, \bibinfo{title}{Latitudinal distribution of carbon monoxide in the deep atmosphere of Venus,} \planss, 41, 487, \dodoi{10.1016/0032-0633(93)90031-V}

\bibitem[{P. {Connes} {et~al.}(1967){Connes}, {Connes}, {Benedict}, \& {Kaplan}}]{Connes1967}
{Connes}, P., {Connes}, J., {Benedict}, W., \& {Kaplan}, L. 1967, \bibinfo{title}{Traces of HCl and HF in the atmosphere of Venus,} \apj, 147, 1230, \dodoi{10.1086/149124}

\bibitem[{P. {Connes} {et~al.}(1968){Connes}, {Connes}, {Kaplan}, \& {Benedict}}]{Connes1968}
{Connes}, P., {Connes}, J., {Kaplan}, L., \& {Benedict}, W. 1968, \bibinfo{title}{Carbon monoxide in the Venus atmosphere,} \apj, 152, 731, \dodoi{10.1086/149590}

\bibitem[{T. {Constantinou} {et~al.}(2025){Constantinou}, {Shorttle}, \& {Rimmer}}]{Constantinou2025}
{Constantinou}, T., {Shorttle}, O., \& {Rimmer}, P.~B. 2025, \bibinfo{title}{{A dry Venusian interior constrained by atmospheric chemistry},} Nature Astronomy, 9, 189, \dodoi{10.1038/s41550-024-02414-5}

\bibitem[{V. {Cottini} {et~al.}(2012{\natexlab{a}}){Cottini}, {Ignatiev}, {Piccioni}, {Drossart}, {Grassi}, \& {Markiewicz}}]{Cottini2012b}
{Cottini}, V., {Ignatiev}, N., {Piccioni}, G., {et~al.} 2012{\natexlab{a}}, \bibinfo{title}{Water vapor near the cloud tops of Venus from Venus Express/VIRTIS dayside data,} \icarus, 217, 561, \dodoi{10.1016/j.icarus.2011.06.018}

\bibitem[{V. {Cottini} {et~al.}(2012{\natexlab{b}}){Cottini}, {Nixon}, {Jennings}, {Anderson}, {Gorius}, {Bjoraker}, {Coustenis}, {Teanby}, {Achterberg}, {Bézard}, {de Kok}, {Lellouch}, {Irwin}, {Flasar}, \& {Bampasidis}}]{Cottini2012a}
{Cottini}, V., {Nixon}, C., {Jennings}, D., {et~al.} 2012{\natexlab{b}}, \bibinfo{title}{Water vapor in Titan’s stratosphere from Cassini CIRS far-infrared spectra,} Icarus, 220, 855–862, \dodoi{10.1016/j.icarus.2012.06.014}

\bibitem[{D.~V. {Cotton} {et~al.}(2012){Cotton}, {Bailey}, {Crisp}, \& {Meadows}}]{Cotton2012}
{Cotton}, D.~V., {Bailey}, J., {Crisp}, D., \& {Meadows}, V. 2012, \bibinfo{title}{The distribution of carbon monoxide in the lower atmosphere of Venus,} \icarus, 217, 570, \dodoi{10.1016/j.icarus.2011.05.020}

\bibitem[{E.~C. {Creecy} {et~al.}(2021){Creecy}, {Li}, {Jiang}, {West}, {Fry}, {Nixon}, {Kenyon}, \& {Seignovert}}]{Creecy2021}
{Creecy}, E.~C., {Li}, L., {Jiang}, X., {et~al.} 2021, \bibinfo{title}{{Titan's Global Radiant Energy Budget During the Cassini Epoch (2004-2017)},} \grl, 48, e95356, \dodoi{10.1029/2021GL095356}

\bibitem[{D. {Crisp}(1986){Crisp}}]{Crisp1986}
{Crisp}, D. 1986, \bibinfo{title}{{Radiative forcing of the Venus mesosphere I. Solar fluxes and heating rates},} \icarus, 67, 484, \dodoi{10.1016/0019-1035(86)90126-0}

\bibitem[{J. {Cui} {et~al.}(2009){Cui}, {Yelle}, {Vuitton}, {Waite}, {Kasprzak}, {Gell}, {Niemann}, {Müller-Wodarg}, {Borggren}, {Fletcher}, {Patrick}, {Raaen}, \& {Magee}}]{Cui2009}
{Cui}, J., {Yelle}, R., {Vuitton}, V., {et~al.} 2009, \bibinfo{title}{Analysis of Titan’s neutral upper atmosphere from Cassini Ion Neutral Mass Spectrometer measurements,} Icarus, 200, 581–615, \dodoi{10.1016/j.icarus.2008.12.005}

\bibitem[{M.~H. {Currie} \& V.~S. {Meadows}(2025){Currie} \& {Meadows}}]{Currie2025}
{Currie}, M.~H., \& {Meadows}, V.~S. 2025, \bibinfo{title}{{There's More to Life in Reflected Light: Simulating the Detectability of a Range of Molecules for High-contrast, High-resolution Observations of Nontransiting Terrestrial Exoplanets},} \psj, 6, 96, \dodoi{10.3847/PSJ/adc004}

\bibitem[{L. {Dai} {et~al.}(2024){Dai}, {Shao}, \& {Sheng}}]{Dai2024}
{Dai}, L., {Shao}, W., \& {Sheng}, Z. 2024, \bibinfo{title}{{An investigation into Venusian atmospheric chemistry based on an open-access photochemistry-transport model at 0{\textendash}112 km},} \aap, 689, A55, \dodoi{10.1051/0004-6361/202450552}

\bibitem[{C. {de Bergh} {et~al.}(1995){de Bergh}, {Bezard}, {Crisp}, {Maillard}, {Owen}, {Pollack}, \& {Grinspoon}}]{DeBergh1995}
{de Bergh}, C., {Bezard}, B., {Crisp}, D., {et~al.} 1995, \bibinfo{title}{Water in the deep atmosphere of Venus from high-resolution spectra of the night side,} Advances in Space Research, 15, 79, \dodoi{10.1016/0273-1177(94)00067-B}

\bibitem[{R. {de Kok} {et~al.}(2007){de Kok}, {Irwin}, {Teanby}, {Lellouch}, {Bézard}, {Vinatier}, {Nixon}, {Fletcher}, {Howett}, {Calcutt}, {Bowles}, {Flasar}, \& {Taylor}}]{DeKok2007}
{de Kok}, R., {Irwin}, P., {Teanby}, N., {et~al.} 2007, \bibinfo{title}{Oxygen compounds in Titan’s stratosphere as observed by Cassini CIRS,} Icarus, 186, 354–363, \dodoi{10.1016/j.icarus.2006.09.016}

\bibitem[{S.~D. {Domagal-Goldman} {et~al.}(2011){Domagal-Goldman}, {Meadows}, {Claire}, \& {Kasting}}]{DomagalGoldman2011}
{Domagal-Goldman}, S.~D., {Meadows}, V.~S., {Claire}, M.~W., \& {Kasting}, J.~F. 2011, \bibinfo{title}{{Using Biogenic Sulfur Gases as Remotely Detectable Biosignatures on Anoxic Planets},} Astrobiology, 11, 419, \dodoi{10.1089/ast.2010.0509}

\bibitem[{T. {Donahue} {et~al.}(1997){Donahue}, {Grinspoon}, {Hartle}, \& {Hodges}}]{Donahue1997}
{Donahue}, T., {Grinspoon}, D., {Hartle}, R., \& {Hodges}, R. 1997, \bibinfo{title}{Ion/neutral Escape of Hydrogen and Deuterium: Evolution of Water,} in Venus II: Geology, Geophysics, Atmosphere, and Solar Wind Environment, ed. S.~W. {Bougher}, D.~M. {Hunten}, \& R.~J. {Phillips}, 385

\bibitem[{T. {Donahue} \& R. {Hodges}(1992){Donahue} \& {Hodges}}]{Donahue1992}
{Donahue}, T., \& {Hodges}, R. 1992, \bibinfo{title}{Past and Present Water Budget of Venus,} \jgr, 97, 6083, \dodoi{10.1029/92JE00343}

\bibitem[{P. {Drossart} {et~al.}(1999){Drossart}, {Fouchet}, {Crovisier}, {Lellouch}, {Encrenaz}, {Feuchtgruber}, \& {Champion}}]{Drossart1999}
{Drossart}, P., {Fouchet}, T., {Crovisier}, J., {et~al.} 1999, \bibinfo{title}{Fluorescence in the 3m bands of methane on Jupiter and Saturn from ISO/SWS observations,} in ESA Special Publication, Vol. 427, The Universe as Seen by ISO, ed. P.~{Cox} \& M.~{Kessler}, 169

\bibitem[{D.~H. {Ehhalt} {et~al.}(1975){Ehhalt}, {Heidt}, {Lueb}, \& {Pollock}}]{Ehhalt1975}
{Ehhalt}, D.~H., {Heidt}, L.~E., {Lueb}, R.~H., \& {Pollock}, W. 1975, \bibinfo{title}{The vertical distribution of trace gases in the stratosphere,} pure and applied geophysics, 113, 389–402, \dodoi{10.1007/bf01592926}

\bibitem[{T. {Encrenaz} {et~al.}(2012){Encrenaz}, {Greathouse}, {Roe}, {Richter}, {Lacy}, {Bézard}, {Fouchet}, \& {Widemann}}]{Encrenaz2012}
{Encrenaz}, T., {Greathouse}, T., {Roe}, H., {et~al.} 2012, \bibinfo{title}{HDO and SO$_{2}$ thermal mapping on Venus: evidence for strong SO$_{2}$ variability,} \aap, 543, A153, \dodoi{10.1051/0004-6361/201219419}

\bibitem[{T. {Encrenaz} {et~al.}(2015){Encrenaz}, {Moreno}, {Moullet}, {Lellouch}, \& {Fouchet}}]{Encrenaz2015}
{Encrenaz}, T., {Moreno}, R., {Moullet}, A., {Lellouch}, E., \& {Fouchet}, T. 2015, \bibinfo{title}{Submillimeter mapping of mesospheric minor species on Venus with ALMA,} \planss, 113, 275, \dodoi{10.1016/j.pss.2015.01.011}

\bibitem[{T. {Encrenaz} {et~al.}(2019){Encrenaz}, {Greathouse}, {Aoki}, {Daerden}, {Giuranna}, {Forget}, {Lefèvre}, {Montmessin}, {Fouchet}, {Bézard}, {Atreya}, {DeWitt}, {Richter}, {Neary}, \& {Viscardy}}]{Encrenaz2019}
{Encrenaz}, T., {Greathouse}, T.~K., {Aoki}, S., {et~al.} 2019, \bibinfo{title}{Ground-based infrared mapping of H2O2 on Mars near opposition,} \aap, 627, A60, \dodoi{10.1051/0004-6361/201935300}

\bibitem[{J. {Evans} \& R. {Ingalls}(1969){Evans} \& {Ingalls}}]{Evans1969}
{Evans}, J., \& {Ingalls}, R. 1969, \bibinfo{title}{Absorption of Radar Signals by the Atmosphere of Venus,} in The Venus Atmosphere, ed. R.~{Jastrow} \& S.~I. {Rasool}, 169

\bibitem[{A. {Fahr} \& A. {Nayak}(1996){Fahr} \& {Nayak}}]{Fahr1996}
{Fahr}, A., \& {Nayak}, A. 1996, \bibinfo{title}{{Temperature dependent ultraviolet absorption cross sections of propylene, methylacetylene and vinylacetylene},} Chemical Physics, 203, 351, \dodoi{10.1016/0301-0104(95)00401-7}

\bibitem[{A. {Fedorova} {et~al.}(2015){Fedorova}, {Bézard}, {Bertaux}, {Korablev}, \& {Wilson}}]{Fedorova2015}
{Fedorova}, A., {Bézard}, B., {Bertaux}, J.-L., {Korablev}, O., \& {Wilson}, C. 2015, \bibinfo{title}{The CO$_{2}$ continuum absorption in the 1.10- and 1.18-m windows on Venus from Maxwell Montes transits by SPICAV IR onboard Venus express,} \planss, 113, 66, \dodoi{10.1016/j.pss.2014.08.010}

\bibitem[{A. {Fedorova} {et~al.}(2016){Fedorova}, {Marcq}, {Luginin}, {Korablev}, {Bertaux}, \& {Montmessin}}]{Fedorova2016}
{Fedorova}, A., {Marcq}, E., {Luginin}, M., {et~al.} 2016, \bibinfo{title}{Variations of water vapor and cloud top altitude in the Venus' mesosphere from SPICAV/VEx observations,} \icarus, 275, 143, \dodoi{10.1016/j.icarus.2016.04.010}

\bibitem[{A. {Fedorova} {et~al.}(2008){Fedorova}, {Korablev}, {Vandaele}, {Bertaux}, {Belyaev}, {Mahieux}, {Neefs}, {Wilquet}, {Drummond}, {Montmessin}, \& {Villard}}]{Fedorova2008}
{Fedorova}, A., {Korablev}, O., {Vandaele}, A., {et~al.} 2008, \bibinfo{title}{HDO and H2O vertical distributions and isotopic ratio in the Venus mesosphere by Solar Occultation at Infrared spectrometer on board Venus Express,} Journal of Geophysical Research: Planets, 113, \dodoi{10.1029/2008je003146}

\bibitem[{B. {Fegley}(2014){Fegley}}]{Fegley2014}
{Fegley}, B. 2014, \bibinfo{title}{Venus,} in Planets, Asteriods, Comets and The Solar System, ed. A.~M. {Davis}, Vol.~2, 127--148

\bibitem[{T. {Ferradaz} {et~al.}(2009){Ferradaz}, {B{\'e}nilan}, {Fray}, {Jolly}, {Schwell}, {Gazeau}, \& {Jochims}}]{Ferradaz2009}
{Ferradaz}, T., {B{\'e}nilan}, Y., {Fray}, N., {et~al.} 2009, \bibinfo{title}{{Temperature-dependent photoabsorption cross-sections of cyanoacetylene and diacetylene in the mid- and vacuum-UV: Application to Titan's atmosphere},} \planss, 57, 10, \dodoi{10.1016/j.pss.2008.10.005}

\bibitem[{J. {Filiberto} {et~al.}(2020){Filiberto}, {Trang}, {Treiman}, \& {Gilmore}}]{Filiberto2020}
{Filiberto}, J., {Trang}, D., {Treiman}, A.~H., \& {Gilmore}, M.~S. 2020, \bibinfo{title}{{Present-day volcanism on Venus as evidenced from weathering rates of olivine},} Science Advances, 6, eaax7445, \dodoi{10.1126/sciadv.aax7445}

\bibitem[{E. {Fischer} {et~al.}(2019){Fischer}, {Mart{\'\i}nez}, {Renn{\'o}}, {Tamppari}, \& {Zent}}]{Fischer2019}
{Fischer}, E., {Mart{\'\i}nez}, G.~M., {Renn{\'o}}, N.~O., {Tamppari}, L.~K., \& {Zent}, A.~P. 2019, \bibinfo{title}{{Relative Humidity on Mars: New Results From the Phoenix TECP Sensor},} Journal of Geophysical Research (Planets), 124, 2780, \dodoi{10.1029/2019JE006080}

\bibitem[{M. {Fortune} {et~al.}(2025){Fortune}, {Gibson}, {Diamond-Lowe}, {Mendon{\c{c}}a}, {Gressier}, {Kitzmann}, {Allen}, {August}, {Ih}, {Meier Vald{\'e}s}, {Zgraggen}, {Buchhave}, {Demory}, {Espinoza}, {Heng}, {Jones}, \& {Rathcke}}]{Fortune2025}
{Fortune}, M., {Gibson}, N.~P., {Diamond-Lowe}, H., {et~al.} 2025, \bibinfo{title}{{Hot Rocks Survey III: A deep eclipse for LHS 1140c and a new Gaussian process method to account for correlated noise in individual pixels},} arXiv e-prints, arXiv:2505.22186.
\newblock \doarXiv{2505.22186}

\bibitem[{T. {Fouchet} {et~al.}(2000){Fouchet}, {Lellouch}, {Bézard}, {Feuchtgruber}, {Drossart}, \& {Encrenaz}}]{Fouchet2000}
{Fouchet}, T., {Lellouch}, E., {Bézard}, B., {et~al.} 2000, \bibinfo{title}{Jupiter's hydrocarbons observed with ISO-SWS: vertical profiles of C$_2$H$_6$ and C$_2$H$_2$, detection of CH$_3$C$_2$H,} \aap, 355, L13, \dodoi{10.48550/arXiv.astro-ph/0002273}

\bibitem[{B. {Funke} {et~al.}(2009){Funke}, {López-Puertas}, {García-Comas}, {Stiller}, {von Clarmann}, {Höpfner}, {Glatthor}, {Grabowski}, {Kellmann}, \& {Linden}}]{Funke2009}
{Funke}, B., {López-Puertas}, M., {García-Comas}, M., {et~al.} 2009, \bibinfo{title}{Carbon monoxide distributions from the upper troposphere to the mesosphere inferred from 4.7 $\mu$m non-local thermal equilibrium emissions measured by MIPAS on Envisat,} Atmospheric Chemistry and Physics, 9, 2387–2411, \dodoi{10.5194/acp-9-2387-2009}

\bibitem[{B. {Gans} {et~al.}(2011){Gans}, {Boy{\'e}-P{\'e}ronne}, {Broquier}, {Delsaut}, {Douin}, {Fellows}, {Halvick}, {Loison}, {Lucchese}, \& {Gauyacq}}]{Gans2011}
{Gans}, B., {Boy{\'e}-P{\'e}ronne}, S., {Broquier}, M., {et~al.} 2011, \bibinfo{title}{{Photolysis of methane revisited at 121.6 nm and at 118.2 nm: quantum yields of the primary products, measured by mass spectrometry},} Physical Chemistry Chemical Physics (Incorporating Faraday Transactions), 13, 8140, \dodoi{10.1039/C0CP02627A}

\bibitem[{J.~B. {Garvin} {et~al.}(2022){Garvin}, {Getty}, {Arney}, {Johnson}, {Kohler}, {Schwer}, {Sekerak}, {Bartels}, {Saylor}, {Elliott}, {Goodloe}, {Garrison}, {Cottini}, {Izenberg}, {Lorenz}, {Malespin}, {Ravine}, {Webster}, {Atkinson}, {Aslam}, {Atreya}, {Bos}, {Brinckerhoff}, {Campbell}, {Crisp}, {Filiberto}, {Forget}, {Gilmore}, {Gorius}, {Grinspoon}, {Hofmann}, {Kane}, {Kiefer}, {Lebonnois}, {Mahaffy}, {Pavlov}, {Trainer}, {Zahnle}, \& {Zolotov}}]{Garvin2022}
{Garvin}, J.~B., {Getty}, S.~A., {Arney}, G.~N., {et~al.} 2022, \bibinfo{title}{{Revealing the Mysteries of Venus: The DAVINCI Mission},} \psj, 3, 117, \dodoi{10.3847/PSJ/ac63c2}

\bibitem[{B. {Gelman} {et~al.}(1979){Gelman}, {Zolotukhin}, {Lamonov}, {Levchuk}, {Mukhin}, {Nenarokov}, {Khotnikov}, {Rotin}, \& {Lipatov}}]{Gelman1979}
{Gelman}, B., {Zolotukhin}, V., {Lamonov}, N., {et~al.} 1979, An analysis of the chemical composition of the atmosphere of Venus on an AMS of the Venera-12 using a gas chromatograph,, An analysis of the chemical composition of the atmosphere of Venus on an AMS of the Venera-12 using a gas chromatograph Transl. into ENGLISH of Analiz Khimicheskogo Sostava Atmosfery Venery na AMS Venera-12 Gazovym Khromatografom (Moscow), 1979 11 p

\bibitem[{H. {Georgii} \& F.~X. {Meixner}(1980){Georgii} \& {Meixner}}]{Georgii1980}
{Georgii}, H., \& {Meixner}, F.~X. 1980, \bibinfo{title}{Measurement of the tropospheric and stratospheric SO2 distribution,} Journal of Geophysical Research: Oceans, 85, 7433–7438, \dodoi{10.1029/jc085ic12p07433}

\bibitem[{F. {Giorgi} \& W.~L. {Chameides}(1985){Giorgi} \& {Chameides}}]{Giorgi1985}
{Giorgi}, F., \& {Chameides}, W.~L. 1985, \bibinfo{title}{{The rainout parameterization in a photochemical model},} \jgr, 90, 7872, \dodoi{10.1029/JD090iD05p07872}

\bibitem[{G. {Gladstone} {et~al.}(1996){Gladstone}, {Allen}, \& {Yung}}]{Gladstone1996}
{Gladstone}, G., {Allen}, M., \& {Yung}, Y. 1996, \bibinfo{title}{Hydrocarbon Photochemistry in the Upper Atmosphere of Jupiter,} Icarus, 119, 1–52, \dodoi{10.1006/icar.1996.0001}

\bibitem[{D.~G. Goodwin {et~al.}(2024)Goodwin, Moffat, Schoegl, Speth, \& Weber}]{Cantera2024}
Goodwin, D.~G., Moffat, H.~K., Schoegl, I., Speth, R.~L., \& Weber, B.~W. 2024, Cantera: An Object-oriented Software Toolkit for Chemical Kinetics, Thermodynamics, and Transport Processes,, \url{https://www.cantera.org} \dodoi{10.5281/zenodo.14455267}

\bibitem[{I.~E. {Gordon} {et~al.}(2017){Gordon}, {Rothman}, {Hill}, {Kochanov}, {Tan}, {Bernath}, {Birk}, {Boudon}, {Campargue}, {Chance}, {Drouin}, {Flaud}, {Gamache}, {Hodges}, {Jacquemart}, {Perevalov}, {Perrin}, {Shine}, {Smith}, {Tennyson}, {Toon}, {Tran}, {Tyuterev}, {Barbe}, {Cs{\'a}sz{\'a}r}, {Devi}, {Furtenbacher}, {Harrison}, {Hartmann}, {Jolly}, {Johnson}, {Karman}, {Kleiner}, {Kyuberis}, {Loos}, {Lyulin}, {Massie}, {Mikhailenko}, {Moazzen-Ahmadi}, {M{\"u}ller}, {Naumenko}, {Nikitin}, {Polyansky}, {Rey}, {Rotger}, {Sharpe}, {Sung}, {Starikova}, {Tashkun}, {Auwera}, {Wagner}, {Wilzewski}, {Wcis{\l}o}, {Yu}, \& {Zak}}]{Gordon2017}
{Gordon}, I.~E., {Rothman}, L.~S., {Hill}, C., {et~al.} 2017, \bibinfo{title}{{The HITRAN2016 molecular spectroscopic database},} \jqsrt, 203, 3, \dodoi{10.1016/j.jqsrt.2017.06.038}

\bibitem[{I.~E. {Gordon} {et~al.}(2022){Gordon}, {Rothman}, {Hargreaves}, {Hashemi}, {Karlovets}, {Skinner}, {Conway}, {Hill}, {Kochanov}, {Tan}, {Wcis{\l}o}, {Finenko}, {Nelson}, {Bernath}, {Birk}, {Boudon}, {Campargue}, {Chance}, {Coustenis}, {Drouin}, {Flaud}, {Gamache}, {Hodges}, {Jacquemart}, {Mlawer}, {Nikitin}, {Perevalov}, {Rotger}, {Tennyson}, {Toon}, {Tran}, {Tyuterev}, {Adkins}, {Baker}, {Barbe}, {Can{\`e}}, {Cs{\'a}sz{\'a}r}, {Dudaryonok}, {Egorov}, {Fleisher}, {Fleurbaey}, {Foltynowicz}, {Furtenbacher}, {Harrison}, {Hartmann}, {Horneman}, {Huang}, {Karman}, {Karns}, {Kassi}, {Kleiner}, {Kofman}, {Kwabia-Tchana}, {Lavrentieva}, {Lee}, {Long}, {Lukashevskaya}, {Lyulin}, {Makhnev}, {Matt}, {Massie}, {Melosso}, {Mikhailenko}, {Mondelain}, {M{\"u}ller}, {Naumenko}, {Perrin}, {Polyansky}, {Raddaoui}, {Raston}, {Reed}, {Rey}, {Richard}, {T{\'o}bi{\'a}s}, {Sadiek}, {Schwenke}, {Starikova}, {Sung}, {Tamassia}, {Tashkun}, {Vander Auwera}, {Vasilenko}, {Vigasin}, {Villanueva}, {Vispoel}, {Wagner}, {Yachmenev}, \& {Yurchenko}}]{hitran2022}
{Gordon}, I.~E., {Rothman}, L.~S., {Hargreaves}, R.~J., {et~al.} 2022, \bibinfo{title}{{The HITRAN2020 molecular spectroscopic database},} \jqsrt, 277, 107949, \dodoi{10.1016/j.jqsrt.2021.107949}

\bibitem[{S. {Gordon} \& B.~J. {McBride}(1994){Gordon} \& {McBride}}]{Gordon1994}
{Gordon}, S., \& {McBride}, B.~J. 1994, {Computer program for calculation of complex chemical equilibrium compositions and applications. Part 1: Analysis}, Tech. rep., NASA.
\newblock \url{https://ui.adsabs.harvard.edu/abs/2024arXiv241021364L}

\bibitem[{R.~J. {Graham} {et~al.}(2021){Graham}, {Lichtenberg}, {Boukrouche}, \& {Pierrehumbert}}]{Graham2021}
{Graham}, R.~J., {Lichtenberg}, T., {Boukrouche}, R., \& {Pierrehumbert}, R.~T. 2021, \bibinfo{title}{{A Multispecies Pseudoadiabat for Simulating Condensable-rich Exoplanet Atmospheres},} \psj, 2, 207, \dodoi{10.3847/PSJ/ac214c}

\bibitem[{D. {Grassi} {et~al.}(2014){Grassi}, {Politi}, {Ignatiev}, {Plainaki}, {Lebonnois}, {Wolkenberg}, {Montabone}, {Migliorini}, {Piccioni}, \& {Drossart}}]{Grassi2014}
{Grassi}, D., {Politi}, R., {Ignatiev}, N., {et~al.} 2014, \bibinfo{title}{The Venus nighttime atmosphere as observed by the VIRTIS-M instrument. Average fields from the complete infrared data set,} Journal of Geophysical Research (Planets), 119, 837, \dodoi{10.1002/2013JE004586}

\bibitem[{J.~S. {Greaves} {et~al.}(2021){Greaves}, {Richards}, {Bains}, {Rimmer}, {Sagawa}, {Clements}, {Seager}, {Petkowski}, {Sousa-Silva}, {Ranjan}, {Drabek-Maunder}, {Fraser}, {Cartwright}, {Mueller-Wodarg}, {Zhan}, {Friberg}, {Coulson}, {Lee}, \& {Hoge}}]{Greaves2021}
{Greaves}, J.~S., {Richards}, A.~M., {Bains}, W., {et~al.} 2021, \bibinfo{title}{Phosphine gas in the cloud decks of Venus,} Nature Astronomy, 5, 655, \dodoi{10.1038/s41550-020-1174-4}

\bibitem[{S.~L. {Grimm} {et~al.}(2021){Grimm}, {Malik}, {Kitzmann}, {Guzm{\'a}n-Mesa}, {Hoeijmakers}, {Fisher}, {Mendon{\c{c}}a}, {Yurchenko}, {Tennyson}, {Alesina}, {Buchschacher}, {Burnier}, {Segransan}, {Kurucz}, \& {Heng}}]{Grimm2021}
{Grimm}, S.~L., {Malik}, M., {Kitzmann}, D., {et~al.} 2021, \bibinfo{title}{{HELIOS-K 2.0 Opacity Calculator and Open-source Opacity Database for Exoplanetary Atmospheres},} \apjs, 253, 30, \dodoi{10.3847/1538-4365/abd773}

\bibitem[{J. Hagege {et~al.}(1968)Hagege, Roberge, \& Vermeil}]{Hagege1968}
Hagege, J., Roberge, P., \& Vermeil, C. 1968, \bibinfo{title}{Collision-induced predissociation in the gas phase. Photolysis of methanol at 1849 {\AA},} Transactions of the Faraday Society, 64, 3288

\bibitem[{R.~J. {Hargreaves} {et~al.}(2020){Hargreaves}, {Gordon}, {Rey}, {Nikitin}, {Tyuterev}, {Kochanov}, \& {Rothman}}]{Hargreaves2020}
{Hargreaves}, R.~J., {Gordon}, I.~E., {Rey}, M., {et~al.} 2020, \bibinfo{title}{{An Accurate, Extensive, and Practical Line List of Methane for the HITEMP Database},} \apjs, 247, 55, \dodoi{10.3847/1538-4365/ab7a1a}

\bibitem[{R.~J. {Hargreaves} {et~al.}(2019){Hargreaves}, {Gordon}, {Rothman}, {Tashkun}, {Perevalov}, {Lukashevskaya}, {Yurchenko}, {Tennyson}, \& {M{\"u}ller}}]{Hargreaves2019}
{Hargreaves}, R.~J., {Gordon}, I.~E., {Rothman}, L.~S., {et~al.} 2019, \bibinfo{title}{{Spectroscopic line parameters of NO, NO$_{2}$, and N$_{2}$O for the HITEMP database},} \jqsrt, 232, 35, \dodoi{10.1016/j.jqsrt.2019.04.040}

\bibitem[{D.~A. {Hauglustaine} {et~al.}(1994){Hauglustaine}, {Granier}, {Brasseur}, \& {M{\'e}Gie}}]{Hauglustaine1994}
{Hauglustaine}, D.~A., {Granier}, C., {Brasseur}, G.~P., \& {M{\'e}Gie}, G. 1994, \bibinfo{title}{{The importance of atmospheric chemistry in the calculation of radiative forcing on the climate system},} \jgr, 99, 1173, \dodoi{10.1029/93JD02987}

\bibitem[{A.~N. {Heays} {et~al.}(2017){Heays}, {Bosman}, \& {van Dishoeck}}]{Heays2017}
{Heays}, A.~N., {Bosman}, A.~D., \& {van Dishoeck}, E.~F. 2017, \bibinfo{title}{{Photodissociation and photoionisation of atoms and molecules of astrophysical interest},} \aap, 602, A105, \dodoi{10.1051/0004-6361/201628742}

\bibitem[{A.~C. Hindmarsh {et~al.}(2005)Hindmarsh, Brown, Grant, Lee, Serban, Shumaker, \& Woodward}]{Hindmarsh2005}
Hindmarsh, A.~C., Brown, P.~N., Grant, K.~E., {et~al.} 2005, \bibinfo{title}{SUNDIALS: Suite of nonlinear and differential/algebraic equation solvers,} ACM Transactions on Mathematical Software (TOMS), 31, 363

\bibitem[{J. {Hoffman} {et~al.}(1980){Hoffman}, {Hodges}, {Donahue}, \& {McElroy}}]{Hoffman1980}
{Hoffman}, J., {Hodges}, R., {Donahue}, T., \& {McElroy}, M. 1980, \bibinfo{title}{Composition of the Venus lower atmosphere from the Pioneer Venus mass spectrometer,} \jgr, 85, 7882, \dodoi{10.1029/JA085iA13p07882}

\bibitem[{R. {Hu} {et~al.}(2012){Hu}, {Seager}, \& {Bains}}]{Hu2012}
{Hu}, R., {Seager}, S., \& {Bains}, W. 2012, \bibinfo{title}{{Photochemistry in Terrestrial Exoplanet Atmospheres. I. Photochemistry Model and Benchmark Cases},} \apj, 761, 166, \dodoi{10.1088/0004-637X/761/2/166}

\bibitem[{W.~F. {Huebner} {et~al.}(1992){Huebner}, {Keady}, \& {Lyon}}]{Huebner1992}
{Huebner}, W.~F., {Keady}, J.~J., \& {Lyon}, S.~P. 1992, \bibinfo{title}{{Solar Photo Rates for Planetary Atmospheres and Atmospheric Pollutants},} \apss, 195, 1, \dodoi{10.1007/BF00644558}

\bibitem[{W.~F. {Huebner} \& J. {Mukherjee}(2015){Huebner} \& {Mukherjee}}]{Huebner2015}
{Huebner}, W.~F., \& {Mukherjee}, J. 2015, \bibinfo{title}{{Photoionization and photodissociation rates in solar and blackbody radiation fields},} \planss, 106, 11, \dodoi{10.1016/j.pss.2014.11.022}

\bibitem[{J. {Ih} {et~al.}(2023){Ih}, {Kempton}, {Whittaker}, \& {Lessard}}]{Ih2023}
{Ih}, J., {Kempton}, E. M.~R., {Whittaker}, E.~A., \& {Lessard}, M. 2023, \bibinfo{title}{{Constraining the Thickness of TRAPPIST-1 b's Atmosphere from Its JWST Secondary Eclipse Observation at 15 {\ensuremath{\mu}}m},} \apjl, 952, L4, \dodoi{10.3847/2041-8213/ace03b}

\bibitem[{E.~C.~Y. {Inn} {et~al.}(1979){Inn}, {Vedder}, {Tyson}, \& {O’Hara}}]{Inn1979}
{Inn}, E. C.~Y., {Vedder}, J.~F., {Tyson}, B.~J., \& {O’Hara}, D. 1979, \bibinfo{title}{COS in the stratosphere,} Geophysical Research Letters, 6, 191–193, \dodoi{10.1029/gl006i003p00191}

\bibitem[{H. {Innes} {et~al.}(2023){Innes}, {Tsai}, \& {Pierrehumbert}}]{Innes2023}
{Innes}, H., {Tsai}, S.-M., \& {Pierrehumbert}, R.~T. 2023, \bibinfo{title}{{The Runaway Greenhouse Effect on Hycean Worlds},} \apj, 953, 168, \dodoi{10.3847/1538-4357/ace346}

\bibitem[{N. {Iwagami} {et~al.}(2008){Iwagami}, {Ohtsuki}, {Tokuda}, {Ohira}, {Kasaba}, {Imamura}, {Sagawa}, {Hashimoto}, {Takeuchi}, {Ueno}, \& {Okumura}}]{Iwagami2008}
{Iwagami}, N., {Ohtsuki}, S., {Tokuda}, K., {et~al.} 2008, \bibinfo{title}{Hemispheric distributions of HCl above and below the Venus’ clouds by ground-based 1.7 m spectroscopy,} \planss, 56, 1424, \dodoi{10.1016/j.pss.2008.05.009}

\bibitem[{R.~B. {Jackson} {et~al.}(2020){Jackson}, {Saunois}, {Bousquet}, {Canadell}, {Poulter}, {Stavert}, {Bergamaschi}, {Niwa}, {Segers}, \& {Tsuruta}}]{Jackson2020}
{Jackson}, R.~B., {Saunois}, M., {Bousquet}, P., {et~al.} 2020, \bibinfo{title}{{Increasing anthropogenic methane emissions arise equally from agricultural and fossil fuel sources},} Environmental Research Letters, 15, 071002, \dodoi{10.1088/1748-9326/ab9ed2}

\bibitem[{W. {Jaeschke} {et~al.}(1976){Jaeschke}, {Schmitt}, \& {Georgii}}]{Jaeschke1976}
{Jaeschke}, W., {Schmitt}, R., \& {Georgii}, H. 1976, \bibinfo{title}{Preliminary results of stratospheric SO2‐measurements,} Geophysical Research Letters, 3, 517–519, \dodoi{10.1029/gl003i009p00517}

\bibitem[{M.~A. {Kahre} {et~al.}(2023){Kahre}, {Haberle}, {Wilson}, {Urata}, {Steakley}, {Brecht}, {Bertrand}, {Kling}, {Batterson}, {Hartwick}, {Harman}, \& {Gkouvelis}}]{Kahre2023}
{Kahre}, M.~A., {Haberle}, R.~M., {Wilson}, R.~J., {et~al.} 2023, \bibinfo{title}{{The NASA Ames legacy Mars global climate model: Radiation code error correction and new baseline water cycle simulation},} \icarus, 400, 115561, \dodoi{10.1016/j.icarus.2023.115561}

\bibitem[{E. {Karkoschka}(1994){Karkoschka}}]{Karkoschka1994}
{Karkoschka}, E. 1994, \bibinfo{title}{{Spectrophotometry of the Jovian Planets and Titan at 300- to 1000-nm Wavelength: The Methane Spectrum},} \icarus, 111, 174, \dodoi{10.1006/icar.1994.1139}

\bibitem[{T. {Karman} {et~al.}(2019){Karman}, {Gordon}, {van der Avoird}, {Baranov}, {Boulet}, {Drouin}, {Groenenboom}, {Gustafsson}, {Hartmann}, {Kurucz}, {Rothman}, {Sun}, {Sung}, {Thalman}, {Tran}, {Wishnow}, {Wordsworth}, {Vigasin}, {Volkamer}, \& {van der Zande}}]{Karman2019}
{Karman}, T., {Gordon}, I.~E., {van der Avoird}, A., {et~al.} 2019, \bibinfo{title}{{Update of the HITRAN collision-induced absorption section},} \icarus, 328, 160, \dodoi{10.1016/j.icarus.2019.02.034}

\bibitem[{J.~F. {Kasting}(1988){Kasting}}]{Kasting1988}
{Kasting}, J.~F. 1988, \bibinfo{title}{{Runaway and moist greenhouse atmospheres and the evolution of Earth and Venus},} \icarus, 74, 472, \dodoi{10.1016/0019-1035(88)90116-9}

\bibitem[{J.~F. {Kasting}(1991){Kasting}}]{Kasting1991}
{Kasting}, J.~F. 1991, \bibinfo{title}{{CO $_{2}$ condensation and the climate of early Mars},} \icarus, 94, 1, \dodoi{10.1016/0019-1035(91)90137-I}

\bibitem[{J.~F. {Kasting} {et~al.}(1979){Kasting}, {Liu}, \& {Donahue}}]{Kasting1979}
{Kasting}, J.~F., {Liu}, S.~C., \& {Donahue}, T.~M. 1979, \bibinfo{title}{{Oxygen levels in the prebiological atmosphere},} \jgr, 84, 3097, \dodoi{10.1029/JC084iC06p03097}

\bibitem[{J.~F. {Kasting} {et~al.}(1984){Kasting}, {Pollack}, \& {Crisp}}]{Kasting1984}
{Kasting}, J.~F., {Pollack}, J.~B., \& {Crisp}, D. 1984, \bibinfo{title}{{Effects of high CO2 levels on surface temperature and atmospheric oxidation state of the early earth},} Journal of Atmospheric Chemistry, 1, 403, \dodoi{10.1007/BF00053803}

\bibitem[{J.~F. {Kasting} {et~al.}(1993){Kasting}, {Whitmire}, \& {Reynolds}}]{Kasting1993}
{Kasting}, J.~F., {Whitmire}, D.~P., \& {Reynolds}, R.~T. 1993, \bibinfo{title}{{Habitable Zones around Main Sequence Stars},} \icarus, 101, 108, \dodoi{10.1006/icar.1993.1010}

\bibitem[{J. Keady \& D. Kilcrease(2002)Keady \& Kilcrease}]{Keady2002}
Keady, J., \& Kilcrease, D. 2002, \bibinfo{title}{Radiation,} in Allen’s Astrophysical Quantities (Springer), 95--120

\bibitem[{R.~J. Kee {et~al.}(2003)Kee, Coltrin, \& Glarborg}]{Kee2003}
Kee, R.~J., Coltrin, M.~E., \& Glarborg, P. 2003, Chemically reacting flow: theory and practice (John Wiley \& Sons)

\bibitem[{H. {Keller-Rudek} {et~al.}(2013){Keller-Rudek}, {Moortgat}, {Sander}, \& {S{\"o}rensen}}]{KellerRudek2013}
{Keller-Rudek}, H., {Moortgat}, G.~K., {Sander}, R., \& {S{\"o}rensen}, R. 2013, \bibinfo{title}{{The MPI-Mainz UV/VIS Spectral Atlas of Gaseous Molecules of Atmospheric Interest},} Earth System Science Data, 5, 365, \dodoi{10.5194/essd-5-365-201310.5194/essdd-6-411-2013}

\bibitem[{B.~N. {Khare} {et~al.}(1984){Khare}, {Sagan}, {Arakawa}, {Suits}, {Callcott}, \& {Williams}}]{Khare1984}
{Khare}, B.~N., {Sagan}, C., {Arakawa}, E.~T., {et~al.} 1984, \bibinfo{title}{{Optical constants of organic tholins produced in a simulated Titanian atmosphere: From soft x-ray to microwave frequencies},} \icarus, 60, 127, \dodoi{10.1016/0019-1035(84)90142-8}

\bibitem[{R.~G. {Knollenberg} \& D.~M. {Hunten}(1980){Knollenberg} \& {Hunten}}]{Knollenberg1980}
{Knollenberg}, R.~G., \& {Hunten}, D.~M. 1980, \bibinfo{title}{{The microphysics of the clouds of Venus - Results of the Pioneer Venus particle size spectrometer experiment},} \jgr, 85, 8039, \dodoi{10.1029/JA085iA13p08039}

\bibitem[{R.~V. {Kochanov} {et~al.}(2016){Kochanov}, {Gordon}, {Rothman}, {Wcis{\l}o}, {Hill}, \& {Wilzewski}}]{Kochanov2016}
{Kochanov}, R.~V., {Gordon}, I.~E., {Rothman}, L.~S., {et~al.} 2016, \bibinfo{title}{{HITRAN Application Programming Interface (HAPI): A comprehensive approach to working with spectroscopic data},} \jqsrt, 177, 15, \dodoi{10.1016/j.jqsrt.2016.03.005}

\bibitem[{R.~K. {Kopparapu} {et~al.}(2013){Kopparapu}, {Ramirez}, {Kasting}, {Eymet}, {Robinson}, {Mahadevan}, {Terrien}, {Domagal-Goldman}, {Meadows}, \& {Deshpande}}]{Kopparapu2013}
{Kopparapu}, R.~K., {Ramirez}, R., {Kasting}, J.~F., {et~al.} 2013, \bibinfo{title}{{Habitable Zones around Main-sequence Stars: New Estimates},} \apj, 765, 131, \dodoi{10.1088/0004-637X/765/2/131}

\bibitem[{T. {Kozakis} {et~al.}(2022){Kozakis}, {Mendon{\c{c}}a}, \& {Buchhave}}]{Kozakis2022}
{Kozakis}, T., {Mendon{\c{c}}a}, J.~M., \& {Buchhave}, L.~A. 2022, \bibinfo{title}{{Is ozone a reliable proxy for molecular oxygen?. I. The O$_{2}$-O$_{3}$ relationship for Earth-like atmospheres},} \aap, 665, A156, \dodoi{10.1051/0004-6361/202244164}

\bibitem[{V.~A. {Krasnopolsky}(2006{\natexlab{a}}){Krasnopolsky}}]{Krasnopolsky2006b}
{Krasnopolsky}, V.~A. 2006{\natexlab{a}}, \bibinfo{title}{{Photochemistry of the martian atmosphere: Seasonal, latitudinal, and diurnal variations},} \icarus, 185, 153, \dodoi{10.1016/j.icarus.2006.06.003}

\bibitem[{V.~A. {Krasnopolsky}(2006{\natexlab{b}}){Krasnopolsky}}]{Krasnopolsky2006}
{Krasnopolsky}, V.~A. 2006{\natexlab{b}}, \bibinfo{title}{A sensitive search for nitric oxide in the lower atmospheres of Venus and Mars: Detection on Venus and upper limit for Mars,} Icarus, 182, 80–91, \dodoi{10.1016/j.icarus.2005.12.003}

\bibitem[{V.~A. {Krasnopolsky}(2007){Krasnopolsky}}]{Krasnopolsky2007}
{Krasnopolsky}, V.~A. 2007, \bibinfo{title}{{Chemical kinetic model for the lower atmosphere of Venus},} \icarus, 191, 25, \dodoi{10.1016/j.icarus.2007.04.028}

\bibitem[{V.~A. {Krasnopolsky}(2008){Krasnopolsky}}]{Krasnopolsky2008}
{Krasnopolsky}, V.~A. 2008, \bibinfo{title}{High-resolution spectroscopy of Venus: Detection of OCS, upper limit to H $_{2}$S, and latitudinal variations of CO and HF in the upper cloud layer,} \icarus, 197, 377, \dodoi{10.1016/j.icarus.2008.05.020}

\bibitem[{V.~A. {Krasnopolsky}(2010{\natexlab{a}}){Krasnopolsky}}]{Krasnopolsky2010b}
{Krasnopolsky}, V.~A. 2010{\natexlab{a}}, \bibinfo{title}{Spatially-resolved high-resolution spectroscopy of Venus 2. Variations of HDO, OCS, and SO $_{2}$ at the cloud tops,} \icarus, 209, 314, \dodoi{10.1016/j.icarus.2010.05.008}

\bibitem[{V.~A. {Krasnopolsky}(2010{\natexlab{b}}){Krasnopolsky}}]{Krasnopolsky2010a}
{Krasnopolsky}, V.~A. 2010{\natexlab{b}}, \bibinfo{title}{Spatially-resolved high-resolution spectroscopy of Venus 1. Variations of CO $_{2}$, CO, HF, and HCl at the cloud tops,} \icarus, 208, 539, \dodoi{10.1016/j.icarus.2010.02.012}

\bibitem[{V.~A. {Krasnopolsky}(2012){Krasnopolsky}}]{Krasnopolsky2012}
{Krasnopolsky}, V.~A. 2012, \bibinfo{title}{{A photochemical model for the Venus atmosphere at 47-112 km},} \icarus, 218, 230, \dodoi{10.1016/j.icarus.2011.11.012}

\bibitem[{V.~A. {Krasnopolsky}(2013){Krasnopolsky}}]{Krasnopolsky2013}
{Krasnopolsky}, V.~A. 2013, \bibinfo{title}{S$_{3}$ and S$_{4}$ abundances and improved chemical kinetic model for the lower atmosphere of Venus,} \icarus, 225, 570, \dodoi{10.1016/j.icarus.2013.04.026}

\bibitem[{V.~A. {Krasnopolsky}(2014){Krasnopolsky}}]{Krasnopolsky2014}
{Krasnopolsky}, V.~A. 2014, \bibinfo{title}{Observations of CO dayglow at 4.7 m, CO mixing ratios, and temperatures at 74 and 104-111 km on Venus,} \icarus, 237, 340, \dodoi{10.1016/j.icarus.2014.04.043}

\bibitem[{V.~A. {Krasnopolsky} \& P.~D. {Feldman}(2001){Krasnopolsky} \& {Feldman}}]{Krasnopolsky2001}
{Krasnopolsky}, V.~A., \& {Feldman}, P.~D. 2001, \bibinfo{title}{Detection of Molecular Hydrogen in the Atmosphere of Mars,} Science, 294, 1914–1917, \dodoi{10.1126/science.1065569}

\bibitem[{V.~A. {Krasnopolsky} \& V.~A. {Parshev}(1981){Krasnopolsky} \& {Parshev}}]{Krasnopolsky1981}
{Krasnopolsky}, V.~A., \& {Parshev}, V.~A. 1981, \bibinfo{title}{{Chemical composition of the atmosphere of Venus},} \nat, 292, 610, \dodoi{10.1038/292610a0}

\bibitem[{J.~R. {Lane} \& H.~G. {Kjaergaard}(2008){Lane} \& {Kjaergaard}}]{Lane2008}
{Lane}, J.~R., \& {Kjaergaard}, H.~G. 2008, \bibinfo{title}{{Calculated Electronic Transitions in Sulfuric Acid and Implications for Its Photodissociation in the Atmosphere},} Journal of Physical Chemistry A, 112, 4958, \dodoi{10.1021/jp710863r}

\bibitem[{P. {Lavvas} {et~al.}(2008){Lavvas}, {Coustenis}, \& {Vardavas}}]{La08}
{Lavvas}, P., {Coustenis}, A., \& {Vardavas}, I. 2008, \bibinfo{title}{Coupling photochemistry with haze formation in Titan’s atmosphere, Part I: Model description,} Planetary and Space Science, 56, 27–66, \dodoi{10.1016/j.pss.2007.05.026}

\bibitem[{P.~P. {Lavvas} {et~al.}(2008){Lavvas}, {Coustenis}, \& {Vardavas}}]{Lavvas2008}
{Lavvas}, P.~P., {Coustenis}, A., \& {Vardavas}, I.~M. 2008, \bibinfo{title}{{Coupling photochemistry with haze formation in Titan's atmosphere, Part I: Model description},} \planss, 56, 27, \dodoi{10.1016/j.pss.2007.05.026}

\bibitem[{J. {Leconte}(2021){Leconte}}]{Leconte2021}
{Leconte}, J. 2021, \bibinfo{title}{{Spectral binning of precomputed correlated-k coefficients},} \aap, 645, A20, \dodoi{10.1051/0004-6361/202039040}

\bibitem[{J. {Leconte} {et~al.}(2024){Leconte}, {Spiga}, {Cl{\'e}ment}, {Guerlet}, {Selsis}, {Milcareck}, {Cavali{\'e}}, {Moreno}, {Lellouch}, {Carri{\'o}n-Gonz{\'a}lez}, {Charnay}, \& {Lef{\`e}vre}}]{Leconte2024}
{Leconte}, J., {Spiga}, A., {Cl{\'e}ment}, N., {et~al.} 2024, \bibinfo{title}{{A 3D picture of moist-convection inhibition in hydrogen-rich atmospheres: Implications for K2-18 b},} \aap, 686, A131, \dodoi{10.1051/0004-6361/202348928}

\bibitem[{A.~Y.~T. {Lee} {et~al.}(2001){Lee}, {Yung}, {Cheng}, {Bahou}, {Chung}, \& {Lee}}]{Lee2001}
{Lee}, A. Y.~T., {Yung}, Y.~L., {Cheng}, B.-M., {et~al.} 2001, \bibinfo{title}{{Enhancement of Deuterated Ethane on Jupiter},} \apjl, 551, L93, \dodoi{10.1086/319827}

\bibitem[{Y.~J. {Lee} {et~al.}(2016){Lee}, {Sagawa}, {Haus}, {Stefani}, {Imamura}, {Titov}, \& {Piccioni}}]{Lee2016}
{Lee}, Y.~J., {Sagawa}, H., {Haus}, R., {et~al.} 2016, \bibinfo{title}{{Sensitivity of net thermal flux to the abundance of trace gases in the lower atmosphere of Venus},} Journal of Geophysical Research (Planets), 121, 1737, \dodoi{10.1002/2016JE005087}

\bibitem[{E. {Lei} \& P. {Molli{\`e}re}(2024){Lei} \& {Molli{\`e}re}}]{Lei2024}
{Lei}, E., \& {Molli{\`e}re}, P. 2024, \bibinfo{title}{{easyCHEM: A Python package for calculating chemical equilibrium abundances in exoplanet atmospheres},} arXiv e-prints, arXiv:2410.21364, \dodoi{10.48550/arXiv.2410.21364}

\bibitem[{R.~J. LeVeque {et~al.}(2002)LeVeque {et~al.}}]{Leveque2002}
LeVeque, R.~J., {et~al.} 2002, Finite volume methods for hyperbolic problems, Vol.~31 (Cambridge university press)

\bibitem[{G. {Li} {et~al.}(2015){Li}, {Gordon}, {Rothman}, {Tan}, {Hu}, {Kassi}, {Campargue}, \& {Medvedev}}]{Li2015}
{Li}, G., {Gordon}, I.~E., {Rothman}, L.~S., {et~al.} 2015, \bibinfo{title}{{Rovibrational Line Lists for Nine Isotopologues of the CO Molecule in the X $^{1}${\ensuremath{\Sigma}}$^{+}$ Ground Electronic State},} \apjs, 216, 15, \dodoi{10.1088/0067-0049/216/1/15}

\bibitem[{L. {Li} {et~al.}(2018){Li}, {Jiang}, {West}, {Gierasch}, {Perez-Hoyos}, {Sanchez-Lavega}, {Fletcher}, {Fortney}, {Knowles}, {Porco}, {Baines}, {Fry}, {Mallama}, {Achterberg}, {Simon}, {Nixon}, {Orton}, {Dyudina}, {Ewald}, \& {Schmude}}]{Li2018}
{Li}, L., {Jiang}, X., {West}, R.~A., {et~al.} 2018, \bibinfo{title}{{Less absorbed solar energy and more internal heat for Jupiter},} Nature Communications, 9, 3709, \dodoi{10.1038/s41467-018-06107-2}

\bibitem[{Q. {Li} {et~al.}(2003){Li}, {Jacob}, {Yantosca}, {Heald}, {Singh}, {Koike}, {Zhao}, {Sachse}, \& {Streets}}]{Li2003}
{Li}, Q., {Jacob}, D.~J., {Yantosca}, R.~M., {et~al.} 2003, \bibinfo{title}{{A global three-dimensional model analysis of the atmospheric budgets of HCN and CH$_{3}$CN: Constraints from aircraft and ground measurements},} Journal of Geophysical Research (Atmospheres), 108, 8827, \dodoi{10.1029/2002JD003075}

\bibitem[{A.~P. {Lincowski} {et~al.}(2018){Lincowski}, {Meadows}, {Crisp}, {Robinson}, {Luger}, {Lustig-Yaeger}, \& {Arney}}]{Lincowski2018}
{Lincowski}, A.~P., {Meadows}, V.~S., {Crisp}, D., {et~al.} 2018, \bibinfo{title}{{Evolved Climates and Observational Discriminants for the TRAPPIST-1 Planetary System},} \apj, 867, 76, \dodoi{10.3847/1538-4357/aae36a}

\bibitem[{A.~P. {Lincowski} {et~al.}(2023){Lincowski}, {Meadows}, {Zieba}, {Kreidberg}, {Morley}, {Gillon}, {Selsis}, {Agol}, {Bolmont}, {Ducrot}, {Hu}, {Koll}, {Lyu}, {Mandell}, {Suissa}, \& {Tamburo}}]{Lincowski2023}
{Lincowski}, A.~P., {Meadows}, V.~S., {Zieba}, S., {et~al.} 2023, \bibinfo{title}{{Potential Atmospheric Compositions of TRAPPIST-1 c Constrained by JWST/MIRI Observations at 15 {\ensuremath{\mu}}m},} \apjl, 955, L7, \dodoi{10.3847/2041-8213/acee02}

\bibitem[{K. {Lodders}(2019){Lodders}}]{Lodders2019}
{Lodders}, K. 2019, \bibinfo{title}{{Solar Elemental Abundances},} arXiv e-prints, arXiv:1912.00844, \dodoi{10.48550/arXiv.1912.00844}

\bibitem[{J.~C. {Loison} {et~al.}(2015){Loison}, {H{\'e}brard}, {Dobrijevic}, {Hickson}, {Caralp}, {Hue}, {Gronoff}, {Venot}, \& {B{\'e}nilan}}]{Loison2015}
{Loison}, J.~C., {H{\'e}brard}, E., {Dobrijevic}, M., {et~al.} 2015, \bibinfo{title}{{The neutral photochemistry of nitriles, amines and imines in the atmosphere of Titan},} \icarus, 247, 218, \dodoi{10.1016/j.icarus.2014.09.039}

\bibitem[{M.~A. {L{\'o}pez-Valverde} {et~al.}(2023){L{\'o}pez-Valverde}, {Funke}, {Brines}, {Stolzenbach}, {Modak}, {Hill}, {Gonz{\'a}lez-Galindo}, {Thomas}, {Trompet}, {Aoki}, {Villanueva}, {Liuzzi}, {Erwin}, {Grabowski}, {Forget}, {L{\'o}pez-Moreno}, {Rodriguez-G{\'o}mez}, {Ristic}, {Daerden}, {Bellucci}, {Patel}, \& {Vandaele}}]{LopezValverde2023}
{L{\'o}pez-Valverde}, M.~A., {Funke}, B., {Brines}, A., {et~al.} 2023, \bibinfo{title}{{Martian Atmospheric Temperature and Density Profiles During the First Year of NOMAD/TGO Solar Occultation Measurements},} Journal of Geophysical Research (Planets), 128, e2022JE007278, \dodoi{10.1029/2022JE007278}

\bibitem[{B. {Maiorov} {et~al.}(2005){Maiorov}, {Ignat'ev}, {Moroz}, {Zasova}, {Moshkin}, {Khatuntsev}, \& {Ekonomov}}]{Maiorov2005}
{Maiorov}, B., {Ignat'ev}, N., {Moroz}, V., {et~al.} 2005, \bibinfo{title}{A New Analysis of the Spectra Obtained by the Venera Missions in the Venusian Atmosphere. I. The Analysis of the Data Received from the Venera-11 Probe at Altitudes Below 37 km in the 0.44 0.66 $\mu$m Wavelength Range,} Solar System Research, 39, 267, \dodoi{10.1007/s11208-005-0042-1}

\bibitem[{M. {Malik} {et~al.}(2019{\natexlab{a}}){Malik}, {Kempton}, {Koll}, {Mansfield}, {Bean}, \& {Kite}}]{Malik2019b}
{Malik}, M., {Kempton}, E. M.~R., {Koll}, D. D.~B., {et~al.} 2019{\natexlab{a}}, \bibinfo{title}{{Analyzing Atmospheric Temperature Profiles and Spectra of M Dwarf Rocky Planets},} \apj, 886, 142, \dodoi{10.3847/1538-4357/ab4a05}

\bibitem[{M. {Malik} {et~al.}(2019{\natexlab{b}}){Malik}, {Kitzmann}, {Mendon{\c{c}}a}, {Grimm}, {Marleau}, {Linder}, {Tsai}, \& {Heng}}]{Malik2019a}
{Malik}, M., {Kitzmann}, D., {Mendon{\c{c}}a}, J.~M., {et~al.} 2019{\natexlab{b}}, \bibinfo{title}{{Self-luminous and Irradiated Exoplanetary Atmospheres Explored with HELIOS},} \aj, 157, 170, \dodoi{10.3847/1538-3881/ab1084}

\bibitem[{M. {Malik} {et~al.}(2017){Malik}, {Grosheintz}, {Mendon{\c{c}}a}, {Grimm}, {Lavie}, {Kitzmann}, {Tsai}, {Burrows}, {Kreidberg}, {Bedell}, {Bean}, {Stevenson}, \& {Heng}}]{Malik2017}
{Malik}, M., {Grosheintz}, L., {Mendon{\c{c}}a}, J.~M., {et~al.} 2017, \bibinfo{title}{{HELIOS: An Open-source, GPU-accelerated Radiative Transfer Code for Self-consistent Exoplanetary Atmospheres},} \aj, 153, 56, \dodoi{10.3847/1538-3881/153/2/56}

\bibitem[{S. {Manabe} \& R.~T. {Wetherald}(1967){Manabe} \& {Wetherald}}]{Manabe1967}
{Manabe}, S., \& {Wetherald}, R.~T. 1967, \bibinfo{title}{{Thermal Equilibrium of the Atmosphere with a Given Distribution of Relative Humidity.},} Journal of the Atmospheric Sciences, 24, 241, \dodoi{10.1175/1520-0469(1967)024<0241:TEOTAW>2.0.CO;2}

\bibitem[{L. {Mancini} {et~al.}(2018){Mancini}, {Esposito}, {Covino}, {Southworth}, {Biazzo}, {Bruni}, {Ciceri}, {Evans}, {Lanza}, {Poretti}, {Sarkis}, {Smith}, {Brogi}, {Affer}, {Benatti}, {Bignamini}, {Boccato}, {Bonomo}, {Borsa}, {Carleo}, {Claudi}, {Cosentino}, {Damasso}, {Desidera}, {Giacobbe}, {Gonz{\'a}lez-{\'A}lvarez}, {Gratton}, {Harutyunyan}, {Leto}, {Maggio}, {Malavolta}, {Maldonado}, {Martinez-Fiorenzano}, {Masiero}, {Micela}, {Molinari}, {Nascimbeni}, {Pagano}, {Pedani}, {Piotto}, {Rainer}, {Scandariato}, {Smareglia}, {Sozzetti}, {Andreuzzi}, \& {Henning}}]{Mancini2018}
{Mancini}, L., {Esposito}, M., {Covino}, E., {et~al.} 2018, \bibinfo{title}{{The GAPS programme with HARPS-N at TNG. XVI. Measurement of the Rossiter-McLaughlin effect of transiting planetary systems HAT-P-3, HAT-P-12, HAT-P-22, WASP-39, and WASP-60},} \aap, 613, A41, \dodoi{10.1051/0004-6361/201732234}

\bibitem[{E. {Marcq} {et~al.}(2008){Marcq}, {Bézard}, {Drossart}, {Piccioni}, {Reess}, \& {Henry}}]{Marcq2008}
{Marcq}, E., {Bézard}, B., {Drossart}, P., {et~al.} 2008, \bibinfo{title}{A latitudinal survey of CO, OCS, H2O, and SO2 in the lower atmosphere of Venus: Spectroscopic studies using VIRTIS‐H,} Journal of Geophysical Research: Planets, 113, \dodoi{10.1029/2008je003074}

\bibitem[{E. {Marcq} {et~al.}(2005){Marcq}, {Bézard}, {Encrenaz}, \& {Birlan}}]{Marcq2005}
{Marcq}, E., {Bézard}, B., {Encrenaz}, T., \& {Birlan}, M. 2005, \bibinfo{title}{Latitudinal variations of CO and OCS in the lower atmosphere of Venus from near-infrared nightside spectro-imaging,} \icarus, 179, 375, \dodoi{10.1016/j.icarus.2005.06.018}

\bibitem[{E. {Marcq} {et~al.}(2006){Marcq}, {Encrenaz}, {Bézard}, \& {Birlan}}]{Marcq2006}
{Marcq}, E., {Encrenaz}, T., {Bézard}, B., \& {Birlan}, M. 2006, \bibinfo{title}{Remote sensing of Venus’ lower atmosphere from ground-based IR spectroscopy: Latitudinal and vertical distribution of minor species,} \planss, 54, 1360, \dodoi{10.1016/j.pss.2006.04.024}

\bibitem[{E. {Marcq} {et~al.}(2015){Marcq}, {Lellouch}, {Encrenaz}, {Widemann}, {Birlan}, \& {Bertaux}}]{Marcq2015}
{Marcq}, E., {Lellouch}, E., {Encrenaz}, T., {et~al.} 2015, \bibinfo{title}{Search for horizontal and vertical variations of CO in the day and night side lower mesosphere of Venus from CSHELL/IRTF 4.53 m observations,} \planss, 113, 256, \dodoi{10.1016/j.pss.2014.12.013}

\bibitem[{E. {Marcq} {et~al.}(2018){Marcq}, {Mills}, {Parkinson}, \& {Vandaele}}]{Marcq2018}
{Marcq}, E., {Mills}, F.~P., {Parkinson}, C.~D., \& {Vandaele}, A.~C. 2018, \bibinfo{title}{Composition and Chemistry of the Neutral Atmosphere of Venus,} \ssr, 214, 10, \dodoi{10.1007/s11214-017-0438-5}

\bibitem[{A. {Marten}(2002){Marten}}]{Marten2002}
{Marten}, A. 2002, \bibinfo{title}{New Millimeter Heterodyne Observations of Titan: Vertical Distributions of Nitriles HCN, HC3N, CH3CN, and the Isotopic Ratio 15N/14N in Its Atmosphere,} Icarus, 158, 532–544, \dodoi{10.1006/icar.2002.6897}

\bibitem[{S.~T. {Massie} \& D.~M. {Hunten}(1981){Massie} \& {Hunten}}]{Massie1981}
{Massie}, S.~T., \& {Hunten}, D.~M. 1981, \bibinfo{title}{Stratospheric eddy diffusion coefficients from tracer data,} Journal of Geophysical Research: Oceans, 86, 9859–9868, \dodoi{10.1029/jc086ic10p09859}

\bibitem[{Y. {Matsumi} {et~al.}(2002){Matsumi}, {Comes}, {Hancock}, {Hofzumahaus}, {Hynes}, {Kawasaki}, \& {Ravishankara}}]{Matsumi2002}
{Matsumi}, Y., {Comes}, F.~J., {Hancock}, G., {et~al.} 2002, \bibinfo{title}{{Quantum yields for production of O($^{1}$D) in the ultraviolet photolysis of ozone: Recommendation based on evaluation of laboratory data},} Journal of Geophysical Research (Atmospheres), 107, 4024, \dodoi{10.1029/2001JD000510}

\bibitem[{C.~P. {McKay} {et~al.}(2001){McKay}, {Coustenis}, {Samuelson}, {Lemmon}, {Lorenz}, {Cabane}, {Rannou}, \& {Drossart}}]{McKay2001}
{McKay}, C.~P., {Coustenis}, A., {Samuelson}, R.~E., {et~al.} 2001, \bibinfo{title}{{Physical properties of the organic aerosols and clouds on Titan},} \planss, 49, 79, \dodoi{10.1016/S0032-0633(00)00051-9}

\bibitem[{C.~P. {McKay} {et~al.}(1989){McKay}, {Pollack}, \& {Courtin}}]{McKay1989}
{McKay}, C.~P., {Pollack}, J.~B., \& {Courtin}, R. 1989, \bibinfo{title}{{The thermal structure of Titan's atmosphere},} \icarus, 80, 23, \dodoi{10.1016/0019-1035(89)90160-7}

\bibitem[{C.~P. {McKay} {et~al.}(1991){McKay}, {Pollack}, \& {Courtin}}]{McKay1991}
{McKay}, C.~P., {Pollack}, J.~B., \& {Courtin}, R. 1991, \bibinfo{title}{{The Greenhouse and Antigreenhouse Effects on Titan},} Science, 253, 1118, \dodoi{10.1126/science.11538492}

\bibitem[{V. {Meadows} \& D. {Crisp}(1996){Meadows} \& {Crisp}}]{Meadows1996}
{Meadows}, V., \& {Crisp}, D. 1996, \bibinfo{title}{Ground-based near-infrared observations of the Venus nightside: The thermal structure and water abundance near the surface,} \jgr, 101, 4595, \dodoi{10.1029/95JE03567}

\bibitem[{V.~S. {Meadows} {et~al.}(2018){Meadows}, {Reinhard}, {Arney}, {Parenteau}, {Schwieterman}, {Domagal-Goldman}, {Lincowski}, {Stapelfeldt}, {Rauer}, {DasSarma}, {Hegde}, {Narita}, {Deitrick}, {Lustig-Yaeger}, {Lyons}, {Siegler}, \& {Grenfell}}]{Meadows2018}
{Meadows}, V.~S., {Reinhard}, C.~T., {Arney}, G.~N., {et~al.} 2018, \bibinfo{title}{{Exoplanet Biosignatures: Understanding Oxygen as a Biosignature in the Context of Its Environment},} Astrobiology, 18, 630, \dodoi{10.1089/ast.2017.1727}

\bibitem[{F.~P. {Mills} \& M. {Allen}(2007){Mills} \& {Allen}}]{Mills2007}
{Mills}, F.~P., \& {Allen}, M. 2007, \bibinfo{title}{{A review of selected issues concerning the chemistry in Venus{\textquoteright} middle atmosphere},} \planss, 55, 1729, \dodoi{10.1016/j.pss.2007.01.012}

\bibitem[{C. {Moeckel} {et~al.}(2023){Moeckel}, {de Pater}, \& {DeBoer}}]{Moeckel2023}
{Moeckel}, C., {de Pater}, I., \& {DeBoer}, D. 2023, \bibinfo{title}{Ammonia Abundance Derived from Juno MWR and VLA Observations of Jupiter,} The Planetary Science Journal, 4, 25, \dodoi{10.3847/psj/acaf6b}

\bibitem[{P. {Molli{\`e}re} {et~al.}(2019){Molli{\`e}re}, {Wardenier}, {van Boekel}, {Henning}, {Molaverdikhani}, \& {Snellen}}]{Molliere2019}
{Molli{\`e}re}, P., {Wardenier}, J.~P., {van Boekel}, R., {et~al.} 2019, \bibinfo{title}{{petitRADTRANS. A Python radiative transfer package for exoplanet characterization and retrieval},} \aap, 627, A67, \dodoi{10.1051/0004-6361/201935470}

\bibitem[{L. {Montabone} {et~al.}(2015){Montabone}, {Forget}, {Millour}, {Wilson}, {Lewis}, {Cantor}, {Kass}, {Kleinb{\"o}hl}, {Lemmon}, {Smith}, \& {Wolff}}]{Montabone2015}
{Montabone}, L., {Forget}, F., {Millour}, E., {et~al.} 2015, \bibinfo{title}{{Eight-year climatology of dust optical depth on Mars},} \icarus, 251, 65, \dodoi{10.1016/j.icarus.2014.12.034}

\bibitem[{R. {Moreno} {et~al.}(2012){Moreno}, {Lellouch}, {Lara}, {Feuchtgruber}, {Rengel}, {Hartogh}, \& {Courtin}}]{Moreno2012}
{Moreno}, R., {Lellouch}, E., {Lara}, L.~M., {et~al.} 2012, \bibinfo{title}{{The abundance, vertical distribution and origin of H$_{2}$O in Titan{\textquoteright}s atmosphere: Herschel observations and photochemical modelling},} \icarus, 221, 753, \dodoi{10.1016/j.icarus.2012.09.006}

\bibitem[{V. {Moroz} {et~al.}(1979){Moroz}, {Moshkin}, {Ekonomov}, {Sanko}, {Parfentev}, \& {Golovin}}]{Moroz1979}
{Moroz}, V., {Moshkin}, B., {Ekonomov}, A., {et~al.} 1979, \bibinfo{title}{Venera 11 and 12 descent-probe spectrophotometry - The Venus dayside sky spectrum,} Soviet Astronomy Letters, 5, 118

\bibitem[{V. {Moroz} {et~al.}(1990){Moroz}, {Spankuch}, {Titov}, {Schafer}, {Dyachkov}, {Dohler}, {Zasova}, {Oertel}, {Linkin}, \& {Nopirakowski}}]{Moroz1990}
{Moroz}, V., {Spankuch}, D., {Titov}, D., {et~al.} 1990, \bibinfo{title}{Water vapor and sulfur dioxide abundances at the Venus cloud tops from the Venera-15 infrared spectrometry data,} Advances in Space Research, 10, 77, \dodoi{10.1016/0273-1177(90)90168-Y}

\bibitem[{J.~J. Moré {et~al.}(1980)Moré, Garbow, \& Hillstrom}]{More1980}
Moré, J.~J., Garbow, B.~S., \& Hillstrom, K.~E. 1980, User Guide for MINPACK-1, Tech. Rep. ANL-80-74, Argonne National Laboratory, Argonne, IL.
\newblock \url{https://www.osti.gov/biblio/6997562}

\bibitem[{J.~I. {Moses} {et~al.}(2005){Moses}, {Fouchet}, {Bézard}, {Gladstone}, {Lellouch}, \& {Feuchtgruber}}]{Moses2005}
{Moses}, J.~I., {Fouchet}, T., {Bézard}, B., {et~al.} 2005, \bibinfo{title}{Photochemistry and diffusion in Jupiter’s stratosphere: Constraints from ISO observations and comparisons with other giant planets,} Journal of Geophysical Research: Planets, 110, \dodoi{10.1029/2005je002411}

\bibitem[{J.~I. {Moses} {et~al.}(2010){Moses}, {Visscher}, {Keane}, \& {Sperier}}]{Moses2010}
{Moses}, J.~I., {Visscher}, C., {Keane}, T.~C., \& {Sperier}, A. 2010, \bibinfo{title}{{On the abundance of non-cometary HCN on Jupiter},} Faraday Discussions, 147, 103, \dodoi{10.1039/c003954c}

\bibitem[{G.~H. {Mount} {et~al.}(1977){Mount}, {Warden}, \& {Moos}}]{Mount1977}
{Mount}, G.~H., {Warden}, E.~S., \& {Moos}, H.~W. 1977, \bibinfo{title}{{Photoabsorption cross section of methane from 1400 to 1850 {\r{A}}.},} \apjl, 214, L47, \dodoi{10.1086/182440}

\bibitem[{S. {Mukherjee} {et~al.}(2023){Mukherjee}, {Batalha}, {Fortney}, \& {Marley}}]{Mukherjee2023}
{Mukherjee}, S., {Batalha}, N.~E., {Fortney}, J.~J., \& {Marley}, M.~S. 2023, \bibinfo{title}{{PICASO 3.0: A One-dimensional Climate Model for Giant Planets and Brown Dwarfs},} \apj, 942, 71, \dodoi{10.3847/1538-4357/ac9f48}

\bibitem[{S. {Mukherjee} {et~al.}(2024){Mukherjee}, {Fortney}, {Wogan}, {Sing}, \& {Ohno}}]{Mukherjee2024}
{Mukherjee}, S., {Fortney}, J.~J., {Wogan}, N.~F., {Sing}, D.~K., \& {Ohno}, K. 2024, \bibinfo{title}{{Effects of Planetary Parameters on Disequilibrium Chemistry in Irradiated Planetary Atmospheres: From Gas Giants to Sub-Neptunes},} arXiv e-prints, arXiv:2410.17169, \dodoi{10.48550/arXiv.2410.17169}

\bibitem[{S. {Mukherjee} {et~al.}(2025){Mukherjee}, {Schlawin}, {Bell}, {Fortney}, {Beatty}, {Greene}, {Ohno}, {Murphy}, {Parmentier}, {Line}, {Welbanks}, {Wiser}, \& {Rieke}}]{Mukherjee2025}
{Mukherjee}, S., {Schlawin}, E., {Bell}, T.~J., {et~al.} 2025, \bibinfo{title}{{A JWST Panchromatic Thermal Emission Spectrum of the Warm Neptune Archetype GJ 436b},} arXiv e-prints, arXiv:2502.17418, \dodoi{10.48550/arXiv.2502.17418}

\bibitem[{L. {Mukhin} {et~al.}(1982){Mukhin}, {Gelman}, {Lamonov}, {Melnikov}, {Nenarokov}, {Okhotnikov}, {Rotin}, \& {Khokhlov}}]{Mukhin1982}
{Mukhin}, L., {Gelman}, B., {Lamonov}, N., {et~al.} 1982, \bibinfo{title}{VENERA-13 and VENERA-14 Gas Chromatography Analysis of the Venus Atmosphere Composition,} Soviet Astronomy Letters, 8, 216

\bibitem[{W.~F. {Murphy}(1977){Murphy}}]{Murphy1977}
{Murphy}, W.~F. 1977, \bibinfo{title}{{The Rayleigh depolarization ratio and rotational Raman spectrum of water vapor and the polarizability components for the water molecule},} \jcp, 67, 5877, \dodoi{10.1063/1.434794}

\bibitem[{C. {Na} {et~al.}(1990){Na}, {Esposito}, \& {Skinner}}]{Na1990}
{Na}, C., {Esposito}, L., \& {Skinner}, T. 1990, \bibinfo{title}{International Ultraviolet Explorer observation of Venus SO$_{2}$ and SO.,} \jgr, 95, 7485, \dodoi{10.1029/JD095iD06p07485}

\bibitem[{J.~M. {Nicovich} {et~al.}(1990){Nicovich}, {Kreutter}, \& {Wine}}]{Nicovich1990}
{Nicovich}, J.~M., {Kreutter}, K.~D., \& {Wine}, P.~H. 1990, \bibinfo{title}{{Kinetics and thermochemistry of ClCO formation from the Cl+CO association reaction},} \jcp, 92, 3539, \dodoi{10.1063/1.457862}

\bibitem[{C.~A. {Nixon} {et~al.}(2013){Nixon}, {Jennings}, {Bézard}, {Vinatier}, {Teanby}, {Sung}, {Ansty}, {Irwin}, {Gorius}, {Cottini}, {Coustenis}, \& {Flasar}}]{Nixon2013}
{Nixon}, C.~A., {Jennings}, D.~E., {Bézard}, B., {et~al.} 2013, \bibinfo{title}{DETECTION OF PROPENE IN TITAN’S STRATOSPHERE,} The Astrophysical Journal, 776, L14, \dodoi{10.1088/2041-8205/776/1/l14}

\bibitem[{H. {Okabe}(1983){Okabe}}]{Okabe1983}
{Okabe}, H. 1983, \bibinfo{title}{{Photochemistry of acetylene at 1849 \&Aring},} \jcp, 78, 1312, \dodoi{10.1063/1.444868}

\bibitem[{J. {Oschlisniok} {et~al.}(2012){Oschlisniok}, {Häusler}, {Pätzold}, {Tyler}, {Bird}, {Tellmann}, {Remus}, \& {Andert}}]{Oschlisniok2012}
{Oschlisniok}, J., {Häusler}, B., {Pätzold}, M., {et~al.} 2012, \bibinfo{title}{Microwave absorptivity by sulfuric acid in the Venus atmosphere: First results from the Venus Express Radio Science experiment VeRa,} \icarus, 221, 940, \dodoi{10.1016/j.icarus.2012.09.029}

\bibitem[{V. {Oyama} {et~al.}(1979){Oyama}, {Carle}, {Woeller}, \& {Pollack}}]{Oyama1979}
{Oyama}, V., {Carle}, G., {Woeller}, F., \& {Pollack}, J. 1979, \bibinfo{title}{Venus Lower Atmospheric Composition: Analysis by Gas Chromatography,} Science, 203, 802, \dodoi{10.1126/science.203.4382.802}

\bibitem[{V. {Oyama} {et~al.}(1980){Oyama}, {Carle}, {Woeller}, {Pollack}, {Reynolds}, \& {Craig}}]{Oyama1980}
{Oyama}, V., {Carle}, G., {Woeller}, F., {et~al.} 1980, \bibinfo{title}{Pioneer Venus gas chromatography of the lower atmosphere of Venus,} \jgr, 85, 7891, \dodoi{10.1029/JA085iA13p07891}

\bibitem[{K.~F. {Palmer} \& D. {Williams}(1975){Palmer} \& {Williams}}]{Palmer1975}
{Palmer}, K.~F., \& {Williams}, D. 1975, \bibinfo{title}{{Optical constants of sulfuric acid; application to the clouds of Venus?},} \ao, 14, 208, \dodoi{10.1364/AO.14.000208}

\bibitem[{M.~R. {Patel} {et~al.}(2021){Patel}, {Sellers}, {Mason}, {Holmes}, {Brown}, {Lewis}, {Rajendran}, {Streeter}, {Marriner}, {Hathi}, {Slade}, {Leese}, {Wolff}, {Khayat}, {Smith}, {Aoki}, {Piccialli}, {Vandaele}, {Robert}, {Daerden}, {Thomas}, {Ristic}, {Willame}, {Depiesse}, {Bellucci}, \& {Lopez‐Moreno}}]{Patel2021}
{Patel}, M.~R., {Sellers}, G., {Mason}, J.~P., {et~al.} 2021, \bibinfo{title}{ExoMars TGO/NOMAD‐UVIS Vertical Profiles of Ozone: 1. Seasonal Variation and Comparison to Water,} Journal of Geophysical Research: Planets, 126, \dodoi{10.1029/2021je006837}

\bibitem[{B.~K.~D. {Pearce} {et~al.}(2020){Pearce}, {Molaverdikhani}, {Pudritz}, {Henning}, \& {H{\'e}brard}}]{Pearce2020}
{Pearce}, B. K.~D., {Molaverdikhani}, K., {Pudritz}, R.~E., {Henning}, T., \& {H{\'e}brard}, E. 2020, \bibinfo{title}{{HCN Production in Titan's Atmosphere: Coupling Quantum Chemistry and Disequilibrium Atmospheric Modeling},} \apj, 901, 110, \dodoi{10.3847/1538-4357/abae5c}

\bibitem[{R. {Penndorf}(1957){Penndorf}}]{Penndorf1957}
{Penndorf}, R. 1957, \bibinfo{title}{{Tables of the Refractive Index for standard Air and the Rayleigh Scattering Coefficient for the Spectral Region between 0.2 and 20.0 mgr and Their Application to Atmospheric Optics},} Journal of the Optical Society of America (1917-1983), 47, 176, \dodoi{10.1364/JOSA.47.000176}

\bibitem[{M.~Y. {Perrin} \& J.~M. {Hartmann}(1989){Perrin} \& {Hartmann}}]{Perrin1989}
{Perrin}, M.~Y., \& {Hartmann}, J.~M. 1989, \bibinfo{title}{{Temperature-dependent measurements and modeling of absorption by CO$_{2}$-N$_{2}$ mixtures in the far line-wings of the 4.3 {\ensuremath{\mu}}m CO$_{2}$ band.},} \jqsrt, 42, 311, \dodoi{10.1016/0022-4073(89)90077-0}

\bibitem[{J. {Pollack} {et~al.}(1993){Pollack}, {Dalton}, {Grinspoon}, {Wattson}, {Freedman}, {Crisp}, {Allen}, {Bezard}, {de Bergh}, {Giver}, {Ma}, \& {Tipping}}]{Pollack1993}
{Pollack}, J., {Dalton}, J., {Grinspoon}, D., {et~al.} 1993, \bibinfo{title}{Near-Infrared Light from Venus' Nightside: A Spectroscopic Analysis,} \icarus, 103, 1, \dodoi{10.1006/icar.1993.1055}

\bibitem[{S. {Ranjan} \& D.~D. {Sasselov}(2017){Ranjan} \& {Sasselov}}]{Ranjan2017}
{Ranjan}, S., \& {Sasselov}, D.~D. 2017, \bibinfo{title}{{Constraints on the Early Terrestrial Surface UV Environment Relevant to Prebiotic Chemistry},} Astrobiology, 17, 169, \dodoi{10.1089/ast.2016.1519}

\bibitem[{S. {Ranjan} {et~al.}(2020){Ranjan}, {Schwieterman}, {Harman}, {Fateev}, {Sousa-Silva}, {Seager}, \& {Hu}}]{Ranjan2020}
{Ranjan}, S., {Schwieterman}, E.~W., {Harman}, C., {et~al.} 2020, \bibinfo{title}{{Photochemistry of Anoxic Abiotic Habitable Planet Atmospheres: Impact of New H$_{2}$O Cross Sections},} \apj, 896, 148, \dodoi{10.3847/1538-4357/ab9363}

\bibitem[{J. {Revels} {et~al.}(2016){Revels}, {Lubin}, \& {Papamarkou}}]{Revels2016}
{Revels}, J., {Lubin}, M., \& {Papamarkou}, T. 2016, \bibinfo{title}{Forward-Mode Automatic Differentiation in {J}ulia,} arXiv:1607.07892 [cs.MS].
\newblock \url{https://arxiv.org/abs/1607.07892}

\bibitem[{P.~B. {Rimmer} \& C. {Helling}(2016){Rimmer} \& {Helling}}]{Rimmer2016}
{Rimmer}, P.~B., \& {Helling}, C. 2016, \bibinfo{title}{{A Chemical Kinetics Network for Lightning and Life in Planetary Atmospheres},} \apjs, 224, 9, \dodoi{10.3847/0067-0049/224/1/9}

\bibitem[{P.~B. {Rimmer} {et~al.}(2021){Rimmer}, {Jordan}, {Constantinou}, {Woitke}, {Shorttle}, {Hobbs}, \& {Paschodimas}}]{Rimmer2021}
{Rimmer}, P.~B., {Jordan}, S., {Constantinou}, T., {et~al.} 2021, \bibinfo{title}{{Hydroxide Salts in the Clouds of Venus: Their Effect on the Sulfur Cycle and Cloud Droplet pH},} \psj, 2, 133, \dodoi{10.3847/PSJ/ac0156}

\bibitem[{T.~D. {Robinson} \& D. {Crisp}(2018){Robinson} \& {Crisp}}]{Robinson2018}
{Robinson}, T.~D., \& {Crisp}, D. 2018, \bibinfo{title}{{Linearized Flux Evolution (LiFE): A technique for rapidly adapting fluxes from full-physics radiative transfer models},} \jqsrt, 211, 78, \dodoi{10.1016/j.jqsrt.2018.03.002}

\bibitem[{P.~N. {Romani} {et~al.}(2008){Romani}, {Jennings}, {Bjoraker}, {Sada}, {McCabe}, \& {Boyle}}]{Romani2008}
{Romani}, P.~N., {Jennings}, D.~E., {Bjoraker}, G.~L., {et~al.} 2008, \bibinfo{title}{Temporally varying ethylene emission on Jupiter,} Icarus, 198, 420–434, \dodoi{10.1016/j.icarus.2008.05.027}

\bibitem[{L.~S. {Rothman} {et~al.}(2010){Rothman}, {Gordon}, {Barber}, {Dothe}, {Gamache}, {Goldman}, {Perevalov}, {Tashkun}, \& {Tennyson}}]{Rothman2010}
{Rothman}, L.~S., {Gordon}, I.~E., {Barber}, R.~J., {et~al.} 2010, \bibinfo{title}{{HITEMP, the high-temperature molecular spectroscopic database},} \jqsrt, 111, 2139, \dodoi{10.1016/j.jqsrt.2010.05.001}

\bibitem[{Z. {Rustamkulov} {et~al.}(2023){Rustamkulov}, {Sing}, {Mukherjee}, {May}, {Kirk}, {Schlawin}, {Line}, {Piaulet}, {Carter}, {Batalha}, {Goyal}, {L{\'o}pez-Morales}, {Lothringer}, {MacDonald}, {Moran}, {Stevenson}, {Wakeford}, {Espinoza}, {Bean}, {Batalha}, {Benneke}, {Berta-Thompson}, {Crossfield}, {Gao}, {Kreidberg}, {Powell}, {Cubillos}, {Gibson}, {Leconte}, {Molaverdikhani}, {Nikolov}, {Parmentier}, {Roy}, {Taylor}, {Turner}, {Wheatley}, {Aggarwal}, {Ahrer}, {Alam}, {Alderson}, {Allen}, {Banerjee}, {Barat}, {Barrado}, {Barstow}, {Bell}, {Blecic}, {Brande}, {Casewell}, {Changeat}, {Chubb}, {Crouzet}, {Daylan}, {Decin}, {D{\'e}sert}, {Mikal-Evans}, {Feinstein}, {Flagg}, {Fortney}, {Harrington}, {Heng}, {Hong}, {Hu}, {Iro}, {Kataria}, {Kempton}, {Krick}, {Lendl}, {Lillo-Box}, {Louca}, {Lustig-Yaeger}, {Mancini}, {Mansfield}, {Mayne}, {Miguel}, {Morello}, {Ohno}, {Palle}, {Petit dit de la Roche}, {Rackham}, {Radica}, {Ramos-Rosado}, {Redfield}, {Rogers}, {Shkolnik}, {Southworth}, {Teske}, {Tremblin}, {Tucker}, {Venot}, {Waalkes}, {Welbanks}, {Zhang}, \& {Zieba}}]{Rustamkulov2023}
{Rustamkulov}, Z., {Sing}, D.~K., {Mukherjee}, S., {et~al.} 2023, \bibinfo{title}{{Early Release Science of the exoplanet WASP-39b with JWST NIRSpec PRISM},} \nat, 614, 659, \dodoi{10.1038/s41586-022-05677-y}

\bibitem[{R.~J. Salawitch {et~al.}(2022)Salawitch, McBride, Thompson, Fleming, McKenzie, H., Doherty, \& Fahey}]{Salawitch2022}
Salawitch, R.~J., McBride, L.~A., Thompson, C.~R., {et~al.} 2022, Twenty Questions and Answers About the Ozone Layer: 2022 Update, Tech. rep., World Meteorological Organization

\bibitem[{B. {Sandel} {et~al.}(2015){Sandel}, {Gröller}, {Yelle}, {Koskinen}, {Lewis}, {Bertaux}, {Montmessin}, \& {Quémerais}}]{Sandel2015}
{Sandel}, B., {Gröller}, H., {Yelle}, R., {et~al.} 2015, \bibinfo{title}{Altitude profiles of O2 on Mars from SPICAM stellar occultations,} Icarus, 252, 154–160, \dodoi{10.1016/j.icarus.2015.01.004}

\bibitem[{R. {Sander}(2015){Sander}}]{Sander2015}
{Sander}, R. 2015, \bibinfo{title}{{Compilation of Henry's law constants (version 4.0) for water as solvent},} Atmospheric Chemistry \& Physics, 15, 4399, \dodoi{10.5194/acp-15-4399-201510.5194/acpd-14-29615-2014}

\bibitem[{B.~J. {Sandor} \& R.~T. {Clancy}(2005){Sandor} \& {Clancy}}]{Sandor2005}
{Sandor}, B.~J., \& {Clancy}, R.~T. 2005, \bibinfo{title}{Water vapor variations in the Venus mesosphere from microwave spectra,} \icarus, 177, 129, \dodoi{10.1016/j.icarus.2005.03.020}

\bibitem[{B.~J. {Sandor} \& R.~T. {Clancy}(2012){Sandor} \& {Clancy}}]{Sandor2012}
{Sandor}, B.~J., \& {Clancy}, R.~T. 2012, \bibinfo{title}{Observations of HCl altitude dependence and temporal variation in the 70-100 km mesosphere of Venus,} \icarus, 220, 618, \dodoi{10.1016/j.icarus.2012.05.016}

\bibitem[{J.~A. {Schmidt} {et~al.}(2013){Schmidt}, {Johnson}, \& {Schinke}}]{Schmidt2013}
{Schmidt}, J.~A., {Johnson}, M.~S., \& {Schinke}, R. 2013, \bibinfo{title}{{Carbon dioxide photolysis from 150 to 210 nm: Singlet and triplet channel dynamics, UV-spectrum, and isotope effects},} Proceedings of the National Academy of Science, 110, 17691, \dodoi{10.1073/pnas.1213083110}

\bibitem[{S. {Schmidt} {et~al.}(1998){Schmidt}, {Benter}, \& {Schindler}}]{Schmidt1998}
{Schmidt}, S., {Benter}, T., \& {Schindler}, R.~N. 1998, \bibinfo{title}{{Photodissociation dynamcis of ClO radicals in the range (237{\ensuremath{\leq}} {\ensuremath{\lambda}}{\ensuremath{\leq}}270) nm and at 205 nm and the velocity distribution of O( $^{1}$D) atoms},} Chemical Physics Letters, 282, 292, \dodoi{10.1016/S0009-2614(97)01302-X}

\bibitem[{S.~P. {Schmidt} {et~al.}(2025){Schmidt}, {MacDonald}, {Tsai}, {Radica}, {Wang}, {Ahrer}, {Bell}, {Fisher}, {Thorngren}, {Wogan}, {May}, {Ferrari}, {Bennett}, {Rustamkulov}, {L{\'o}pez-Morales}, \& {Sing}}]{Schmidt2025}
{Schmidt}, S.~P., {MacDonald}, R.~J., {Tsai}, S.-M., {et~al.} 2025, \bibinfo{title}{{A Comprehensive Reanalysis of K2-18 b's JWST NIRISS+NIRSpec Transmission Spectrum},} arXiv e-prints, arXiv:2501.18477, \dodoi{10.48550/arXiv.2501.18477}

\bibitem[{E.~W. Schwieterman {et~al.}(2018)Schwieterman, Kiang, Parenteau, Harman, DasSarma, Fisher, Arney, Hartnett, Reinhard, Olson, Meadows, Cockell, Walker, Grenfell, Hegde, Rugheimer, Hu, \& Lyons}]{Schwieterman2018}
Schwieterman, E.~W., Kiang, N.~Y., Parenteau, M.~N., {et~al.} 2018, \bibinfo{title}{Exoplanet biosignatures: A review of remotely detectable signs of life,} Astrobiol., 18, 663

\bibitem[{E.~W. {Schwieterman} {et~al.}(2022){Schwieterman}, {Olson}, {Pidhorodetska}, {Reinhard}, {Ganti}, {Fauchez}, {Bastelberger}, {Crouse}, {Ridgwell}, \& {Lyons}}]{Schwieterman2022}
{Schwieterman}, E.~W., {Olson}, S.~L., {Pidhorodetska}, D., {et~al.} 2022, \bibinfo{title}{{Evaluating the Plausible Range of N$_{2}$O Biosignatures on Exo-Earths: An Integrated Biogeochemical, Photochemical, and Spectral Modeling Approach},} \apj, 937, 109, \dodoi{10.3847/1538-4357/ac8cfb}

\bibitem[{A. {Segura} {et~al.}(2005){Segura}, {Kasting}, {Meadows}, {Cohen}, {Scalo}, {Crisp}, {Butler}, \& {Tinetti}}]{Segura2005}
{Segura}, A., {Kasting}, J.~F., {Meadows}, V., {et~al.} 2005, \bibinfo{title}{{Biosignatures from Earth-Like Planets Around M Dwarfs},} Astrobiology, 5, 706, \dodoi{10.1089/ast.2005.5.706}

\bibitem[{A. {Seiff} {et~al.}(1985){Seiff}, {Schofield}, {Kliore}, {Taylor}, {Limaye}, {Revercomb}, {Sromovsky}, {Kerzhanovich}, {Moroz}, \& {Marov}}]{Seiff1985}
{Seiff}, A., {Schofield}, J.~T., {Kliore}, A.~J., {et~al.} 1985, \bibinfo{title}{{Models of the structure of the atmosphere of Venus from the surface to 100 kilometers altitude},} Advances in Space Research, 5, 3, \dodoi{10.1016/0273-1177(85)90197-8}

\bibitem[{A. {Seiff} {et~al.}(1998){Seiff}, {Kirk}, {Knight}, {Young}, {Mihalov}, {Young}, {Milos}, {Schubert}, {Blanchard}, \& {Atkinson}}]{Seiff1998}
{Seiff}, A., {Kirk}, D.~B., {Knight}, T. C.~D., {et~al.} 1998, \bibinfo{title}{{Thermal structure of Jupiter's atmosphere near the edge of a 5-{\ensuremath{\mu}}m hot spot in the north equatorial belt},} \jgr, 103, 22857, \dodoi{10.1029/98JE01766}

\bibitem[{J. Seinfeld {et~al.}(2006)Seinfeld, Seinfeld, \& Pandis}]{Seinfeld2006}
Seinfeld, J., Seinfeld, P., \& Pandis, S. 2006, Atmospheric Chemistry and Physics: From Air Pollution to Climate Change, Wiley-Interscience publication (Wiley)

\bibitem[{J.~H. Seinfeld \& S.~N. Pandis(2016)Seinfeld \& Pandis}]{Seinfeld2016}
Seinfeld, J.~H., \& Pandis, S.~N. 2016, Atmospheric chemistry and physics: from air pollution to climate change (John Wiley \& Sons)

\bibitem[{F. {Selsis} {et~al.}(2023){Selsis}, {Leconte}, {Turbet}, {Chaverot}, \& {Bolmont}}]{Selsis2023}
{Selsis}, F., {Leconte}, J., {Turbet}, M., {Chaverot}, G., \& {Bolmont}, {\'E}. 2023, \bibinfo{title}{{A cool runaway greenhouse without surface magma ocean},} \nat, 620, 287, \dodoi{10.1038/s41586-023-06258-3}

\bibitem[{D.~E. {Shemansky}(1972){Shemansky}}]{Shemansky1972}
{Shemansky}, D.~E. 1972, \bibinfo{title}{{CO$_{2}$ Extinction Coefficient 1700-3000 {\r{A}}},} \jcp, 56, 1582, \dodoi{10.1063/1.1677408}

\bibitem[{T.~G. {Slanger} \& G. {Black}(1982){Slanger} \& {Black}}]{Slanger1982}
{Slanger}, T.~G., \& {Black}, G. 1982, \bibinfo{title}{{Photodissociative channels at 1216 {\r{A}} for H$_{2}$O, NH$_{3}$, and CH$_{4}$},} \jcp, 77, 2432, \dodoi{10.1063/1.444111}

\bibitem[{N. Smith {et~al.}(1998)Smith, B{\'e}nilan, \& Bruston}]{Smith1998}
Smith, N., B{\'e}nilan, Y., \& Bruston, P. 1998, \bibinfo{title}{The temperature dependent absorption cross sections of C4H2 at mid ultraviolet wavelengths,} Planetary and space science, 46, 1215

\bibitem[{L.~E. {Sohl} {et~al.}(2024){Sohl}, {Fauchez}, {Domagal-Goldman}, {Christie}, {Deitrick}, {Haqq-Misra}, {Harman}, {Iro}, {Mayne}, {Tsigaridis}, {Villanueva}, {Young}, \& {Chaverot}}]{Sohl2024}
{Sohl}, L.~E., {Fauchez}, T.~J., {Domagal-Goldman}, S., {et~al.} 2024, \bibinfo{title}{{The CUISINES Framework for Conducting Exoplanet Model Intercomparison Projects, Version 1.0},} \psj, 5, 175, \dodoi{10.3847/PSJ/ad5830}

\bibitem[{R. Steudel(2003)Steudel}]{Steudel2003}
Steudel, R., ed. 2003, Topics in Current Chemistry, Vol. 230, Elemental Sulfur and Sulfur‑Rich Compounds I (Springer), \dodoi{10.1007/b12115}

\bibitem[{L.~J. {Stief} {et~al.}(1975){Stief}, {Payne}, \& {Klemm}}]{Stief1975}
{Stief}, L.~J., {Payne}, W.~A., \& {Klemm}, R.~B. 1975, \bibinfo{title}{{A flasch photolysis-resonance fluorescence study of the formation of O($^{1}$D) in the photolysis of water and the reaction of O($^{1}$D) with H$_{2}$, Ar, and He -},} \jcp, 62, 4000, \dodoi{10.1063/1.430323}

\bibitem[{I.~A. {Surkov} {et~al.}(1987){Surkov}, {Shcheglov}, {Ryvkin}, {Sheinin}, \& {Davydov}}]{Surkov1987}
{Surkov}, I.~A., {Shcheglov}, O., {Ryvkin}, M., {Sheinin}, D., \& {Davydov}, N. 1987, \bibinfo{title}{Water vapour distribution in the middle and lower Venus atmosphere.,} Kosmicheskie Issledovaniia, 25, 678

\bibitem[{Y. {Surkov} {et~al.}(1982){Surkov}, {Ivanova}, {Pudov}, {Pavlenko}, {Davydov}, \& {Shejnin}}]{Surkov1982}
{Surkov}, Y., {Ivanova}, V., {Pudov}, A., {et~al.} 1982, \bibinfo{title}{Venera 13 and Venera 14 measurements of the water vapor content in the Venus atmosphere,} Pisma v Astronomicheskii Zhurnal, 8, 411

\bibitem[{N. {Teanby} {et~al.}(2013){Teanby}, {Irwin}, {Nixon}, {Courtin}, {Swinyard}, {Moreno}, {Lellouch}, {Rengel}, \& {Hartogh}}]{Teanby2013}
{Teanby}, N., {Irwin}, P., {Nixon}, C., {et~al.} 2013, \bibinfo{title}{Constraints on Titan’s middle atmosphere ammonia abundance from Herschel/SPIRE sub-millimetre spectra,} Planetary and Space Science, 75, 136–147, \dodoi{10.1016/j.pss.2012.11.008}

\bibitem[{M.~A. {Thompson} {et~al.}(2022){Thompson}, {Krissansen-Totton}, {Wogan}, {Telus}, \& {Fortney}}]{Thompson2022}
{Thompson}, M.~A., {Krissansen-Totton}, J., {Wogan}, N., {Telus}, M., \& {Fortney}, J.~J. 2022, \bibinfo{title}{{The case and context for atmospheric methane as an exoplanet biosignature},} Proceedings of the National Academy of Science, 119, e2117933119, \dodoi{10.1073/pnas.2117933119}

\bibitem[{G. {Thuillier} {et~al.}(2004){Thuillier}, {Floyd}, {Woods}, {Cebula}, {Hilsenrath}, {Hers{\'e}}, \& {Labs}}]{Thuillier2004}
{Thuillier}, G., {Floyd}, L., {Woods}, T.~N., {et~al.} 2004, \bibinfo{title}{{Solar Irradiance Reference Spectra},} in Solar Variability and its Effects on Climate. Geophysical Monograph 141, ed. J.~M. {Pap}, P.~{Fox}, C.~{Frohlich}, H.~S. {Hudson}, J.~{Kuhn}, J.~{McCormack}, G.~{North}, W.~{Sprigg}, \& S.~T. {Wu}, Vol. 141, 171, \dodoi{10.1029/141GM13}

\bibitem[{M.~G. {Tomasko} {et~al.}(2008{\natexlab{a}}){Tomasko}, {B{\'e}zard}, {Doose}, {Engel}, {Karkoschka}, \& {Vinatier}}]{Tomasko2008b}
{Tomasko}, M.~G., {B{\'e}zard}, B., {Doose}, L., {et~al.} 2008{\natexlab{a}}, \bibinfo{title}{{Heat balance in Titan's atmosphere},} \planss, 56, 648, \dodoi{10.1016/j.pss.2007.10.012}

\bibitem[{M.~G. {Tomasko} {et~al.}(2008{\natexlab{b}}){Tomasko}, {Doose}, {Engel}, {Dafoe}, {West}, {Lemmon}, {Karkoschka}, \& {See}}]{Tomasko2008a}
{Tomasko}, M.~G., {Doose}, L., {Engel}, S., {et~al.} 2008{\natexlab{b}}, \bibinfo{title}{{A model of Titan's aerosols based on measurements made inside the atmosphere},} \planss, 56, 669, \dodoi{10.1016/j.pss.2007.11.019}

\bibitem[{M.~G. {Tomasko} {et~al.}(1980){Tomasko}, {Doose}, {Smith}, \& {Odell}}]{Tomasko1980}
{Tomasko}, M.~G., {Doose}, L.~R., {Smith}, P.~H., \& {Odell}, A.~P. 1980, \bibinfo{title}{{Measurements of the flux of sunlight in the atmosphere of Venus},} \jgr, 85, 8167, \dodoi{10.1029/JA085iA13p08167}

\bibitem[{O.~B. {Toon} {et~al.}(1989){Toon}, {McKay}, {Ackerman}, \& {Santhanam}}]{Toon1989}
{Toon}, O.~B., {McKay}, C.~P., {Ackerman}, T.~P., \& {Santhanam}, K. 1989, \bibinfo{title}{{Rapid calculation of radiative heating rates and photodissociation rates in inhomogeneous multiple scattering atmospheres},} \jgr, 94, 16287, \dodoi{10.1029/JD094iD13p16287}

\bibitem[{M.~G. {Trainer} {et~al.}(2019){Trainer}, {Wong}, {McConnochie}, {Franz}, {Atreya}, {Conrad}, {Lefèvre}, {Mahaffy}, {Malespin}, {Manning}, {Martín‐Torres}, {Martínez}, {McKay}, {Navarro‐González}, {Vicente‐Retortillo}, {Webster}, \& {Zorzano}}]{Trainer2019}
{Trainer}, M.~G., {Wong}, M.~H., {McConnochie}, T.~H., {et~al.} 2019, \bibinfo{title}{Seasonal Variations in Atmospheric Composition as Measured in Gale Crater, Mars,} Journal of Geophysical Research: Planets, 124, 3000–3024, \dodoi{10.1029/2019je006175}

\bibitem[{S.-M. {Tsai} {et~al.}(2017){Tsai}, {Lyons}, {Grosheintz}, {Rimmer}, {Kitzmann}, \& {Heng}}]{Tsai2017}
{Tsai}, S.-M., {Lyons}, J.~R., {Grosheintz}, L., {et~al.} 2017, \bibinfo{title}{{VULCAN: An Open-source, Validated Chemical Kinetics Python Code for Exoplanetary Atmospheres},} \apjs, 228, 20, \dodoi{10.3847/1538-4365/228/2/20}

\bibitem[{S.-M. {Tsai} {et~al.}(2021){Tsai}, {Malik}, {Kitzmann}, {Lyons}, {Fateev}, {Lee}, \& {Heng}}]{Tsai2021}
{Tsai}, S.-M., {Malik}, M., {Kitzmann}, D., {et~al.} 2021, \bibinfo{title}{{A Comparative Study of Atmospheric Chemistry with VULCAN},} \apj, 923, 264, \dodoi{10.3847/1538-4357/ac29bc}

\bibitem[{S.-M. {Tsai} {et~al.}(2023){Tsai}, {Lee}, {Powell}, {Gao}, {Zhang}, {Moses}, {H{\'e}brard}, {Venot}, {Parmentier}, {Jordan}, {Hu}, {Alam}, {Alderson}, {Batalha}, {Bean}, {Benneke}, {Bierson}, {Brady}, {Carone}, {Carter}, {Chubb}, {Inglis}, {Leconte}, {Line}, {L{\'o}pez-Morales}, {Miguel}, {Molaverdikhani}, {Rustamkulov}, {Sing}, {Stevenson}, {Wakeford}, {Yang}, {Aggarwal}, {Baeyens}, {Barat}, {de Val-Borro}, {Daylan}, {Fortney}, {France}, {Goyal}, {Grant}, {Kirk}, {Kreidberg}, {Louca}, {Moran}, {Mukherjee}, {Nasedkin}, {Ohno}, {Rackham}, {Redfield}, {Taylor}, {Tremblin}, {Visscher}, {Wallack}, {Welbanks}, {Youngblood}, {Ahrer}, {Batalha}, {Behr}, {Berta-Thompson}, {Blecic}, {Casewell}, {Crossfield}, {Crouzet}, {Cubillos}, {Decin}, {D{\'e}sert}, {Feinstein}, {Gibson}, {Harrington}, {Heng}, {Henning}, {Kempton}, {Krick}, {Lagage}, {Lendl}, {Lothringer}, {Mansfield}, {Mayne}, {Mikal-Evans}, {Palle}, {Schlawin}, {Shorttle}, {Wheatley}, \& {Yurchenko}}]{Tsai2023}
{Tsai}, S.-M., {Lee}, E. K.~H., {Powell}, D., {et~al.} 2023, \bibinfo{title}{{Photochemically produced SO$_{2}$ in the atmosphere of WASP-39b},} \nat, 617, 483, \dodoi{10.1038/s41586-023-05902-2}

\bibitem[{C.~C.~C. {Tsang} {et~al.}(2008){Tsang}, {Irwin}, {Wilson}, {Taylor}, {Lee}, {de Kok}, {Drossart}, {Piccioni}, {Bezard}, \& {Calcutt}}]{Tsang2008}
{Tsang}, C. C.~C., {Irwin}, P. G.~J., {Wilson}, C.~F., {et~al.} 2008, \bibinfo{title}{Tropospheric carbon monoxide concentrations and variability on Venus from Venus Express/VIRTIS‐M observations,} Journal of Geophysical Research: Planets, 113, \dodoi{10.1029/2008je003089}

\bibitem[{G.~L. {Vaghjiani}(1993){Vaghjiani}}]{Vaghjiani1993}
{Vaghjiani}, G.~L. 1993, \bibinfo{title}{{Ultraviolet absorption cross sections for N$_{2}$H$_{4}$ vapor between 191-291 nm and H($^{2}$S) quantum yield in 248 nm photodissociation at 296 K},} \jcp, 98, 2123, \dodoi{10.1063/1.464190}

\bibitem[{I.~M. {Vardavas} \& J.~H. {Carver}(1984){Vardavas} \& {Carver}}]{Vardavas1984}
{Vardavas}, I.~M., \& {Carver}, J.~H. 1984, \bibinfo{title}{{Solar and terrestrial parameterizations for radiative-convective models},} \planss, 32, 1307, \dodoi{10.1016/0032-0633(84)90074-6}

\bibitem[{A.~A. {Viggiano} \& F. {Arnold}(1981){Viggiano} \& {Arnold}}]{Viggiano1981}
{Viggiano}, A.~A., \& {Arnold}, F. 1981, \bibinfo{title}{Extended sulfuric acid vapor concentration measurements in the stratosphere,} Geophysical Research Letters, 8, 583–586, \dodoi{10.1029/gl008i006p00583}

\bibitem[{C. {Visscher} \& J.~I. {Moses}(2011){Visscher} \& {Moses}}]{Visscher2011}
{Visscher}, C., \& {Moses}, J.~I. 2011, \bibinfo{title}{{Quenching of Carbon Monoxide and Methane in the Atmospheres of Cool Brown Dwarfs and Hot Jupiters},} \apj, 738, 72, \dodoi{10.1088/0004-637X/738/1/72}

\bibitem[{C. {Visscher} {et~al.}(2010){Visscher}, {Moses}, \& {Saslow}}]{Visscher2010}
{Visscher}, C., {Moses}, J.~I., \& {Saslow}, S.~A. 2010, \bibinfo{title}{{The deep water abundance on Jupiter: New constraints from thermochemical kinetics and diffusion modeling},} \icarus, 209, 602, \dodoi{10.1016/j.icarus.2010.03.029}

\bibitem[{V. {Vuitton} {et~al.}(2006){Vuitton}, {Yelle}, \& {Anicich (Retired)}}]{Vuitton2006}
{Vuitton}, V., {Yelle}, R.~V., \& {Anicich (Retired)}, V.~G. 2006, \bibinfo{title}{The Nitrogen Chemistry of Titan’s Upper Atmosphere Revealed,} The Astrophysical Journal, 647, L175–L178, \dodoi{10.1086/507467}

\bibitem[{J.~H. {Waite} {et~al.}(2013){Waite}, {Bell}, {Lorenz}, {Achterberg}, \& {Flasar}}]{Waite2013}
{Waite}, J.~H., {Bell}, J., {Lorenz}, R., {Achterberg}, R., \& {Flasar}, F.~M. 2013, \bibinfo{title}{{A model of variability in Titan's atmospheric structure},} \planss, 86, 45, \dodoi{10.1016/j.pss.2013.05.018}

\bibitem[{I.~C. {Walker} {et~al.}(2007){Walker}, {Holland}, {Shaw}, {McEwen}, \& {Guest}}]{Walker2007}
{Walker}, I.~C., {Holland}, D.~M.~P., {Shaw}, D.~A., {McEwen}, I.~J., \& {Guest}, M.~F. 2007, \bibinfo{title}{{The electronic states of cyclopropane studied by VUV absorption and ab initio multireference configuration interaction calculations},} Journal of Physics B Atomic Molecular Physics, 40, 1875, \dodoi{10.1088/0953-4075/40/10/021}

\bibitem[{F.~M. White \& J. Majdalani(2006)White \& Majdalani}]{White2006}
White, F.~M., \& Majdalani, J. 2006, Viscous fluid flow, Vol.~3 (McGraw-Hill New York)

\bibitem[{W. {Wilson} {et~al.}(1981){Wilson}, {Klein}, {Kahar}, {Gulkis}, {Olsen}, \& {Ho}}]{Wilson1981}
{Wilson}, W., {Klein}, M., {Kahar}, R., {et~al.} 1981, \bibinfo{title}{Venus I. Carbon monoxide distribution and molecular-line searches,} \icarus, 45, 624, \dodoi{10.1016/0019-1035(81)90029-4}

\bibitem[{J.~D. {Windsor} {et~al.}(2023){Windsor}, {Robinson}, {Kopparapu}, {Young}, {Trilling}, \& {LLama}}]{Windsor2023}
{Windsor}, J.~D., {Robinson}, T.~D., {Kopparapu}, R.~k., {et~al.} 2023, \bibinfo{title}{{A Radiative-convective Model for Terrestrial Planets with Self-consistent Patchy Clouds},} \psj, 4, 94, \dodoi{10.3847/PSJ/acbf2d}

\bibitem[{N. Wogan(2025{\natexlab{a}})Wogan}]{Differentia2025}
Wogan, N. 2025{\natexlab{a}}, Differentia, v0.1.4 Zenodo, \dodoi{10.5281/zenodo.15678369}

\bibitem[{N. Wogan(2025{\natexlab{b}})Wogan}]{SolarSystemExploration2025}
Wogan, N. 2025{\natexlab{b}}, SolarSystemExploration, v1.0.0 Zenodo, \dodoi{10.5281/zenodo.16509802}

\bibitem[{N. Wogan(2025{\natexlab{c}})Wogan}]{Photochem2025}
Wogan, N. 2025{\natexlab{c}}, Photochem, v0.6.7 Zenodo, \dodoi{10.5281/zenodo.16322107}

\bibitem[{N. Wogan(2025{\natexlab{d}})Wogan}]{Clima2025}
Wogan, N. 2025{\natexlab{d}}, Clima, v0.5.11 Zenodo, \dodoi{10.5281/zenodo.15786151}

\bibitem[{N. Wogan(2025{\natexlab{e}})Wogan}]{Equilibrate2025}
Wogan, N. 2025{\natexlab{e}}, Equilibrate, v0.2.0 Zenodo, \dodoi{10.5281/zenodo.16508386}

\bibitem[{N. Wogan(2025{\natexlab{f}})Wogan}]{PhotochemClimaData2025}
Wogan, N. 2025{\natexlab{f}}, photochem\_clima\_data, v0.3.0 Zenodo, \dodoi{10.5281/zenodo.15785405}

\bibitem[{N. Wogan(2025{\natexlab{g}})Wogan}]{PlanetaryAtmosphereObservations2025}
Wogan, N. 2025{\natexlab{g}}, Planetary Atmosphere Observations, v0.0.3 Zenodo, \dodoi{10.5281/zenodo.16509950}

\bibitem[{N.~F. {Wogan} {et~al.}(2024){Wogan}, {Batalha}, {Zahnle}, {Krissansen-Totton}, {Tsai}, \& {Hu}}]{Wogan2024}
{Wogan}, N.~F., {Batalha}, N.~E., {Zahnle}, K.~J., {et~al.} 2024, \bibinfo{title}{{JWST Observations of K2-18b Can Be Explained by a Gas-rich Mini-Neptune with No Habitable Surface},} \apjl, 963, L7, \dodoi{10.3847/2041-8213/ad2616}

\bibitem[{N.~F. {Wogan} {et~al.}(2022){Wogan}, {Catling}, {Zahnle}, \& {Claire}}]{Wogan2022}
{Wogan}, N.~F., {Catling}, D.~C., {Zahnle}, K.~J., \& {Claire}, M.~W. 2022, \bibinfo{title}{{Rapid timescale for an oxic transition during the Great Oxidation Event and the instability of low atmospheric O$_{2}$},} Proceedings of the National Academy of Science, 119, e2205618119, \dodoi{10.1073/pnas.2205618119}

\bibitem[{N.~F. {Wogan} {et~al.}(2023){Wogan}, {Catling}, {Zahnle}, \& {Lupu}}]{Wogan2023}
{Wogan}, N.~F., {Catling}, D.~C., {Zahnle}, K.~J., \& {Lupu}, R. 2023, \bibinfo{title}{{Origin-of-life Molecules in the Atmosphere after Big Impacts on the Early Earth},} \psj, 4, 169, \dodoi{10.3847/PSJ/aced83}

\bibitem[{N.~F. {Wogan} {et~al.}(2025){Wogan}, {Mang}, {Batalha}, {Zahnle}, {Mukherjee}, {Visscher}, {Fortney}, {Marley}, \& {Morley}}]{Wogan2025}
{Wogan}, N.~F., {Mang}, J., {Batalha}, N.~E., {et~al.} 2025, \bibinfo{title}{{The Sonora Substellar Atmosphere Models. V. A Correction to the Disequilibrium Abundance of CO$_{2}$ for Sonora Elf Owl},} Research Notes of the American Astronomical Society, 9, 108, \dodoi{10.3847/2515-5172/add407}

\bibitem[{M.~J. {Wolff} {et~al.}(2009){Wolff}, {Smith}, {Clancy}, {Arvidson}, {Kahre}, {Seelos}, {Murchie}, \& {Savij{\"a}rvi}}]{Wolff2009}
{Wolff}, M.~J., {Smith}, M.~D., {Clancy}, R.~T., {et~al.} 2009, \bibinfo{title}{{Wavelength dependence of dust aerosol single scattering albedo as observed by the Compact Reconnaissance Imaging Spectrometer},} Journal of Geophysical Research (Planets), 114, E00D04, \dodoi{10.1029/2009JE003350}

\bibitem[{F. {Wunderlich} {et~al.}(2023){Wunderlich}, {Grenfell}, \& {Rauer}}]{Wunderlich2023}
{Wunderlich}, F., {Grenfell}, J.~L., \& {Rauer}, H. 2023, \bibinfo{title}{{Uncertainty in phosphine photochemistry in the Venus atmosphere prevents a firm biosignature attribution},} \aap, 676, A135, \dodoi{10.1051/0004-6361/202142548}

\bibitem[{Q. {Xue} {et~al.}(2024){Xue}, {Bean}, {Zhang}, {Mahajan}, {Ih}, {Eastman}, {Lunine}, {Mansfield}, {Coy}, {Kempton}, {Koll}, \& {Kite}}]{Xue2024}
{Xue}, Q., {Bean}, J.~L., {Zhang}, M., {et~al.} 2024, \bibinfo{title}{{JWST Thermal Emission of the Terrestrial Exoplanet GJ 1132b},} \apjl, 973, L8, \dodoi{10.3847/2041-8213/ad72e9}

\bibitem[{L. {Young}(1972){Young}}]{Young1972}
{Young}, L. 1972, \bibinfo{title}{High resolution spectra of Venus—A review,} \icarus, 17, 632, \dodoi{10.1016/0019-1035(72)90029-2}

\bibitem[{W. {Yu} \& M. {Blair}(2013){Yu} \& {Blair}}]{Yu2013}
{Yu}, W., \& {Blair}, M. 2013, \bibinfo{title}{{DNAD, a simple tool for automatic differentiation of Fortran codes using dual numbers},} Computer Physics Communications, 184, 1446, \dodoi{10.1016/j.cpc.2012.12.025}

\bibitem[{Y.~L. {Yung} \& W.~B. {Demore}(1982){Yung} \& {Demore}}]{Yung1982}
{Yung}, Y.~L., \& {Demore}, W.~B. 1982, \bibinfo{title}{{Photochemistry of the stratosphere of Venus: Implications for atmospheric evolution},} \icarus, 51, 199, \dodoi{10.1016/0019-1035(82)90080-X}

\bibitem[{K. {Zahnle} {et~al.}(2008){Zahnle}, {Haberle}, {Catling}, \& {Kasting}}]{Zahnle2008}
{Zahnle}, K., {Haberle}, R.~M., {Catling}, D.~C., \& {Kasting}, J.~F. 2008, \bibinfo{title}{{Photochemical instability of the ancient Martian atmosphere},} Journal of Geophysical Research (Planets), 113, E11004, \dodoi{10.1029/2008JE003160}

\bibitem[{K. {Zahnle} {et~al.}(2016){Zahnle}, {Marley}, {Morley}, \& {Moses}}]{Zahnle2016}
{Zahnle}, K., {Marley}, M.~S., {Morley}, C.~V., \& {Moses}, J.~I. 2016, \bibinfo{title}{{Photolytic Hazes in the Atmosphere of 51 Eri b},} \apj, 824, 137, \dodoi{10.3847/0004-637X/824/2/137}

\bibitem[{K.~J. {Zahnle}(1986){Zahnle}}]{Zahnle1986}
{Zahnle}, K.~J. 1986, \bibinfo{title}{{Photochemistry of methane and the formation of hydrocyanic acid (HCN) in the Earth's early atmosphere},} \jgr, 91, 2819, \dodoi{10.1029/JD091iD02p02819}

\bibitem[{L. {Zasova} {et~al.}(1993){Zasova}, {Moroz}, {Esposito}, \& {Na}}]{Zasova1993}
{Zasova}, L., {Moroz}, V., {Esposito}, L., \& {Na}, C. 1993, \bibinfo{title}{SO $_{2}$ in the Middle Atmosphere of Venus: IR Measurements from Venera-15 and Comparison to UV Data,} \icarus, 105, 92, \dodoi{10.1006/icar.1993.1113}

\bibitem[{X. {Zhang} {et~al.}(2012){Zhang}, {Liang}, {Mills}, {Belyaev}, \& {Yung}}]{Zhang2012}
{Zhang}, X., {Liang}, M.~C., {Mills}, F.~P., {Belyaev}, D.~A., \& {Yung}, Y.~L. 2012, \bibinfo{title}{{Sulfur chemistry in the middle atmosphere of Venus},} \icarus, 217, 714, \dodoi{10.1016/j.icarus.2011.06.016}

\end{thebibliography}
\bibliographystyle{aasjournalv7}

\end{document}